\pdfoutput=1
\documentclass{aa}  
\usepackage{graphicx}
\usepackage{txfonts}
\usepackage{subcaption}
\usepackage{chngpage}
\usepackage{natbib}
\bibpunct{(}{)}{;}{a}{}{,}
\begin{document}

	\title{A Radio Census of the Massive Stellar Cluster Westerlund 1}
	
	%\subtitle{}
	
	\author{H. Andrews,
		\inst{1}\thanks{e-mail: holly.andrews.16@ucl.ac.uk (UCL)}
		D. Fenech\inst{2},
		R. K. Prinja\inst{1},
		J. S. Clark\inst{3}, \and
		L. Hindson\inst{4}
	}
	
	\institute{University College London, Gower St, Bloomsbury, London WC1E 6BT
		\and
		Cavendish Laboratory, JJ Thomson Avenue, Cambridge CB3 0HE
		\and
		Open University, Walton Hall, Kents Hill, Milton Keynes MK7 6AA
		\and
		University of Hertfordshire, Hatfield, Hertfordshire, AL10 9AB
	}
	
	\date{Received 8th July, 2019; accepted 4th September, 2019}
	
	% \abstract{}{}{}{}{} 
	% 5 {} token are mandatory
	
	\abstract
	%  leave it empty if necessary  
	{Massive stars and their stellar winds are important for a number of feedback processes. The mass lost in the stellar wind can help determine the end-point of the star as a NS or a BH. However, the impact of mass-loss on the post-Main Sequence evolutionary stage of massive stars is not well understood. Westerlund 1 is an ideal astrophysical laboratory in which to study massive stars and their winds in great detail over a large range of different evolutionary phases.}
	{We aim to study the radio emission from Westerlund 1, in order to measure radio fluxes from the population of massive stars, and determine mass-loss rates and spectral indices where possible.}
	{Observations were carried out in 2015 and 2016 with the Australia telescope compact array (ATCA) at 5.5 and 9$\,$GHz using multiple configurations, with maximum baselines ranging from 750$\,$m  to 6$\,$km.}
	{30 stars were detected in the radio from the fully concatenated dataset, 10 of which were WRs (predominantly late type WN stars), 5 YHGs, 4 RSGs, 1 LBV star, the sgB[e] star W9, and several O and B supergiants. New source detections in the radio were found for 5 WR stars, and 5 OB supergiants. These detections have led to evidence for 3 new OB supergiant binary candidates, inferred from derived spectral index limits.}
	{Spectral indices and index limits were determined for massive stars in Westerlund 1. For cluster members found to have partially optically thick emission, mass-loss rates were calculated. Under the approximation of a thermally emitting stellar wind and a steady mass-loss rate, clumping ratios were then estimated for 8 WRs. Diffuse radio emission was detected throughout the cluster. Detections of knots of radio emission with no known stellar counterparts indicate the highly clumped structure of this intra-cluster medium, likely shaped by a dense cluster wind.}
	
	
	\titlerunning{A Radio Census of Westerlund 1}
	\authorrunning{H. Andrews et al.}
	\maketitle
	%
	%________________________________________________________________
	
	\section{Introduction}  
	\defcitealias{Dougher2010}{D10}
	\defcitealias{Fenech2018}{F18}
	Massive stars and the outflows formed from their stellar winds are responsible for a number of important feedback processes, of chemical and mechanical origin, both to their immediate surroundings as well as further afield. Massive stars are predominantly found to form in massive stellar clusters or associations, and so understanding not only the stars themselves and their individual evolution, but also the impact on and from their environment is important for quantifying all the physics involved. This includes the chemical and mechanical feedback from individual stars to their circumstellar environment, as well as larger scale feedback processes, such as the production of cosmic rays, or galactic superwinds. Understanding the cluster environments can also help us to understand how outflows from these massive clusters may trigger or inhibit star formation in nearby regions.
	
	These massive stars are also progenitors of the some of the most exotic endpoints in astrophysics, ending their lives as either a neutron star (NS), or directly collapsing to a black hole (BH), with many massive stars also experiencing a corresponding supernovae (SNe) explosion. This is of particular importance when considering the recent discovery of gravitational waves, which occur due to the mergers of these important astrophysical objects, with progenitors of these mergers believed to be of very large masses themselves \citep{Abbott2016}. These final endpoints are characterised primarily by the initial mass of the star, and the mass lost through its stellar lifetime. The characterisation of the SNe itself can also be found to be influenced by the surrounding environment. By understanding the cluster environments that many of these massive stars live and die in, this can help us to understand the origins of the geometries of SNe remnants observed today, and any associated asymmetries. 
	
	Despite the clear motivations behind understanding this field of astrophysics, the evolutionary pathways of these massive stars are not well understood. Even the consideration of the relative phases of stars beyond the main-sequence, where the most massive stars are able to become hydrogen stripped Wolf-Rayet (WR) stars, are not well understood. The range of evolutionary stages experienced by massive stars have direct consequences on their final endpoints.
	
	Massive stars are impacted by factors such as mass-loss, rotation speeds, and the possible effect of magnetism, as well as binarity. Quantifying the mass loss through stellar winds can help determine the possible evolutionary pathways for different initial stellar masses, helping to constrain and determine possible progenitors for each type of final stellar endpoint. 
	
	Mass-loss rates are not always consistent between those assumed in evolutionary codes and from observations, especially for the case of early-type O and B stars, where discrepancies of up to a factor of 10 have recorded$\,$\citep{Puls2006,Fullerton2006}. One of the ways in which this is impacted is the presence of clumping in the wind, which has a direct impact on the final value of mass-loss determined in the wind, with the necessary inclusion of a clumping factor in calculations quenching the observed mass-loss rates$\,$\citep{Prinja2010,Prinja2013,Sundqvist2011,Surlan2012}.
	
	Another factor that significantly impacts stellar evolution is binarity. Many WR stars have been determined to be located in binary systems. Almost 70\% of massive stars are believed to reside in binary systems$\,$ \citep{Sana2012,Sana2013}, and interactions between the primary and secondary stars, especially for those with close separations, could cause significant levels of mass transfer, affecting the evolution of both stellar components. There are several diagnostic factors by which you can determine the presence of binarity for a stellar system, from radial velocity variations to the presence of hard X-rays. Another way of discovering a binary candidate is by measuring a negative spectral index for the star in the radio regime, indicating the presence of non-thermal emission which may be attributed to colliding stellar winds$\,$\citep{Blomme2010}.
	
	Binarity is also of importance in terms of providing a possible source, via the colliding winds of binaries, as a source for cosmic rays. Single massive stars, especially stars with dense winds, such as WRs, are also a potential source for generating cosmic rays, via shocks between the stellar winds and their circumstellar environments$\,$\citep{Cesarsky1983}. The generation of cosmic rays in the stellar winds of massive stars are believed to be possible in the locations of open clusters in particular, where the regions of interstellar material and a possible strong radiation field can help to sustain cosmic rays produced by stellar sources, either via their winds or from the resultant SNe$\,$\citep{Bednarek2014}.
	
	Westerlund 1$\,$(Wd1) provides a unique astrophysical laboratory in order to investigate the stellar evolution of massive stars at a multitude of evolutionary stages, with hundreds of O and B stars as well as the largest coeval population of yellow hypergiants (YHGs) in our galaxy. Westerlund 1 is the most massive stellar cluster in the Milky Way, discovered in 1961 by Bengt Westerlund$\,$\citep{Westerlund1961}. It has a total mass estimate of $\sim$ 10$^{5}$ M$_{\odot}\,$\citep{Clark2005}. It has a current age estimate of 5$\,$Myr and a distance estimate of 5$\,$kpc. Recent estimates from Gaia DR2 have suggested a smaller distance of $\sim$ 3$\,$kpc may be more appropriate$\,$\citep{Aghakhanloo2019}, but other discussions of the cluster distance in light of Gaia data have suggested significant limitations in the use of Gaia to determine Wd1's distance, due to the extended structure of some of the cluster's brighter members$\,$\citep{Clark2018}. For this paper, the estimate of 5$\,$kpc is used.  
	
	Wd1 has been previously observed across most of the electromagnetic spectrum, with observations carried out in the IR, optical, millimetre, and X-ray \citep{Clark2005,Negueruela2010,Crowther2006,Bonanos2007,Clark2008,Damineli2016,Muno2006}. It has been determined to be a source of highly energetic cosmic rays from the observations of the products of CRs, TeV $\gamma$-rays, observed from the cluster$\,$\citep{Abramowski2012,Aharonian2019}. It has also previously been observed in the radio with the use of the Australia telescope compact array (ATCA) \citep[\citealt{Clark1998}; ][henceforth \citetalias{Dougher2010}]{Dougher2010}, and most recently, in the millimetre \citep[][henceforth \citetalias{Fenech2018}]{Fenech2018}. The observations discussed in this paper provide a direct follow-on from the prior radio observations, with improved sensitivity and resolution.
	
	Section 2 details the observations, as well as the data reduction. Section 3 describes the data analysis then carried out in order to determine fluxes and spatial extents. Section 4 introduces the result of the diffuse radio emission detected throughout the cluster. In Section 5 we introduce the new radio source detections found, separated into stellar sources, uncatalogued sources found in the previous mm-observations, and further previously uncatalogued radio sources determined (with no optical or mm counterpart). Section 6 goes into detail about the results of the stellar sources, considering the WR population, the cool supergiant and hypergiant population, and other stellar sources, including Wd1-9, the luminous blue variable (LBV) Wd1-243 and the O and B supergiants. The extended emission of the cluster and its possible origin is discussed in section 7. A summary of our conclusions and possible future avenues for investigation is then provided in section 8. 
	
	 \begin{figure*}
		\includegraphics[width=\textwidth]{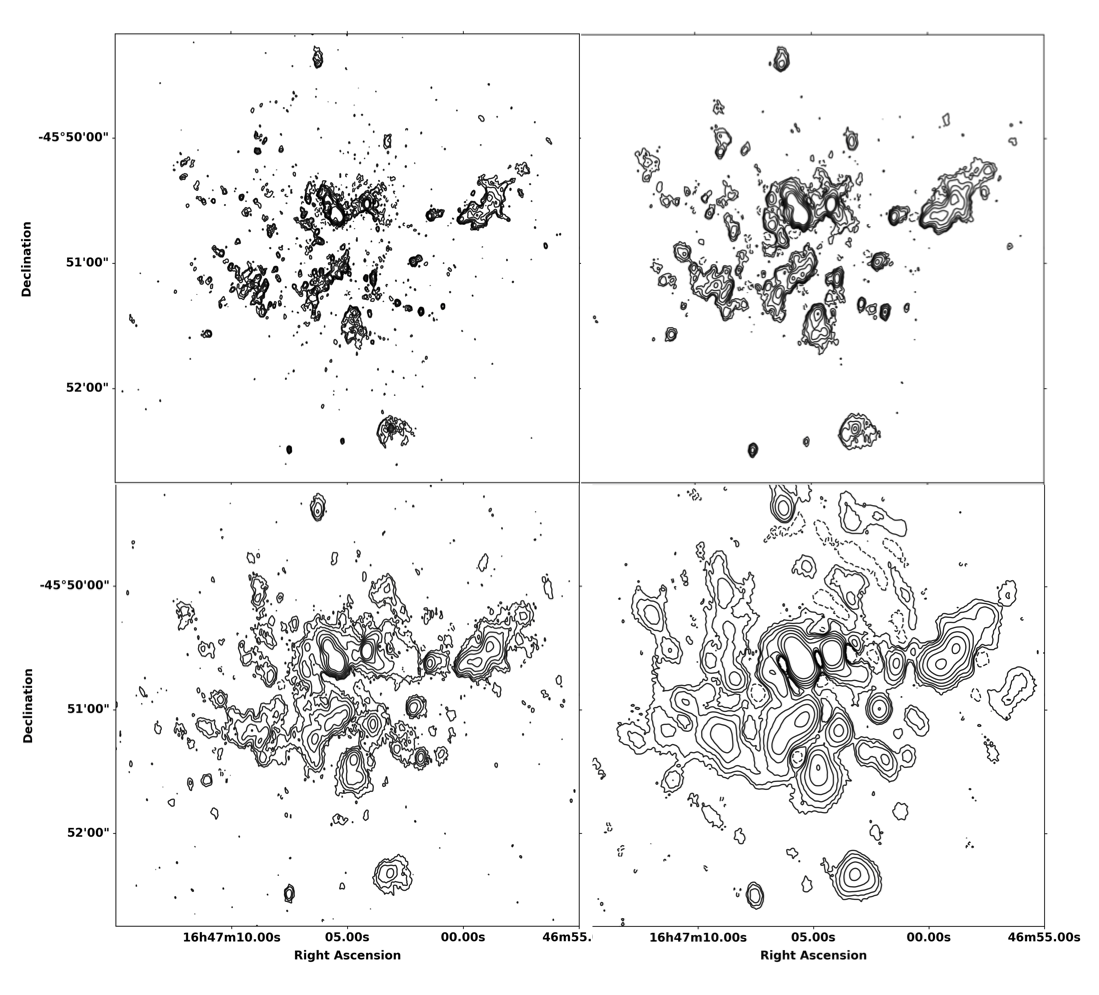}
		\caption{This figure shows combined frequency ATCA contours, from non-primary beam corrected images, separated by configuration, with 6A$\,$(\textit{top left}), 1.5A$\,$(\textit{top right}), 1.5B$\,$(\textit{bottom left}), and 750C$\,$(\textit{bottom right}). The contour levels are set at -3, 3, 6, 9, 12, 24, 48 and 192 $\times$ $\sigma$, with $\sigma$ = 0.02, 0.03, 0.06 and 0.06$\,$mJy for 6A, 1.5A, 1.5B and 750C respectively. }
		\label{fig:fig1}
	\end{figure*}

	\begin{figure*}
	\includegraphics[width=\textwidth]{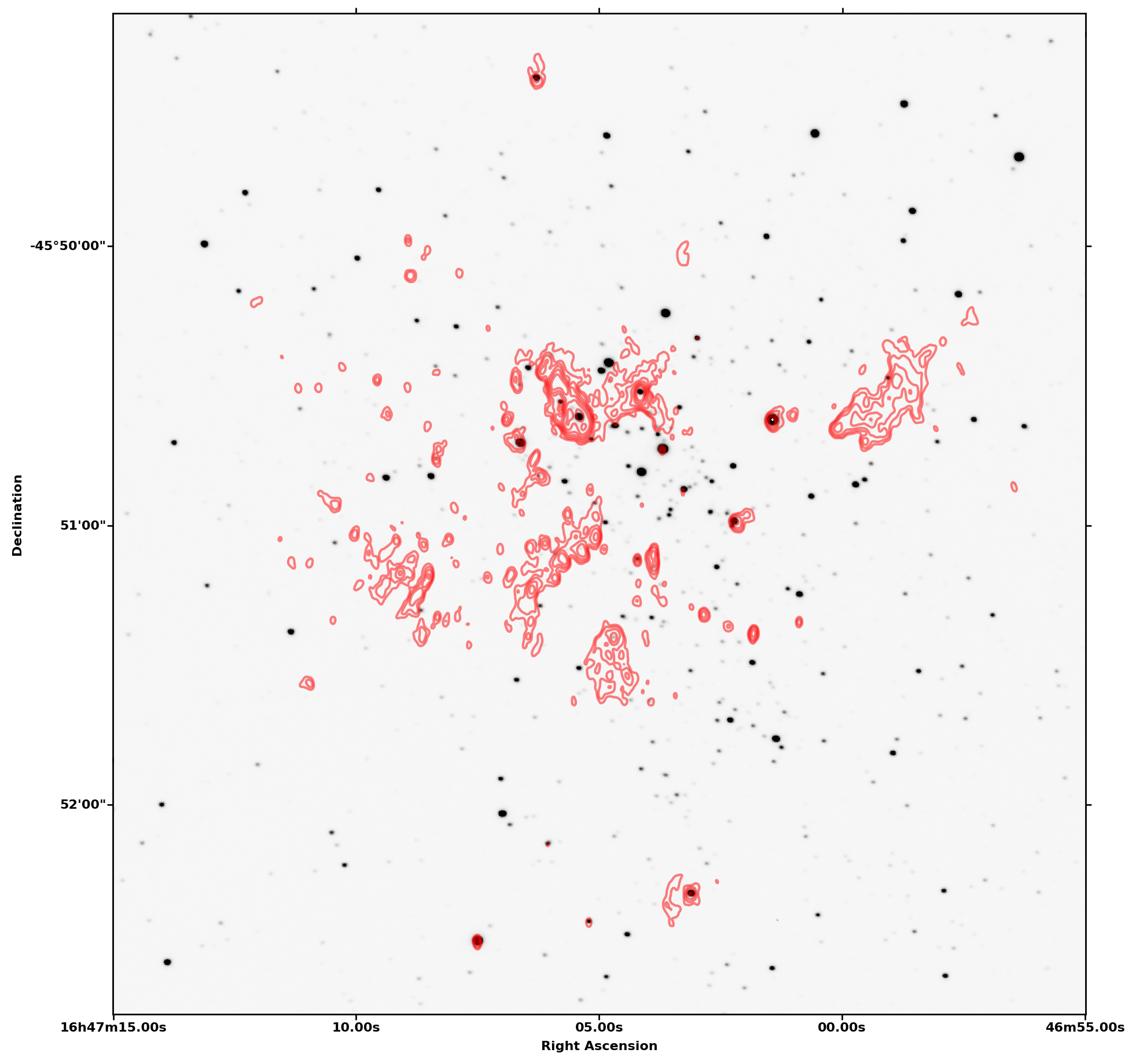}
	\caption{ATCA contours (in red) from the \textsc{FullConcat} dataset (using a non-primary beam corrected image) overlaid on the FORS R-band image. The R-band image has a limiting magnitude of 17.5. The contour levels are set at -3, 3, 6, 9, 12, 24, 48, 96 and 192 $\sigma$, with $\sigma$ set to be 0.015$\,$mJy. This contour levels are adjusted in value with comparison to the contour levels shown in Figure \ref{fig:fig1} in order to better show identification of radio emission with the relevant optical sources.}
	\label{fig:fig2}
	\end{figure*}

	\begin{figure*}
	\includegraphics[width=\textwidth]{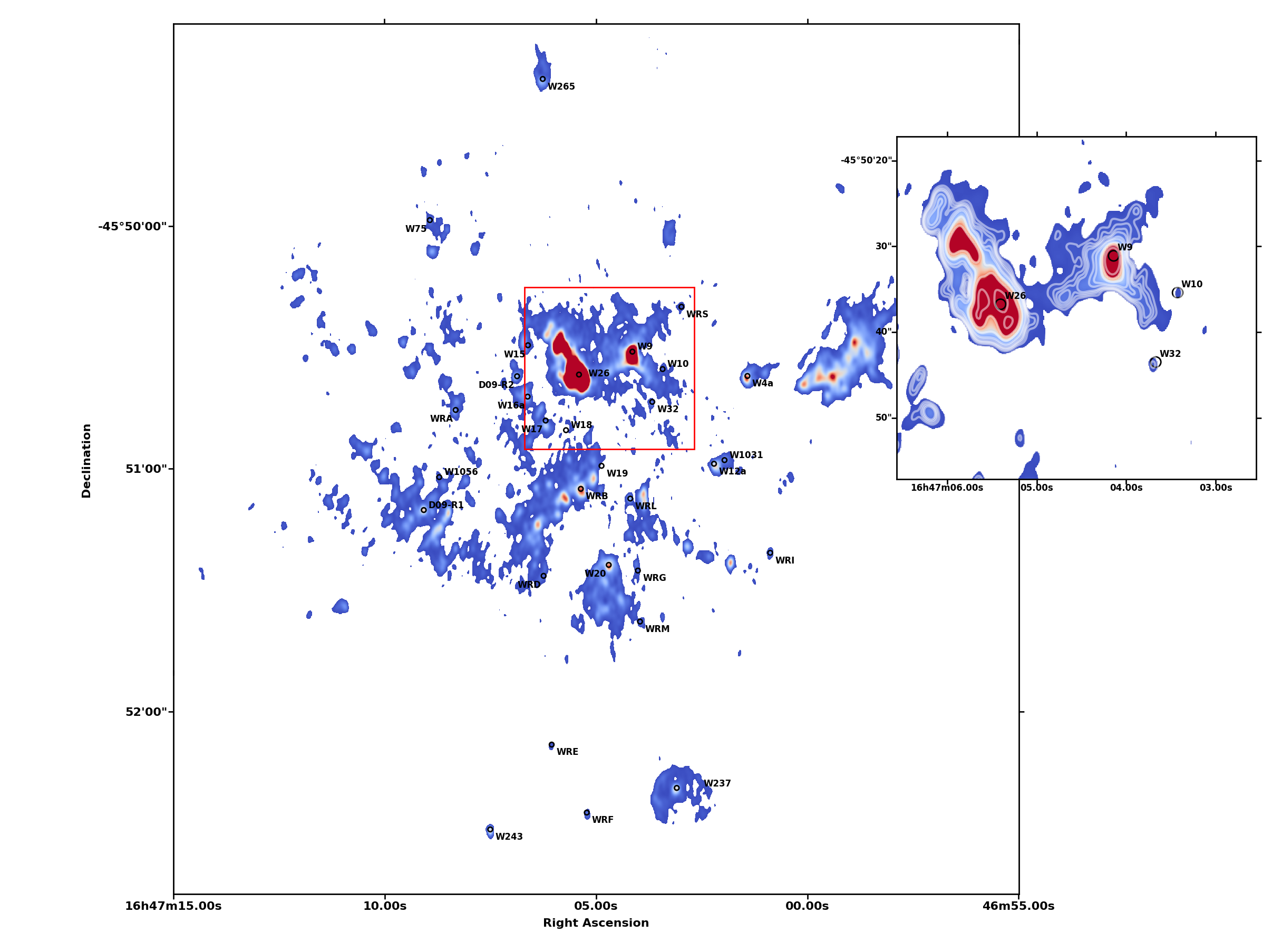}
	\caption{ATCA colour-scale  image (from the non-primary beam corrected image) of the \textsc{FullConcat} dataset. The colour-scale is set from 0.6 - 2.0$\,$mJy beam$^{-1}$. The subplot is a zoomed-in image of the central region, with a slightly adjusted colour-scale of 0.15 - 5.0$\,$mJy beam$^{-1}$, and overlaid contours at levels of -3,3,6,8,10,12,18,24,48,96,192 $\times$ 45$\,\mu$Jy beam$^{-1}$. Stellar sources have been identified by comparing the list of known sources from previous observations \citep[\citetalias{Dougher2010}; \citetalias{Fenech2018}; ][]{Clark2005,Clark2019b} to the SEAC detections from the \textsc{FullConcat} dataset.}
	\label{fig:fig3}
	\end{figure*}
	
	\section{Observations and data reduction}
	
	\begin{table}
		\caption{Summary of observations taken by ATCA.}
		\label{table:obs_details}
		\centering
		\begin{tabular}{p{1cm}p{2.5cm}p{2.1cm}p{2cm}}
			\hline \hline
			Array  & Observing & Obs. Duration    & Beam   \\
			Config.  & Dates & (hours) & Size \\
			\hline
			6A  & 27th$\,$-$\,$29th  & 31.77 & 1.71" $\times$ 0.99"  \\
			(6$\,$km) & October 2015 &  & (0.790$^{\circ}$) \\
			&&&\\
			1.5A & 25th$\,$-$\,$27th & 16.45 &  3.59" $\times$ 1.99" \\
			(1.5$\,$km) & November 2015 & & (-7.23$^{\circ}$) \\
			&&&\\
			1.5B  & 3rd June 2016 & 15.76 & 4.40" $\times$ 2.41"  \\
			(1.5$\,$km) & & & (-2.73$^{\circ}$)  \\
			&&& \\
			750C  & 14th$\,$-$\,$15th  & 28.37 & 6.98" $\times$ 4.44" \\
			(750$\,$m) & December 2015 &  & (6.28$^{\circ}$) \\
			\hline
		\end{tabular}
		\tablefoot{In the first column, the maximum baseline for each configuration is given in parentheses, and in the last column, the position angle for the beam size is given in the parentheses.}
	\end{table}	
	
	Observations were made with the use of ATCA over 4 different observing periods from October 2015 to June 2016. ATCA contains 6 radio dishes with a diameter of 22$\,$m each. Data were taken in four different configurations, with maximum baselines set at 750$\,$m (750C), 1.5$\,$km  (1.5A, 1.5B) and 6$\,$km (6A), with details listed in Table \ref{table:obs_details}. Two of these configurations were set at a largest baseline of 1.5$\,$km, due to a repetition of observations carried out to account for bad weather on the first set of 1.5$\,$km observations. The data were collected over two spectral windows, with central frequencies 5.5$\,$GHz, and 9$\,$GHz, and a bandwidth of 2$\,$GHz for each band. Each spectral window contained 2048 channels, with bandwidths of 1$\,$MHz. Integration time on source per pointing was $\sim$16$\,$hours. J1636-4101 was used as a phase calibrator for the 6$\,$km configuration and 1600-48 was used as a phase calibrator for all other configurations. 1934-638 was observed as a bandpass calibrator for all configurations and was also used for flux amplitude calibration. A secondary calibrator, 0823-500, was also observed. 

	Data reduction and calibration was carried out with the use of \textsc{miriad}$\,$\citep[][]{Miriad}, with additional flagging of data carried out in astronomical image processing software (\textsc{aips}) with the use of scripted e-MERLIN RFI mitigation pipeline for interferometry (SERPent$\,$\footnote{https://github.com/daniellefenech/SERPent}) on the 9$\,$GHz data \citep[][]{Peck2013}. This involved the flagging of radio frequency interference (RFI) and any erroneous data, as well as calibrating for the bandpass, phase, flux density and any polarisation leakage, following standard procedures as laid out in the Miriad User Guide \citep[][]{MiriadGuide}. After calibration, data were concatenated in several forms, discussed in detail below, with a run-through of both phase and amplitude self-calibration then additionally carried out on these final datasets. This involved the use of \textsc{casa}$\,$\citep[][]{Casa}, where the functions \textsc{gaincal} and \textsc{applycal} were carried out alongside the use of the deconvolution imaging tool \textsc{tclean}, cleaned in the mode `multi-scale multi-frequency synthesis' with Briggs weighting applied (with the robust parameter set to 0). The psf mode was set to `clark'. The resolution was set to be equivalent to the smallest primary beam size present in each dataset. For the fully concatenated dataset, this beam size was taken from the 6$\,$km observations, 1.71" $\times$ 0.99" (position angle 0.790$^{\circ}$), as calculated automatically in \textsc{CASA}, and the cell size was set to 0.3$\times$0.3$\,$arcseconds. The multiple deconvolution scales applied were at the size of the beam, as well as additional scales set at 5, 10, 12, 15, 18, 25, 47 pixels, in order to consider the varying resolutions when concatenating datasets from different configurations and spectral windows.
	
	This led to a final set of fits files containing fully calibrated and cleaned images. For data analysis, \textsc{pbcor} (primary beam correction) was applied to mitigate the effects of attenuation from the primary beam for sources further from the pointing centre. 

	\subsection{A note on different datasets}
	
	Several datasets were considered for the analysis of the radio observations. The definitive fluxes of radio sources from the cluster were taken from the fully integrated dataset containing all configurations and all spectral windows observed by ATCA, hereby referred to as the dataset \textsc{FullConcat}. Fluxes were also considered for comparative use from datasets split into the 5.5$\,$GHz and 9$\,$GHz, combined over all array configurations, \textsc{Full5} and \textsc{Full9}. These datasets are used for considering the mass-loss rates from each spectral window for thermally emitting sources, and the \textsc{Full9} is used to compare flux values to the prior radio observations taken of Wd1 by \citetalias{Dougher2010}.
	
	Additional datasets generated from the ATCA observations are tapered datasets of both the 5.5 and 9$\,$GHz observations, \textsc{Taper5} and \textsc{Taper9}. The datasets of tapered visibilities were created in order to compare datasets at different wavelength regimes more accurately, by comparing emission detected only from common ranges of u-v visibilities for both the radio and the millimetre ALMA observations \citepalias{Fenech2018}. These datasets were used for the calculation of spectral indices, as well as mass-loss rates, and is discussed in more detail in Section 3.5 and Section 6.1.1. The corresponding mass-loss rates were then used to calculate clumping gradients, as discussed in Section 6.1.2. The tapered millimetre observations are referred to as \textsc{TaperALMA}, in order to distinguish from the results presented in \citetalias{Fenech2018}, which contains all the full u-v range from the ALMA observations.
	
	\section{Analysis}
	
	After data reduction was completed, analysis was carried out on the datasets using SEAC\footnote{ https://github.com/daniellefenech/SEAC.}, a source extraction software tool \citep{Peck2014,Morford2019}. This tool involves the use of \textsc{parseltongue}, a way to use a python interface with the software \textsc{aips}. The data was converted for use from fits files into the AIPS environment. A consideration of the possible offset of sources to previous datasets was determined following the procedure of previous radio and millimetre wavelength measurements \citepalias{Dougher2010,Fenech2018}. The ATCA datasets were run through the \textsc{aips} task \textsc{hgeom}, where the images were aligned to the same geometry as the FORS and ALMA dataset. The data was then run through \textsc{jmfit} to check any possible offset for the most radio luminous source W9, and only a small offset of the peak position was found. Figure $\ref{fig:fig2}$ shows the aligned dataset (without primary-beam correction) overlaid as contours on a FORS camera R-band image (655$\,$nm)$\,$\citepalias[][]{Dougher2010}, where contours are scaled by the overall rms uncertainty, $\sigma$, taken to be 0.015$\,$mJy beam$^{-1}$.
	
	The SEAC software utilises a floodfill algorithm. This algorithm selects initial pixels as possible sources which have values above a specified `seed' threshold. An `island' is then generated by appending adjacent pixels to an array containing just the 'seed' pixel with a peak value initially, until the flux values of these surrounding pixels no longer reach above a specified `flood' threshold. These were set at 5$\sigma$ for the seed threshold and 3$\sigma$ for the flood threshold. The value of $\sigma$ is the rms value of the local spatial region. Local rms values were determined from a generated noise map, created from dividing the full image into a grid with a user-specified number of cells. The rms level is then calculated within each individual cell, and the corresponding cell gives the value of the local noise level, $\sigma$, for each source. This allows for changes in the background radio emission to be considered, especially with regards to possible effects from extended emission of the particularly radio luminous objects W9 and W26, which otherwise could have affected the rms of the total image. 
	
	\subsection{Selection of source fluxes}
	
	With the use of SEAC, fluxes were determined from each dataset, for each source. The definitive source fluxes were determined from the detections made in the \textsc{FullConcat} dataset. Many of the sources detected are surrounded by significant levels of radio emission, as discussed further in section 4. Due to this, a measurement of the core components of the sources was carried out using an additional segmentation tool within SEAC. This involved the implementation of the Watershed algorithm$\,$\citep[][]{Watershed}.  SEAC allows the use of two different options in applying this algorithm, either involving a Gaussian filter approach, or using a noise elevation map. Both versions of the segmentation tool were applied to the datasets when trying to pick out accurate core and extended components of the sources, leading to 3 source output lists for each dataset.
	
	\begin{figure}
		\centering
		\begin{subfigure}{0.4\textwidth}
			\resizebox{\hsize}{!}{\includegraphics{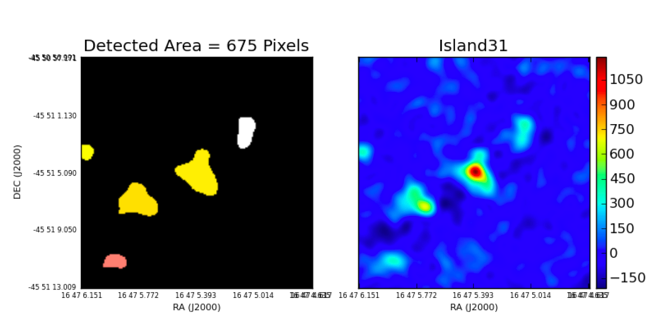}}
			\caption{Image of WR B from the \textsc{TaperALMA} dataset.}
			 \label{fig:fig4a}
		\end{subfigure}
			\begin{subfigure}{0.4\textwidth}
		\resizebox{\hsize}{!}{\includegraphics{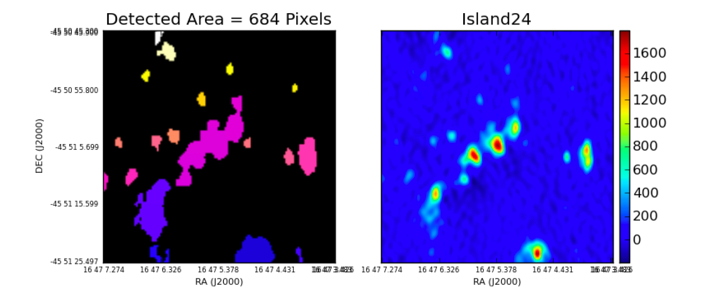}}
		\caption{Image of WR B from the \textsc{Taper9} dataset.}
		\label{fig:fig4b}
		\end{subfigure}
				\begin{subfigure}{0.4\textwidth}
			\resizebox{\hsize}{!}{\includegraphics{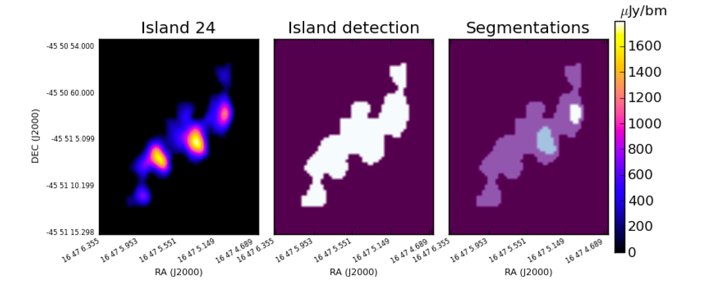}}
		\caption{\textsc{Noise} Segmentation.}
		\label{fig:fig4c}
		\end{subfigure}
				\begin{subfigure}{0.4\textwidth}
		\resizebox{\hsize}{!}{\includegraphics{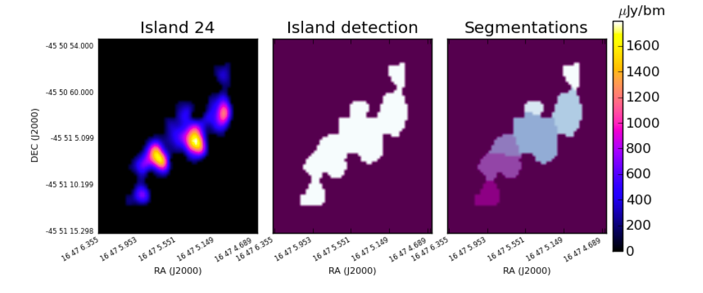}}
		\caption{\textsc{Gauss} Segmentation.}
		\label{fig:fig4d}
		\end{subfigure}
			\setlength{\belowcaptionskip}{-10pt}
		\caption{SEAC output for WR B from the u-v tapered images created from the \textsc{TaperALMA} dataset (\textit{a}), and the \textsc{Taper9} dataset, with the non-segmented source detection from SEAC (\textit{b}), the result of the noise segmentation (\textit{c}), and the result of the gauss segmentation (\textit{d}). Although the noise segmentation results in a better selection of the core WR B component, the gauss segmentation allows for a more appropriate size of emission from the source to be selected and compared to the ALMA resolved source. }
		\label{fig:fig4}
	\end{figure}

	The two possible segmentation algorithms led to different representations of source emission. Figure \ref{fig:fig4} demonstrates how applying the two segmentation approaches have clearly different results, shown here for the source WR B, with a non-segmented image taken from the \textsc{Taper9} dataset, shown in Figure \ref{fig:fig4b}. The \textsc{Noise} segmentation uses two user-defined thresholds to select the strength of the flux that defines a core component. The `bottom' threshold defines the background noise level to dismiss, and the `top' threshold defines the level of emission above which peaks can be detected. This results in the selection of core components of emission embedded within a larger general area of extended emission, as shown by the brighter shading of the two core components detected in Figure \ref{fig:fig4c}. 
	
	The \textsc{Gauss} segmentation picks out local maxima of peaks of radio emission, and then applies a Gaussian filter. These Gaussians are then smoothed by a user-defined level. This leads to the segmentation of large areas of diffuse emission into smaller groups, helping to segregate out sources which may be effected by crowding, as is seen clearly in Figure \ref{fig:fig4d}. This method leads to the inclusion of a larger proportion of extended emission than when determining the core flux for a source from the \textsc{Noise} segmentation. A higher level of smoothing will lead to a smaller number of final segments considered, and will attribute larger levels of surrounding extended emission to the source.
	
	Segmentation was carried out when considering the definitive fluxes of sources taken from \textsc{FullConcat}, as well as when measuring the fluxes to compare the radio observations to the millimetre observations, taken from \textsc{Taper5} and \textsc{Taper9}. It allowed for the core components of sources to be detected, and allowed for further detection of sources embedded in larger regions of continuous diffuse emission surrounding several of the known stellar sources.  Segmentation was not carried out for the \textsc{TaperALMA}, as source structure in ALMA was considered sufficiently compact for all sources, as previously seen in the analysis of the full dataset$\,$\citepalias{Fenech2018}. 
	
	\subsection{Confirmation of ATCA detected Sources}
	
	The definitive flux density values were determined from the \textsc{FullConcat} dataset. This was the dataset with the lowest resultant noise level, and the largest integrated flux surrounding each of the sources, so gave the best chance of determining the most complete number of source detections. The decision of final flux density values took into consideration the size of the emission area associated with the source, picked out by SEAC - this required a thorough visual inspection of the SEAC results, for all segmentation options, on each dataset. After source identification was carried out and flux density values were confirmed, further analysis could be applied to the results to calculate physical quantities associated with the sources, including spatial sizes, spectral indices and mass-loss values. 
	
	Errors on the flux densities are given from the combination of the error found by the calculation of integrated flux density in SEAC and the error given for the flux amplitude calibrator 1934-638 \citep[][online calibrator database]{ATCAcaldatabase}. Calibrator errors were listed as 0.1$\,$\% at 5.5$\,$GHz, and 0.2$\,$\% at 9$\,$GHz. These two values were then combined in quadrature to give the error from amplitude calibration for the \textsc{FullConcat} dataset. The combined amplitude calibrator error was then combined in quadrature with the SEAC flux error to give final errors for the flux densities. Errors for the fluxes given from the \textsc{TaperALMA} measurements used a combination in quadrature of the errors found from the flux determination in SEAC and a 5\% error for the absolute amplitude calibration error, following the prescription in \citetalias{Fenech2018}. However, due the presence of the diffuse emission, it is unclear whether the boundary measured is the true boundary between emission truly related to nearby stellar sources, and emission from the diffuse background. Errors quoted in this paper are therefore potentially underestimates of the true uncertainty on these values, and should be considered as conservative. 	

	\subsection{Spatial extent}
	
	The determination of spatial sizes follows a similar prescription to the method carried out in \citetalias{Fenech2018}. Source sizes were measured from the \textsc{FullConcat} dataset. Measurements of the spatial extents were made using \textsc{jmfit} in \textsc{aips} to determine a Gaussian fit for the source. Peak flux values and pixel locations for sources as calculated from SEAC were used as central positions from which to measure the source structure and provide a Gaussian model fit to the data. A threshold of surrounding pixels above 3$\sigma$ of the local rms was applied. Convolved and deconvolved source sizes were then determined from this. For sources where there were large levels of surrounding extended emission, the size of the core component was given an initial estimate manually, instead of using the 3$\sigma$ threshold. There were still difficulties experienced in applying a Gaussian fit due to the extended levels of diffuse radio background throughout the cluster. For sources that were found to be clearly non-Gaussian, a largest angular size (LAS) was determined instead, via visual inspection of the goodness of fit. 
	
	Both convolved and deconvolved sizes are given in Table \ref{table:atca_known_Sizes}, when applicable. For non-Gaussian sources, only the convolved size is given. For sizes determined from a Gaussian fit, errors are taken directly for the convolved sizes, with errors from the associated minimum and maximum ranges of the deconvolved size used to calculate the deconvolved spatial dimension errors. Sources that are fully resolved can all be separated into point-like sources that fit well to a Gaussian, or sources with significant asymmetric extended emission.  
	
	Most of the sources were found to have resolved, extended emission. The large de-convolved sizes that were measured can also be considered in comparison to the prior ALMA observations in \citetalias{Fenech2018} where many of the stellar detections were unexpectedly found to be resolved, especially many of the WRs in the cluster. The diffuse radio emission detected throughout the cluster may have impacted the measurements. The impact of the extended emission is discussed further in Section 4.
	
	\subsection{Comparison to Do10 measurements}

	The preliminary results revealed the presence of diffuse emission throughout the cluster. The possible astrophysical origin of this and its impact on our conclusions about sources present in the cluster is discussed further in section 4 and 7. Motivated by this diffuse emission, and as a sanity check on the possible effect of this onto the source fluxes, a consideration of the source size and flux was carried out on the \textsc{Full9} dataset in comparison to the previous 8.6$\,$GHz radio observations$\,$\citepalias{Dougher2010}. This consideration involved the use of altering the user-parameters applied in the segmentation tool in SEAC to adjust the final size of the emission measured for core components of each source. The number of pixels with which flux was measured was adjusted indirectly with the alteration of the user-parameters, so that the relative sizes were found to be as similar in the output as possible, using the \citetalias{Dougher2010} images as a reference. This involved the consideration of the relative pixel sizes of each image, and was carried out on a source by source basis. 
	
	In general, this comparison showed that we could make conservative measurements of sources by selecting core components of the source from SEAC, with increased fluxes still typically found for sources in the new radio observations. A small number of sources that were exceptions to this were found to be so due to their location within the cluster, where they were in close proximity to much brighter radio sources and surrounded by large extended regions of emission. Other than the effect of larger components being picked up for the total emission regions, including extended components, the increase in flux seen for the \textsc {Full9} dataset is likely due to increased u-v coverage, resulting from the use of a wider bandwidth in these more recent ATCA observations. The increased bandwidth will allow for a larger number of u-v visibilities to be sampled, allowing for the \textsc{Full9} dataset to be sensitive to a larger number of spatial scales within the same emission region. We can therefore conclude that the Do10 and Full9 datasets provide source fluxes that are broadly consistent with each other. 
	
	The comparative fluxes determined from this analysis are listed in Table \ref{table:atca_do10comp}, with a comparative plot also presented in Figure \ref{fig:figB2}. A note was also made of whether sources were particularly isolated with respect to other sources and the diffuse emission, or were located in more crowded regions within the cluster. Where possible, both fluxes from core components and total source fluxes, including an extended component, were measured. 
	
	\subsection{Spectral indices}
	
	\begin{figure}
				\begin{subfigure}{0.4\textwidth}
			\resizebox{\hsize}{!}{\includegraphics{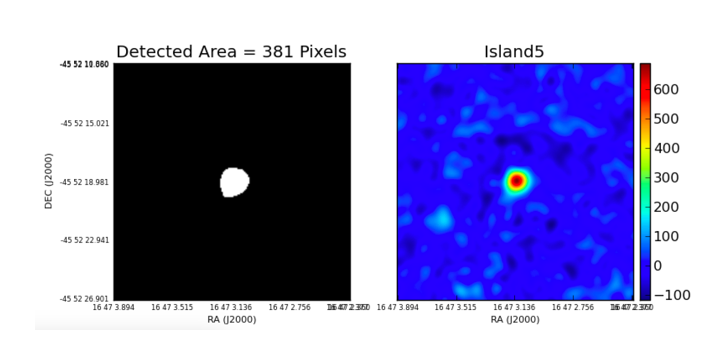}}
			\caption{Image of W237 from the \textsc{TaperALMA} dataset.}
			\label{fig:fig5a}
		\end{subfigure}
		\begin{subfigure}{0.4\textwidth}
			\resizebox{\hsize}{!}{\includegraphics{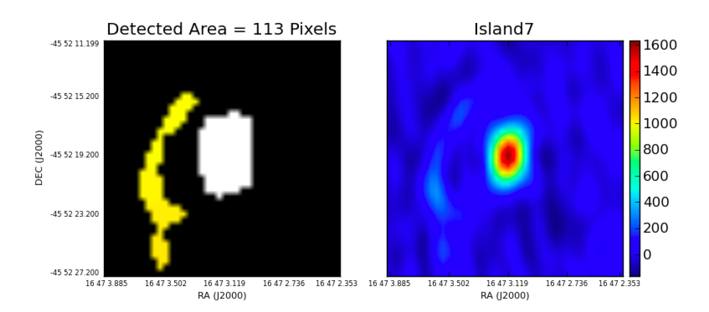}}
			\caption{Image of W237 from the \textsc{Taper5} dataset.}
			\label{fig:fig5b}
		\end{subfigure}
		\begin{subfigure}{0.4\textwidth}
			\resizebox{\hsize}{!}{\includegraphics{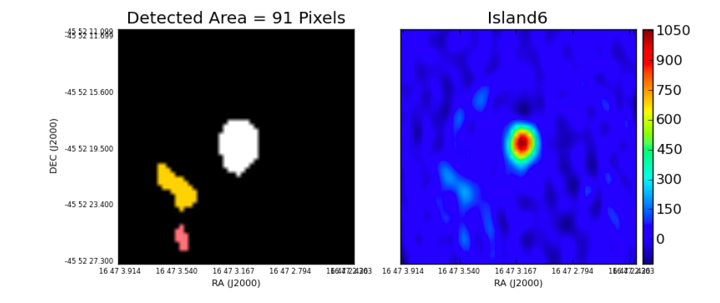}}
			\caption{Image of W237 from the \textsc{Taper9} dataset..}
			\label{fig:fig5c}
		\end{subfigure}
		\begin{subfigure}{0.4\textwidth}
			\resizebox{\hsize}{!}{\includegraphics{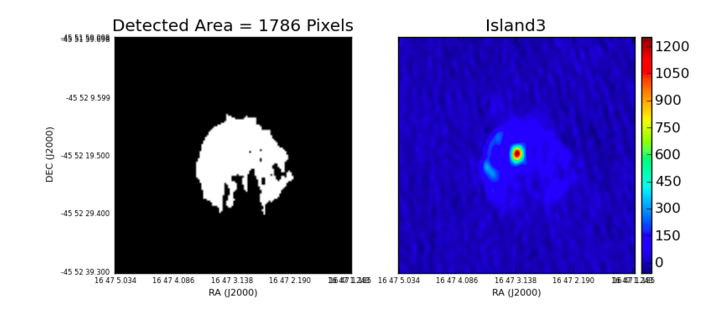}}
			\caption{Image of W237 from the \textsc{FullConcat} dataset.}
			\label{fig:fig5d}
		\end{subfigure}
		\centering
		\caption{SEAC output for W237 from the u-v tapered dataset \textsc{TaperALMA} (\textit{a}), and the \textsc{Taper5} (\textit{b}) and \textsc{Taper9} (\textit{c}) observations, alongside the \textsc{FullConcat} dataset (\textit{d}). This shows that the source size measured by SEAC from the datasets including a tapered range of u-v visibilities is necessary for comparison between the ATCA and ALMA observations, than the source detection from the \textsc{FullConcat} dataset, in order to isolate the core emission.}
		\label{fig:fig5}
	\end{figure}
	
	The spectral indices of all prior known sources were calculated by comparing source fluxes at three frequencies; 5.5$\,$GHz, 9$\,$GHz and 100$\,$GHz. In order to compare the observations for calculating spectral indices, tapering was applied to datasets of ATCA and ALMA observations, so that only equivalent u-v ranges were considered. However, this led to a reduction in the absolute fluxes found for the sources due to the removal of several short spacings in u-v space from the data considered for the image. The reduction in absolute flux density values found from these images was not considered critical to the spectral index derived, as the final value calculated was reliant on the relative flux values at each frequency, rather than the absolute values. 
	
	Using tapered visibilities allowed for similar source sizes to be considered from the ALMA and ATCA observations, as seen in Figure \ref{fig:fig5}. The W237 images from \textsc{Taper5} and \textsc{Taper9}, as shown in Figures \ref{fig:fig5b} and \ref{fig:fig5c} show only the compact bright emission detected for W237, with no inclusion of dimmer more extended emission from the longer spatial scales, that can be seen around W237 in the \textsc{FullConcat} dataset, shown in Figure \ref{fig:fig5d}. However, an initial run-through of SEAC on these datasets still led to inconsistencies in spectral variance when considering the relationship found between the two radio spectral windows, and the overall relationship calculated in the radio-mm spectral index. 
	
	Upon further investigation on a source-by-source basis, the initial inconsistencies in spectral variance were found to be due to differing source sizes determined for the core component of the emission. In order to give the closest comparison possible of the flux when considering the spectral index, a variety of segmentation thresholds were applied to the data, in order to measure core components of the sources. This involved editing the user-defined noise levels in the \textsc{Noise} segmentation tool, and adjusting the smoothing parameter in the \textsc{Gauss} segmentation, until the source size determined for known stellar objects with previous detections were as similar in size as possible, across the different datasets (separated by spectral window). As the pixel sizes of the ALMA and ATCA datasets differed, this was taken into consideration when finding smaller source sizes representative of core emission. The revision of comparative fluxes and spectral indices was not found to significantly affect the overall spectral index determined for the sources from the comparison across all 3 spectral windows at 5.5$\,$GHz, 9$\,$GHz, and 100$\,$GHz. However, it did reduce the level of conflict found when comparing the spectral variance across the two radio wavelengths, 5.5$\,$GHz and 9$\,$GHz, to the index measured across all three frequency bands.

	\section{Extended emission}
		
	One of the main results to become immediately apparent when viewing the images generated by these observations was the large level of diffuse radio emission present throughout the cluster, as seen clearly in Figures \ref{fig:fig1} - \ref{fig:fig3}. The presence of this pervasive diffuse emission calls into question the origin of the material (and therefore the resulting flux measured) surrounding the majority of sources, where the radio emission is found to be resolved and extended. It is unclear whether this material can be truly attributed to these stellar sources due to their stellar wind, or whether the material is actually part of the diffuse background that has become captured by, or is simply interacting with, far more compact nebulae surrounding the sources. If these extended nebulae are truly part of the diffuse background, it is unclear why the majority of this emission is still seen to be directly surrounding the stellar objects, with the most significantly extended structures found around cluster members that are expected to have dense winds and prior periods of extreme mass-loss, such as the cool supergiants, hypergiants and the WR stars.  The possible origin of this diffuse emission is discussed further in Section 7. 
		
	Due to the high levels of diffuse radio emission observed, we had to acknowledge a caveat in our analysis of these observations that radio fluxes measured may include radio emission that has been misattributed to stellar sources. The definitive radio flux densities of the sources, given in Table \ref{table:atca_known}, were calculated from the image generated from the \textsc{FullConcat} dataset, as shown in Figure \ref{fig:fig3}, where the inclusion of the full set of data allowed for the highest sensitivity.

	This diffuse emission may have affected the fluxes measured, the spatial extents found for sources, and all parameters derived from these measurements, including the spectral indices and the mass-loss rates. We were able to mitigate for the impact of this with the use of conservative flux measurements, and by considering only the core components of emission from sources wherever possible. This is discussed in more detail in Section 3. 
	
	The spectral indices and mass-loss rates used radio and ALMA measurements in comparison. When calculating the spectral indices, the effect from the extended emission could lead to over-estimates for the radio measurements. As this background flux is believed to be optically-thin in nature, it would be expected to have a flat spectra, and so have a greater impact on the ATCA flux values than on the higher frequency ALMA values. It would also preferentially effect the ATCA radio observations, due to the lower relative fluxes of the stellar sources, if they are expected to emit thermal emission, and be impacted by the lower resolution of the radio observations.	 Accounting for all possible impacts led to the consideration of source fluxes from comparative source sizes across the ALMA and the ATCA tapered datasets when calculating spectral indices and mass-loss values, as was described in Section 3.5. 
	
	 We can also consider the possible impact of external ionisation on the resultant spectral indices. Previous observations of externally ionised winds of hot stars, including the wind of the red supergiant (RSG) IRS 7, and the YHG HR 8752, have shown spectral indices that are consistent with values found for internally ionised stellar winds. This indicates that although the stellar winds of these cool evolved stars are ionised by surrounding hot stars (in the case of hot stars in the nearby nuclear cluster for the RSG, and the case of an early B companion for the YHG \citep{Stickland1978}), the external ionisation doesn't have a significant effect on the spectral indices \citep{Higgs1978,YusefZadeh1991}.
	 	
	 As the extended emission can be seen to be asymmetric in structure over the cluster, with the strongest illumination orientated towards the cluster centre, it is also of importance to consider whether this would have an impact on the resulting stellar winds. The Pistol star, an LBV located in the Quintuplet cluster, has one hemisphere ionised by its host cluster, though is surrounded by a shell of cold dust, as has been seen in the infra-red \citep[][]{Lau2014}. The spectral index of this star has been found to be consistent with free-free emission from optically thin ionised plasma, which is evidence that even with only part of the shell externally ionised, the end result can be seen to be thermal emission \citep[][]{YusefZadeh1989}.

	The result of over-estimated fluxes would lead to flattened spectral indices. This means any impact on the spectral indices would be to make thermal sources appear less thermal. For the majority of sources, especially in the case of the WRs, these objects are already thermal, as shown in Section 6.1, and so any removal of this extended emission would only lead to spectral indices that could still be considered to be in line with the canonical value for a thermally-emitting stellar wind. It would also lead to flattened spectral indices for several of the cool hypergiants and supergiants, which are already significantly deviated from the canonical wind value, in line with expectations of composite emission, containing optically thick and thin components. The extended emission would also have led to effects causing over-estimates of the spatial extent of sources, and over-estimates of mass-loss rates. In order to counteract any possible impact of this, the mass-loss rates are calculated using only core components of flux from the stellar sources as a conservative estimate, and the resulting mass-loss rates are compared throughout the discussion to rates calculated for analogous sources of each spectral type. 

	In order to try to quantify the impact of the extended emission on our results, a comparison was carried out between the prior 8.6$\,$GHz \citetalias{Dougher2010} observations and the 9$\,$GHz results from the new radio census, with the outcome shown in Table \ref{table:atca_do10comp} and Figure \ref{fig:figB2}. This was discussed previously in Section 3.4, which concluded that the fluxes were broadly consistent between the observations, with slight increases generally found for the new radio results due to the larger u-v coverage resulting in higher sensitivity to more spatial scales. Any decrease in fluxes was found to be due to the location of a source embedded within a diffuse region, and so only a core component could be selected from within the background, resulting in conservative source fluxes measured. 
	
	One example of source consideration is the ATCA flux value attributed to one of the most radio luminous sources in the cluster, W9, found to be significantly higher at 9$\,$GHz in the more recent results, with a flux of 30.47$\,\pm\,$0.09$\,$mJy, in comparison to the core flux component measured by \citetalias{Dougher2010}, 24.9$\,\pm\,$2.5$\,$mJy. For the measurement including the extended component, vastly different fluxes were found, with our flux measurement of 80.8$\,\pm\,$0.5$\,$mJy, versus the \citetalias{Dougher2010} detection of 30.5$\,\pm\,$3.0$\,$mJy. This is due in part to a larger amount of surrounding radio emission detected around W9. However, it was found that due to the increased sensitivity to a range of spatial scales of emission, despite core components for the \textsc{Full9} dataset picked out by SEAC typically resulting in comparatively \textit{smaller} sizes than the core components selected in \citetalias{Dougher2010}, the core region measured in the \textsc{Full9} still contained a much larger level of flux.

\section{Results of the radio census}	

\subsection{Stellar sources}

\begin{table}
	\caption{Summary of the radio detections of different types of cluster members from the \textsc{FullConcat} dataset.}
	\label{table:detection_summary}
	\centering
	\begin{tabular}{p{2.5cm}p{2.5cm}}
		\hline\hline
		Category &Source Number   \\
		\hline
		ATCA+optical & 30 \\
		ATCA+ALMA & 30  \\
		ATCA-only  & 53  \\
		\hline
		Total & 113 \\
		\hline
		WRs & 10  \\
		YHGs & 5  \\
		RSGs & 4 \\
		BSGs & 2 \\ 
		LBV & 1 \\
		sgB[e] & 1 \\ 
		OB supergiants & 7 \\
		\hline
		Total & 30 \\
		\hline
	\end{tabular}
\end{table}

These radio observations led to 30 radio detections of known stellar sources in Westerlund 1, summarised in Table \ref{table:detection_summary}. Out of the 30 stellar sources with confirmed radio fluxes found in \textsc{FullConcat}, 5 of the 10 detections of WRs and 5 of the 7 detections of OB supergiants were new radio detections. The WRs newly detected in the radio were WR D and WR G, WN7o stars, WR I, a WN8o star, and WR E and M, both WC9 stars. OB supergiants with new detections were W10, W18 and W19, all early B supergiants, and W1031 and W1056, both late O-type giants. Flux densities for stellar sources with optical counterparts are given in Table$\,$\ref{table:atca_known}, with flux densities taken from the integrated island flux found by SEAC. Sources with core and total flux components as detected by SEAC are listed with both flux values, specified by the superscript \textit{c} and \textit{t} respectively.  

By investigating the \textsc{Full9} dataset, we could compare our radio detections directly to the previous 8.6$\,$GHz radio observations of Westerlund 1 cluster members$\,$\citepalias{Dougher2010}. Of the 21 sources detected by \citetalias{Dougher2010} at 8.6$\,$GHz, all but 1 source was also detected in the \textsc{Full9} dataset. The lack of detection for this source, WR V, can be linked to its location embedded within an area of diffuse emission in the cluster (discussed in more detail in section 6.1).

Spectral indices and spectral index limits were determined for all sources where at least one flux density from the tapered datasets \textsc{TaperALMA} , \textsc{Taper5} or \textsc{Taper9} could be found. These are presented in Table \ref{table:spec_index}. This table also includes spectral index limits found for the new detections of OB supergiants with no corresponding millimetre detections, with the radio flux density taken from \textsc{FullConcat} due to the faintness of the sources, and using the flux density limit from the full ALMA observations \citepalias[][]{Fenech2018}. 

The spectral index limits derived from the new radio detections of OB supergiants led to the conclusion that these stars were binary candidates. The spectral index limits were found to constrain the spectral indices to negative values, implying the presence of non-thermal emission. These spectral index limits included the cluster members currently categorised as OB supergiants, W15, W18, and W1031, as well as OB supergiants with new radio detections, W10, W19, and W1056. This is discussed in more detail in section 6.3.1. 

\subsection{Non-Stellar sources}

\subsubsection{ALMA sources}

Previous observations made in the millimetre \citepalias[][]{Fenech2018}, detected a large number of sources that were not associated with any catalogued stellar sources. 30 of these objects have now been found to have associated detections in the radio. Flux densities were determined, with the positions and fluxes listed in Table \ref{table:alma_uncat}, with sources presented in a corresponding plot of the cluster, in Figure \ref{fig:figB3}. The number assigned to these sources is the FCP18 source number, as given in \citetalias{Fenech2018}.

One of the sources can clearly be attributed to a knot of emission associated with the extended nebulae around the YHG W4a. 4 of the sources are found to be aligned with a significant area of extended emission to the west of Wd1 (to the RHS side of the figure). A large number of the rest of the sources can be found clustered within other regions of diffuse radio background or nearby radio luminous known stellar sources, such as W9, WR B and D09-R1. At least 7 of the sources detected were found to be in isolated positions and not linked to any known stellar sources in the cluster.

15 of these sources were strong enough in the radio for spectral indices to be calculated from the tapered datasets, \textsc{Taper5}, \textsc{Taper9} and \textsc{TaperALMA}, as shown in Table \ref{table:spec_index_almaonly}. Overwhelmingly, all of these sources were found to have clearly non-thermal spectral indices. Some of the sources that are seen to have a negative spectral index are clearly embedded within a larger diffuse radio background. This indicates that shocks within the cluster wind may be responsible for these knots of emission \citep[][]{Bell1978}. Some of the uncategorised sources are in the near vicinity of stellar sources and the impact of extended emission from these nearby stars may influence the spectral behaviour observed \citep[][]{YusefZadeh2003}. 

However, some sources cannot be linked to either of these external influences, and so must experience a non-thermal spectral index for another reason, most notably in the case of sources FCP18-16, -22, -27, and -95. Possible explanations for this spectral behaviour are the presence of synchrotron emission in the radio resulting from an as of yet undetected binary stellar system, where there is an interacting shock between two stellar winds. Another possible explanation is that the knots of radio and millimetre emission could be of extra-galactic origin rather than from Wd1 itself, though this likelihood was discussed and largely dismissed previously in \citetalias{Fenech2018}.

\subsubsection{ATCA-only sources}

Alongside the detection of previously observed sources in other wave-bands, 53 additional sources of radio emission were detected through the use of SEAC, listed in Table \ref{table:atca_uncat}, with the source number assigned from the full SEAC output list, ordered by RA. A conservative limit in determining the number of these radio-only sources meant that only knots of radio emission detected without the use of the segmentation tools within SEAC were considered. Some of these sources may be related to extended radio emission from nearby radio luminous sources, including HA19-23 in the nearby vicinity of W12a, and HA19-43 in the close vicinity of the large extended tail of W20. A large number can also be seen (as displayed in Figure \ref{fig:figB4}) to be in isolated regions away from any of the previously determined sources in Wd1. Additionally, many of these isolated radio sources display a geometry that indicate the emission cannot be considered an extension to other known sources, due to the compact, point-source like morphology of the emission, including examples such as HA19-65 to the north of the cluster, HA19-3, and HA19-4 to the west of the cluster, and HA19-85, -88 and -89 to the east.

		\section{Stellar sources}
		
		\begin{table*}
			\caption{Flux densities for known sources from the \textsc{FullConcat} dataset.}
			\label{table:atca_known}
			\centering
			\begin{tabular}{p{1.2cm}p{2.5cm}p{2.5cm}*2{p{2.8cm}}p{2.2cm}}
				\hline \hline
				Source  & Spectral Type & RA & DEC  & Flux$_{\textsc{FullConcat}}$ (mJy)  \\
				\hline
				WR A & WN7b+OB? & 16 47 8.36412  &  -45 50 45.5948 & 0.89 $\pm$0.06 \\
				&&&& 1.38 $\pm$ 0.09 \\
				WR D  &  WN7o & 16 47 6.29738  &  -45 51 26.6985 & 0.71	$\pm$ 0.05 \\
				WR B & WN7o+OB? & 16 47 5.34956   &  -45 51 5.3995 & 3.38 $\pm$ 0.03$^{c}$ \\
				&&&& 16.53 $\pm$ 0.15$^{t}$ \\
				WR G & WN7o & 16 47 4.05743   &  -45 51 23.7000 & 0.34 $\pm$ 0.04   \\
				WR P & WN7o & 16 47 1.61624 & -45 51 45.5984 & 0.05$\pm$0.02$^{m}$ \\
				WR I & WN8o & 16 47 0.89863   &  -45 51 20.6974 & 0.42	$\pm$ 0.02 \\
				WR L  & WN9h+OB? & 16 47 4.20100 &   -45 51 7.5000 & 0.55 $\pm$	0.04 \\
				WR S & WN10-11h/BHG & 16 47 2.96652   &  -45 50 19.7997 & 0.16 $\pm$ 0.04 \\
				WR E & WC9 & 16 47 6.06807  &   -45 52 8.6988 &  0.13 $\pm$ 0.02 \\
				WR F & WC9d+OB? & 16 47 5.20647  &   -45 52 25.1996 & 0.29 $\pm$ 0.02 \\
				WR M & WC9d  & 16 47 3.94256  &    -45 51 38.1000 & 0.24 $\pm$	0.04 \\
				\hline 
				W16a & A5Ia$^{+}$& 16 47 6.69880  &   -45 50 41.0980 & 2.70 $\pm$ 0.11 \\
				W12a  &F1Ia$^{+}$& 16 47 2.19106 &     -45 50 59.3991 & 1.76 $\pm$ 0.03$^{c}$ \\
				&&&&   3.77 $\pm$ 0.07$^{t}$ \\
				W4a &F3Ia$^{+}$ & 16 47 1.44479  &   -45 50 37.4982 & 3.18 $\pm$ 0.04$^{c}$ \\
				&&&& 4.72 $\pm$ 0.07$^{t}$ \\
				W32 &F5Ia$^{+}$ & 16 47 3.71289   & -45 50 43.8000 & 0.18 $\pm$	0.04 \\
				W265 &F5Ia$^{+}$ &  16 47 6.29597  &   -45 49 24.2985  & 1.02 $\pm$ 0.04$^{c}$  \\
				&&&& 3.34	$\pm$ 0.11$^{t}$ \\
				W237 & M3Ia & 16 47 3.10953 & -45 52 19.1998 & 1.31 $\pm$ 0.02$^{c}$ \\	
				&&&& 9.57 $\pm$	0.19$^{t}$  \\
				W75 &M4Ia & 16 47 8.93719   &  -45 49 58.7933 & 0.34 $\pm$ 0.03 \\
				W20  &M5Ia& 16 47 4.68920 &  -45 51 24.2999 & 2.54 $\pm$ 0.03$^{c}$  \\
				&&&& 20.65 $\pm$ 0.23$^{t}$  \\
				W26  &M5-6Ia&  16 47 5.40679   &  -45 50 36.5994 & 153.11 $\pm$ 0.17 \\ 
				\hline
				W17 & O9Iab & 16 47 6.18210  &   -45 50 49.4987  &  0.98 $\pm$ 0.04$^{c}$ \\
				&&&& 2.04$\pm$0.08$^{t}$ \\
				W243 & A2Ia (LBV) & 16 47 7.50460  &   -45 52 29.3966  &  1.38 $\pm$	0.04  \\
				\hline
				W9 &sgB[e]& 16 47 4.14355  &   -45 50 31.5000 & 27.38 $\pm$ 0.06$^{c}$ \\
				&&&& 83.14 $\pm$ 0.41$^{t}$ \\
				D09-R1 & BSG & 16 47 9.08253 &  -45 51 10.1929  & 1.29 $\pm$ 0.03$^{c}$ \\
				&&&&5.44  $\pm$ 0.10 $^{t}$ \\ 
				D09-R2 &BSG& 16 47 6.89972  &    -45 50 37.1977 & 0.96 $\pm$ 0.06 \\
				\hline	
				W15  & O9Ib & 16 47 6.72735  &    -45 50 28.7980 &  1.65 $\pm$	0.09  \\
				W10 & B0.5I + OB & 16 47 3.42580   &  -45 50 35.3999 &  0.16 $\pm$	0.05   \\
				W18  & B0.5Ia &  16 47 5.60787  &   -45 50 50.3993  & 0.36 $\pm$ 0.06 \\
				W19  &B1Ia&  16 47 4.86139  &    -45 50 57.8998  & 0.08 $\pm$ 0.03  \\
				W1031 & O9III & 16 47 1.96137  &   -45 50 57.8988 & 0.90	$\pm$ 0.05  \\
				W1056  & O9.5II & 16 47 8.70904  &    -45 51 2.0939 & 0.09 $\pm$ 0.02 \\
				\hline
			\end{tabular}
		\tablefoot{Flux densities are from the \textsc{FullConcat} dataset, measured by SEAC (with flood threshold, $\sigma_{f}$ = 3, and seed threshold, $\sigma_{s}$ = 5), with consideration of segmentation where necessary. Flux densities are given in $mJy$. For extended sources, where possible, core flux densities are denoted by the superscript, $^{c}$, and total flux densities are denoted by the superscript, $^{t}$. A marginal flux detection is indicated with the superscript $^{m}$. Spectral type identifications are taken from \cite{Crowther2006} for the WR stars, \cite{Clark2010} for the YHGs and RSGs, \citetalias{Dougher2010} for D09-R1 and D09-R2, and \cite{Clark2019b} or the OB stars.}
		\end{table*}	 
	
	\begin{table*}
		\caption{Spectral indices and index limits of stellar sources.}
		\label{table:spec_index}
		\centering
		\hspace*{-1cm}
		\begin{tabular}{p{1.2cm}*2{p{3.2cm}}p{3cm}p{2.8cm}} 
			\hline \hline
			Source & Flux$_{\textsc{Taper5}}$ (mJy) & Flux$_{\textsc{Taper9}}$ (mJy) & Flux$_{\textsc{TaperALMA}}$ (mJy) &  Spectral Index ($\alpha$) \\
			\hline	
			WR O & $<$ 0.10 & $<$ 0.10& 0.28 $\pm$ 0.04 & $>$ 0.37 \\
			WR U & $<$ 0.19 & $<$ 0.15  & 0.17 $\pm$ 0.05 & $>$ -0.01 \\
			WR Q &$<$ 0.12 & $<$ 0.10 &  0.18 $\pm$ 0.05 & $>$ 0.17  \\
			WR A & 0.62 $\pm$ 0.08 & 0.94 $\pm$ 0.05 & 3.89 $\pm$ 0.22 & 0.61 $\pm$ 0.04 \\
			WR D & $<$0.17 & 0.17 $\pm$ 0.05 & 0.58 $\pm$ 0.07 & 0.50 $\pm$ 0.16 \\
			WR B &  3.51 $\pm$ 0.12 &  2.96 $\pm$ 0.09 & 2.62 $\pm$ 0.23 & -0.10 $\pm$ 0.03 \\
			WR G & $<$0.16 & 0.13 $\pm$ 0.04 &  0.72 $\pm$ 0.08 & 0.71 $\pm$ 0.14 \\
			WR P &$<$ 0.12 & $<$ 0.12  & 0.50 $\pm$ 0.06 & $>$ 0.52  \\
			WR I & 0.16 $\pm$ 0.04 & 0.47 $\pm$ 0.04 & 2.02 $\pm$ 0.11 & 0.79 $\pm$ 0.08 \\
			WR V & $<$ 0.24 & $<$ 0.28  & 0.90 $\pm$ 0.09  & $>$ 0.47 \\
			WR L & 0.24 $\pm$ 0.06 & 0.48 $\pm$ 0.04 & 3.37 $\pm$ 0.18	&  0.88 $\pm$ 0.07 \\
			WR S  & $<$ 0.13 & $<$ 0.09 & 0.36 $\pm$ 0.07 & $>$ 0.42 \\	
			W13  & $<$ 0.20 & $<$ 0.21 & 0.18 $\pm$ 0.05  & $>$  -0.05 \\
			WR K &$<$ 0.16 & $<$ 0.14 & 0.16 $\pm$ 0.05 & $>$  0.02 \\
			WR E  & $<$ 0.16 &$<$ 0.10 & 0.81 $\pm$ 0.06 &  $>$ 0.66 \\
			WR F &  $<$0.12 & 0.37 $\pm$ 0.04 & 1.26 $\pm$  0.09 & 0.51 $\pm$ 0.07 \\ 
			WR C  &$<$ 0.17 & $<$ 0.11  & 0.22 $\pm$ 0.04 &  $>$ 0.15  \\
			WR H   &$<$ 0.12 & $<$ 0.09  & 0.28 $\pm$ 0.06 & $>$ 0.35 \\  
			WR M & $<$0.16 & 0.11 $\pm$ 0.03 & 0.39 $\pm$ 0.05 &	0.53 $\pm$ 0.12	\\ \hline
			W16a & 1.59 $\pm$0.11 & 1.67 $\pm$ 0.11 & 1.23 $\pm$ 0.15 & -0.10 $\pm$ 0.05 \\
			W12a & 1.77 $\pm$ 0.05& 1.70 $\pm$ 0.08 & 0.78 $\pm$ 0.13 & -0.30 $\pm$ 0.07 \\
			W4 & 1.85 $\pm$ 0.04  & 1.84 $\pm$ 0.04 & 1.78 $\pm$ 0.16 & -0.01 $\pm$ 0.04  \\
			W265 & 1.00 $\pm$ 0.04 & 0.94 $\pm$ 0.08 & - & -0.13 $\pm$ 0.23 \\ 
			W237 & 	1.07 $\pm$ 0.06 & 1.03 $\pm$ 0.05 & 1.05 $\pm$ 0.08  & 0.00$\pm$ 0.04 \\
			W75 & $<$0.13 & 0.23 $\pm$ 0.05 & 0.22 $\pm$ 0.05 & -0.01 $\pm$ 0.12  \\
			W20 &  2.23 $\pm$ 0.07 &  2.10 $\pm$ 0.06 & 1.99 $\pm$ 0.16 & -0.03 $\pm$ 0.03  \\
			W26 & 106.3 $\pm$ 0.2 &  117.7 $\pm$ 0.4 & 103.3 $\pm$ 5.2  &  -0.02 $\pm$ 0.02 \\
			\hline
			W17 & 0.81 $\pm$ 0.09  & 0.86 $\pm$ 0.09 & 0.76 $\pm$ 0.12 & -0.03 $\pm$ 0.07  \\
			W46a   &$<$ 0.11 &$<$ 0.09  & 0.18 $\pm$ 0.04 & $>$ 0.21 \\
			W243 & 0.82 $\pm$ 0.05 & 1.54 $\pm$ 0.07 & 9.69 $\pm$ 0.49 &  0.82 $\pm$ 0.03 \\ 
			\hline
			W7  & $<$ 0.09 & $<$ 0.11 & 0.20 $\pm$ 0.04& $>$ 0.27  \\
			W9 & 22.10 $\pm$ 0.07 & 30.34 $\pm$ 0.11 &  158.6 $\pm$ 7.9 &  0.68 $\pm$ 0.02 \\
			D09-R1 & 0.82 $\pm$ 0.07 & 0.96 $\pm$ 0.07 &  0.85 $\pm$ 0.14 & -0.01 $\pm$ 0.07 \\
			D09-R2 & 0.54 $\pm$ 0.06 &  0.59 $\pm$ 0.07 &  0.51 $\pm$ 0.07 & -0.03 $\pm$ 0.07 \\
			\hline
		 	Source && Flux$_{\textsc{FullConcat}}$ (mJy)  & Flux$_{\textsc{ALMA}}$  (mJy) &  Spectral Index ($\alpha$) \\
			\hline
			W15  && 1.65$\pm$0.09  & $<$ 0.10 &  $<$ -1.15 \\
			W10 && 0.16$\pm$0.05  & $<$ 0.15 & $<$ -0.04    \\
			W18  && 0.36$\pm$0.06 &$<$ 0.127 & $<$ -0.43   \\
			W19  && 0.08$\pm$0.03 &$<$ 0.104& $<$ 0.10 \\
			W1031 && 0.90$\pm$ 0.05& $<$ 0.09 & $<$  -0.95  \\
			W1056   && 0.09$\pm$0.02  & $<$ 0.100 & $<$ 0.02  \\
			\hline
		\end{tabular}
	\tablefoot{The (\textit{top}) panel shows spectral indices and index limits derived from  \textsc{TaperALMA}, \textsc{Taper5} and \textsc{Taper9}  for stellar sources with an ALMA detection. All data are tapered to contain the same range of u-v visibilities, and are set with thresholds of $\sigma_{f}$ = 3, $\sigma_{s}$ = 5. For extended sources, the core fluxes are used for a better comparison with the spatial region that is considered with the ALMA emission. The detection limits on the ATCA data were assigned by determining the rms, $\sigma$, in the local area of each source that was not detected, using a circular region of 6 arcseconds around the source location. The flux density limits from ATCA were set at 3$\sigma$. The (\textit{bottom}) panel shows spectral index limits using flux densities from the full datasets of ATCA observations, split by frequency, for stellar sources with no ALMA detection. The millimetre flux limits are taken from Table A.2 of \citetalias{Fenech2018}, where the detection limit is 3$\sigma$, with $\sigma$ the rms from a circular region of 2$\,$arcseconds around the optical position of the stars.}
	\end{table*}   

\begin{table*}
	\caption{Spatial extents measured from the \textsc{FullConcat} dataset.}
	\label{table:atca_known_Sizes}
	\centering
	\begin{tabular}{p{1.2cm}p{1.6cm}p{1.8cm}p{0.7cm}*4{p{2.2cm}}} \hline \hline
		Source & RA (16 47)& DEC (-45) & Offset & \multicolumn{2}{c}{Size Convolved} & \multicolumn{2}{c}{Size Deconvolved}   \\
		&&&&Major axis (")  &Minor axis (") &Major axis (") &Minor axis (") \\
		\hline
		WR A & 8.36412  &  50 45.5948 & 0.47 & 2.49 $\pm$ 0.07 & 1.34 $\pm$ 0.04 & 1.82 $\pm$ 0.11 & 0.89 $\pm$ 0.08  \\ 
		WR D  & 6.29738  &  51 26.6985 & 0.65 & 4.96 & LAS & - &  -  \\
		WR B & 5.34956   &  51 5.3995 & 0.40  &3.23 $\pm$ 0.03 & 2.32 $\pm$ 0.02 & 2.77 $\pm$ 0.04 & 2.06 $\pm$ 0.03  \\
		&&&& 6.98 & LAS & - &-  \\
		WR G & 4.05743   &  51 23.7000 & 1.57 & 3.05 $\pm$ 0.19 & 1.19 $\pm$ 0.08  & 2.53 $\pm$  0.24 & 0.63 $\pm$ 0.18   \\
		WR I & 0.89863   &  51 20.6974 & 0.22 & 1.86 $\pm$ 0.07 & 1.05 $\pm$ 0.04 & 0.80 $\pm$ 0.21 & -  \\
		WR L  & 4.20100 &   51 7.5000 & 0.14 & 1.76 $\pm$ 0.05 & 1.13 $\pm$ 0.03 & 0.55 $\pm$ 0.34 & 0.39 $\pm$ 0.17   \\
		WR S & 2.96652   &  50 19.7997 & 0.25 &  2.25 $\pm$  0.26 & 1.96 $\pm$ 0.22 &  1.74 $\pm$ 0.51 &  1.42 $\pm$ 0.65   \\ 
		WR E & 6.06807  &   52 8.6988 &  0.54 & 1.78 $\pm$ 0.19 & 1.10 $\pm$ 0.12 & 0.51 $\pm$ 0.53 & 0.47 $\pm$ 0.41   \\
		WR F & 5.20647  &   52 25.1996  & 0.25 & 1.89 $\pm$ 0.10 & 1.06 $\pm$ 0.06  & 0.81 $\pm$ 0.30 & 0.37 $\pm$ 0.28  \\
		WR M   & 3.94256  &   51 38.1000 & 0.36  & 2.18 $\pm$ 0.17 & 1.61 $\pm$ 0.12  & 1.49 $\pm$ 0.33 & 1.10 $\pm$ 0.4  \\
		\hline 
		W16a &  6.69880  &  50 41.0980 & 1.37 & 3.66 $\pm$ 0.09  &  2.41 $\pm$ 0.06 & 3.25 $\pm$ 0.10 & 2.18 $\pm$ 0.07 \\
		W12a  & 2.19106 &  50 59.3991 & 0.61 & 2.39 $\pm$ 0.03 & 2.05 $\pm$ 0.03 & 1.80 $\pm$ 0.05 & 1.67 $\pm$ 0.06  \\
		&&& & 4.81 & LAS & - & -    \\
		W4a & 1.44479  &  50 37.4982 & 0.47 & 2.35 $\pm$ 0.02 & 1.55 $\pm$ 0.01  & 1.62 $\pm$ 0.03 & 1.19 $\pm$ 0.02  \\
		&&&& 6.91 &LAS & -  & -   \\
		W32 & 3.71289   & 50 43.8000 & 0.50 &  1.78 $\pm$ 0.09 & 1.04 $\pm$ 0.05 & 0.50 $\pm$ 0.4 & 0.31 $\pm$ 0.28  \\
		W265  &  6.29597  &  49 24.2985  & 0.72 & 2.46 $\pm$ 0.05 & 1.92 $\pm$ 0.04 & 1.94 $\pm$ 0.09 & 1.44 $\pm$ 0.10\\
		&&&& 12.59 &LAS & - & -\\
		W237 & 3.10953 & 52 19.1998 & 0.43 & 2.60 $\pm$ 0.04 & 1.94 $\pm$ 0.03 & 1.97 $\pm$ 0.06 & 1.64 $\pm$  0.04  \\	
		&&&&15.36 &LAS & - & -  \\
		W75 & 8.93719   &  49 58.7933 & 0.22  & 2.74 $\pm$ 0.15 & 1.46 $\pm$ 0.08 & 2.16 $\pm$ 0.19 & 1.05 $\pm$ 0.13  \\
		W20  & 4.68920 &  51 24.2999 & 0.50 & 2.89 $\pm$ 0.03 & 2.21 $\pm$ 0.02 & 2.35 $\pm$ 0.04 & 1.96 $\pm$ 0.03   \\
		&&& & 21.05 & LAS &- & -  \\
		W26  & 5.40679   &  50 36.5994 & 0.18 & 17.04 & LAS & - & -    \\ 
		\hline
		W17 & 6.18210  &  50 49.4987  & 0.76 & 2.52 $\pm$ 0.05  & 2.04 $\pm$ 0.04  & 2.18 $\pm$ 0.13 & 1.38 $\pm$ 0.11  \\ 
		W243 & 7.50460  &   52 29.3966  & 0.25  & 1.75 $\pm$ 0.02 & 1.00 $\pm$ 0.01 & 0.40 $\pm$ 0.11 &  - \\
		\hline
		W9 & 4.14355  &  50 31.5000 & 0.58 & 2.212 $\pm$ 0.001 &  1.235 $\pm$ 0.001 & 1.405 $\pm$ 0.002 & 0.735 $\pm$ 0.001  \\
		&&& & 9.39 & LAS & -  &  - \\
		D09-R1 & 9.08253 &  51 10.1929  &  0.07  & 2.37 $\pm$ 0.03 &  1.88 $\pm$ 0.02 & 1.92 $\pm$ 0.05 & 1.24 $\pm$ 0.07\\ 
		&&&& 12.85  &LAS & - & -  \\ 
		D09-R2 & 6.89972  &  50 37.1977 & 0.07 & 2.13 $\pm$ 0.05  & 1.49 $\pm$ 0.04  & 1.45 $\pm$ 0.10 & 0.86 $\pm$ 0.13 \\
		\hline	
		W15  & 6.72735  &     50 28.7980 & 1.40 & 3.95 $\pm$ 0.10 & 1.61 $\pm$ 0.04  & 3.55 $\pm$  0.12 & 1.27 $\pm$  0.06 \\
		W10 & 3.42580   &  50 35.3999 & 1.15  & 2.55 $\pm$ 0.22 &  1.41$\pm$ 0.12 & 1.90 $\pm$ 0.23 & 0.98 $\pm$ 0.22 \\
		W18  &  5.60787  &  50 50.3993  & 1.08 &3.86 $\pm$ 0.67 & 1.79 $\pm$ 0.31  & 3.47 $\pm$ 0.8 & 1.49 $\pm$  0.4 \\
		W19  & 4.86139  &  50 57.8998  & 1.19 & 3.75 $\pm$ 0.16 &  2.39 $\pm$ 0.10 & 3.35 $\pm$ 0.19 & 2.15 $\pm$ 0.13  \\
		W1031 &  1.96137  &  50 57.8988 & 1.55  & 2.90 $\pm$ 0.13  & 2.53 $\pm$ 0.11 &  2.73 $\pm$ 0.15 & 1.87 $\pm$ 0.17    \\
		W1056 & 8.70904  &  51 2.0939 &  1.01 &2.90 $\pm$ 0.28 & 1.28 $\pm$ 0.12 &2.36 $\pm$ 0.36  & 0.74 $\pm$ 0.29   \\
		\hline
	\end{tabular}
\tablefoot{Spatial extents are given in terms of convolved and deconvolved sizes. Gaussian fits were applied to the stellar sources using the task \textsc{jmfit} in the software package \textsc{aips}. When Gaussian fits could not be applied, a largest angular scale (LAS) was found instead.}
\end{table*}	

		\subsection{Wolf-Rayet stars}

\begin{figure*}	
	\begin{subfigure}{0.36\textwidth}
		\resizebox{\hsize}{!}{\includegraphics{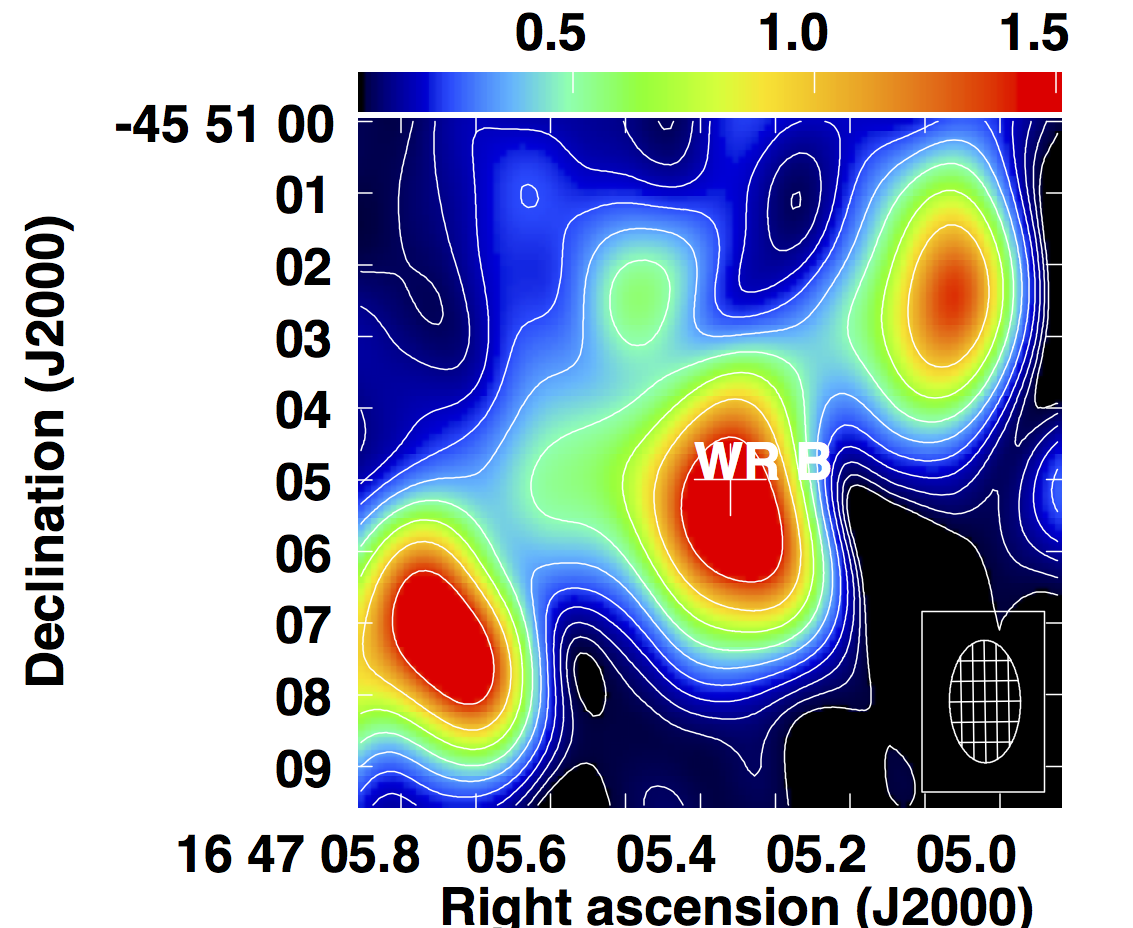}}
		\caption{WR B.}
		\label{fig:WRB_postage}
	\end{subfigure}
	\begin{subfigure}{0.28\textwidth}
		\resizebox{\hsize}{!}{\includegraphics{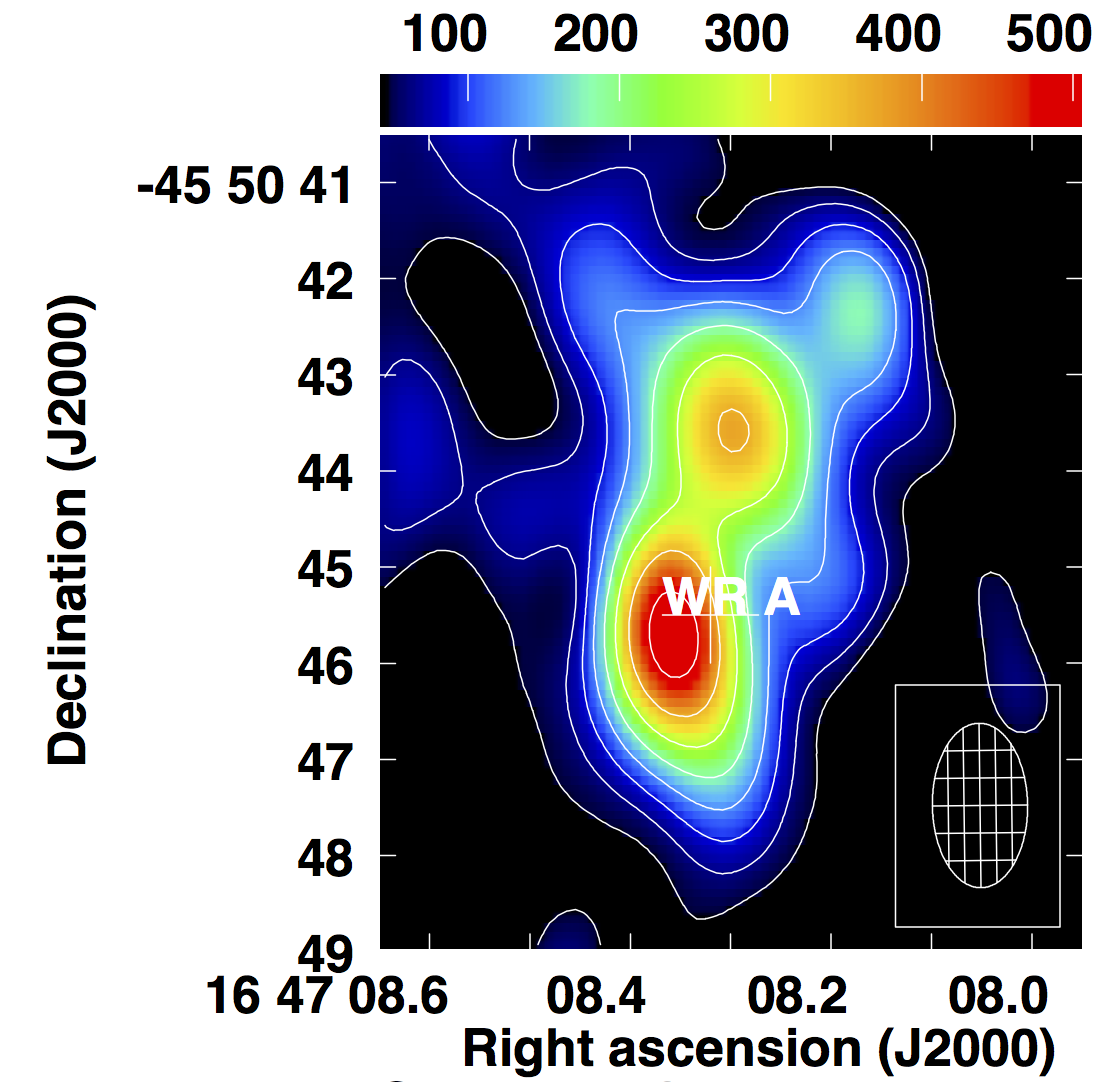}}
		\caption{WR A.}
		\label{fig:WRA_postage}
	\end{subfigure}
	\begin{subfigure}{0.32\textwidth}
		\resizebox{\hsize}{!}{\includegraphics{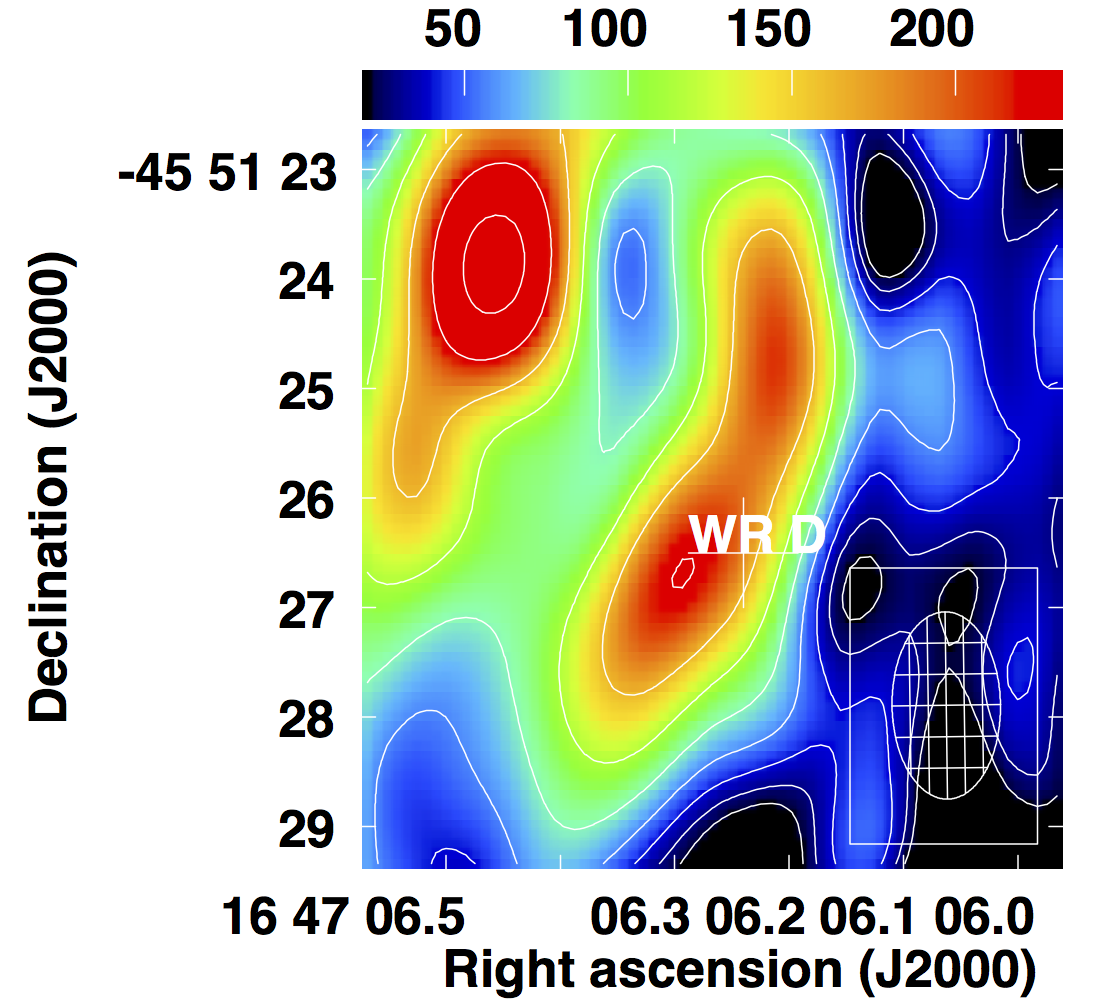}}
		\caption{WR D.}
		\label{fig:WRD_postage}
	\end{subfigure}
	\label{fig:WR_postage}
	\caption{Non-primary beam corrected images of Wolf-Rayets B, A and D, from the \textsc{FullConcat} dataset. Contours are plotted at 1,1.4,2,2.8,4,5.7,8,11.3,16,22.6,32,45.25,64,90.50 $\times$ $\sigma$, the local rms value.}
\end{figure*}

Westerlund 1 contains 24 known WR stars, comprising almost 4\% of the galactic population$\,$\citep[online catalogue,][]{Crowther2019}.  10 WR stars were detected, with 5 WRs not previously detected in the radio. WR stars were detected across a variety of subtypes, with many late-type WN stars detected (WN7, WN8, WN9), as well as half of the WC WRs in the cluster. 4 of the WN stars detected have prior evidence suggesting they may be binary systems, and all of the detected WC stars detected have some evidence of binarity from other observations. Of the 24 WR stars present in Westerlund 1, 3 were outside of the field of view, meaning that almost half of the available WR stars were detected with associated radio emission (10 detections and 1 marginal detection of 21 available WR stars). 

WR P was detected in the ATCA observations, but this result is not presented in Table \ref{table:atca_known} as it was a marginal detection, with a flux of 0.05$\,\pm\,$0.02$\,$mJy. Due to a low SNR value of $\sim$ 2.8, it cannot be listed as a confirmed source, though it's worth noting the caveat that with a slightly different defined background noise level (for example, by considering a sub-image created of this source from the full field of view, which leads to a different absolute cell-size measured in generating the background noise level), the flux reaches above the SNR limit of $\sim$ 3. 

Of the WR stars not detected but within the central field of view, many of these were located in a nearby proximity to a stellar source with high levels of surrounding extended radio emission. Obscuration from this surrounding emission is highly likely so that any radio flux emitted by the fainter stars was not separable from the extended emission. This is a likely cause of the non-detection of WR V, located close to W9, one of the most luminous radio sources present in the cluster. At the location of WR V, the peak flux was found to be 0.3$\,$mJy, with a mean value over the immediate surrounding area of the source of $\sim$0.1$\,$mJy, on the same order as the level of 3$\sigma$ in that region, with $\sigma$ the rms value of the local spatial region. This is especially of note considering this source was detected in previous radio observations$\,$\citepalias[][]{Dougher2010}. 

Other sources affected by this include WR U, located near to W16a, WR R, located near to W26, and WR J, located near W12a.  The other WRs not detected by ATCA, WR W, O, Q, K, C and  H, may emit radio fluxes, but any emission was too faint to be detected above the noise level of the observations. This is especially likely if we assume a thermal spectral index for these sources. 

Fluxes of the WRs from the \textsc{Full9} dataset were compared to \citetalias{Dougher2010}, as shown in Table  \ref{table:atca_do10comp}. Of sources detected in both datasets, consistent fluxes were found for all sources that were not surrounded by extended emission, or in the proximity of nearby radio luminous sources. For sources surrounded by extended emission, the conservative choice of flux determination on these sources (as discussed in section 3), lead to lower flux values, as seen for the source WR S. In other cases, the fluxes were found to be typically slightly higher, due to the increased sensitivity of the more recent observations. 

Spectral indices and limits were determined from the tapered datasets, \textsc{Taper5}, \textsc{Taper9} and \textsc{TaperALMA}, with spectral index values found for 8 WRs and spectral index limits found for 11 WRs. The majority of the WR stars were found to have a spectral index in line with the canonical value for a thermal stellar wind. This indicates that even though a majority of the WRs detected have evidence of binarity from other observations \citep[][]{Clark2019b}, thermal emission is still the dominant mechanism from the outer wind regions. Any colliding winds present must be located at close separations, likely inside of the radio photospheres of the stars. Spectral index limits determined all indicated flattened or positive spectral indices, consistent with expectations of single stars with thermal wind emission.

Out of the WRs, only WR B was found to have a negative spectral index, found to be -0.10$\pm$0.03 as shown in \ref{table:spec_index}. This  indicates the presence of non-thermal emission. WR B also shows the most complex morphology of the WRs, and is surrounded by a large region of extended emission, as seen in Figure \ref{fig:WRB_postage}. The core component of the source was used to determine the spectral index. The negative spectral index determined, $\alpha$ = -0.10 is in line with that found in previous millimetre and radio comparisons$\,$\citepalias[][]{Fenech2018}. 

WR A, as shown in Figure \ref{fig:WRA_postage}, was found to have a positive spectral index of 0.61$\,\pm\,$0.04. This revised spectral index is flatter than the previous spectral index found in \citepalias[][]{Fenech2018}, more in line with the canonical value for partially optically thick free-free emission. WR D, as shown in Figure \ref{fig:WRD_postage}, was found to have a slightly flattened positive spectral index of $\sim$ 0.50, also indicating partially optically thick and thin emission. A revised spectral index limit on WR V , $\alpha$ > 0.47, is consistent with leading to the canonical value for a stellar wind, containing partially optically thick and thin emission. This is in rough agreement with the radio-millimetre spectral index found in \citepalias[][]{Fenech2018}. 

WR G was found to have a highly thermal spectral index $\alpha \sim$ 0.71, and the WN8 and WN9+OB? subtypes WR I and WR L were found to have spectral indices $\sim$ 0.79 and $\sim$ 0.88 respectively. These values are suggestive of a partially optically thick stellar wind with the increased optically thick component possibly due to colliding wind regions. This is because for close in binaries with short period, the material from the wind collision zone (WCZ) can become cooled and cause optically thick thermal emission to dominate the spectrum, causing a steepening of the index over the mm-radio continuum \citep{Pittard2010,Stevens1992}. The suggestion is supported by previous evidence of binarity for sources WR G and L, although WR I has no current evidence indicative of binarity. 

WR S (WN10-11h/BHG) is detected, but it is not bright enough to be seen in the \textsc{Taper9} or \textsc{Taper5} datasets, so can only be given a spectral index limit. The spectral index limit, $\alpha$ > 0.42, is consistent with the canonical stellar wind value, suggestive of partially optically thick emission, and is higher than the previous radio-millimetre spectral index value determined in \citetalias{Fenech2018}. The flux value for WR S, much fainter than in the \citetalias{Dougher2010} observations, was likely affected by its proximity to the radio luminous object W9. 

WR F (WC9d+OB) and WR M (WC9d) both have spectral indices slightly flattened with comparison to the canonical stellar wind value, of $\sim$ 0.51 and $\sim$ 0.53 respectively. This could be due to a higher presence of optically thin or non-thermal components, and is likely related to the binarity of these sources.  Spectral index limits are also derived for a further 11 WR stars, all in line with either a slightly flattened spectral index or the canonical wind spectral index value. WR E, C, and H are all binary candidates, but no evidence is found of a significant non-thermal component for these sources.  For 9 of the WRs with definitive flux detections in the radio, only excepting WR B, the spectral indices and spectral index limits are in line with the value expected for stellar wind emission, making it a reasonable assumption to use their fluxes to determine mass-loss rates.

\subsubsection{Mass-loss rates}

\begin{table*}
	\caption{Mass-loss rate determinations of WRs.}
	\label{table:mass_loss}
	\centering
	\begin{tabular}{p{1.5cm}p{2.5cm}*3{p{2.5cm}}}
		\hline \hline
		Source & Spectral Type & \multicolumn{3}{c}{ATCA $\dot{M}f_{cl}^{\frac{1}{2}}$} (M$_{\odot}$$\,$yr$^{-1}$)  \\
		& & All & 5.5GHz & 9 GHz   \\
		\hline
		WR A & WN7b + OB? & 7.02$\pm$0.35e-05 & - & 4.01$\pm$2.50e-05 \\
		WR D & WN7o  & 5.92$\pm$0.31e-05 & 2.97$\pm$0.74e-05 & 3.09$\pm$0.41e-05   \\
		WR G & WN7o & 3.41$\pm$0.30e-05 & - &  - \\
		WR I & WN8o  & 3.10$\pm$0.11e-05 & 1.62$\pm$0.32e-05 & 2.96$\pm$0.15e-05 \\
		WR L & WN9h+OB?  & 2.72$\pm$0.15e-05 & 1.55$\pm$0.32e-05  & 2.35$\pm$0.16e-05 \\
		WR S & WN10-11h/BHG &0.63$\pm$0.12e-05 & - & 0.34$\pm$0.10e-05   \\
		WR E & WC9   & 3.62$\pm$0.42e-05 & - & 2.90$\pm$0.40e-05  \\
		WR F & WC9d+OB? & 6.62$\pm$0.34e-05 & 3.62$\pm$0.74e-05  & 6.76$\pm$0.44e-05  \\
		WR M & WC9d   & 5.74$\pm$0.71e-05 &- & 3.84$\pm$0.72e-05 \\
		\hline
	\end{tabular}
\tablefoot{Mass-loss rates are calculated from fluxes given by the \textsc{FullConcat}, \textsc{Full5} and \textsc{Full9} datasets. These mass-loss rates are calculated for objects where an approximation of a thermally emitting stellar wind is considered valid, using the Wright and Barlow equation$\,$\citep[][]{WrightBarlow1975}.}
\end{table*}

For all sources with spectral indices found to be consistent with optically thick or thin free-free emission ($\alpha$ > 0), mass-loss rates were determined from their associated fluxes. The Wright and Barlow equation was applied to determine radio mass-loss rates$\,$\citep[][]{WrightBarlow1975}. This equation includes the assumptions that the stellar wind is spherically symmetric and only emits thermal free-free emission. The mass-loss rate, $\dot{M}$ (combined with the clumping factor, $f_{cl}$) can be determined from the source flux as follows, 

\begin{equation}
\dot{M}f_{cl}^{\frac{1}{2}}= 0.095 \times \frac{\mu v_{\infty} S_{\nu}^{\frac{3}{4}}D^{\frac{3}{2}}}{Z \gamma^{\frac{1}{2}} g_{ff}^{\frac{1}{2}} \nu^{\frac{1}{2}} } \quad \textrm{M$_{\odot}$$\,$yr$^{-1}$}
\label{eq:wb}
\end{equation} 

where $\dot{M}$ is the mass-loss rate in M$_{\odot}\,$$\,$yr$^{-1}$, $\mu$ is the mean molecular weight per ion, $v_{\infty}$ is the terminal velocity of the wind (in km$\,$s$^{-1}$), $S_{\nu}$ is the observed flux (in mJy), measured at the frequency $\nu$ (in Hz), $D$ is the distance (in kpc), $Z$ is the ratio of electron to ion density, $\gamma$ is the mean number of electrons per ion, and $g_{ff}$ is the gaunt factor, defined by,

\begin{equation}
g_{ff} \sim 9.77 \left( 1 + 0.13 \log\left(T_{e}^{\frac{3}{2}}/\nu \sqrt{(\bar{Z^{2}})} \right) \right) 
\label{eq:gaunt}
\end{equation} 

where $T_{e}$ is the electron temperature$\,$\citep{LeithererRobert1991}. $f_{cl}$ is the clumping factor, defined by $f_{cl} = < \rho^{2} > / < \rho >^{2}$, where $\rho$ is the density of the wind. This demonstrates the underlying relationship between the mass-loss rate derived from the wind and the intrinsic density. If $f_{cl}$ is taken to be 1, it represents a smooth-wind, with larger values indicating the presence of structure, suggesting a clumped wind. 

For the different stellar types considered, different assumptions are made about the various parameters, $T_{e}, Z, \mu$ and $\gamma$. Many of these follow the assumptions made in \citetalias{Fenech2018}. For OB stars, $\mu$ was taken to be 1.4, and $Z$ and $\gamma$ were both taken to  be 1.0. This relates to a chemistry where H is fully ionised, the dominant form of He is singly ionised, and the He abundance is n$_{He}$ / n$_{H}$ = 0.1. $Z$ and $\gamma$ were also taken to be 1.0 for WRs, and $\mu$ was taken as 4 for WN6 and earlier spectral types, with 2.0 used for spectral types later than WN6 and 4.7 used for WC8 and WC9 stars. For RSGs, LBVs and YHGs, $\mu$ was assumed to be 1.4, $Z$ was taken to be 0.9, and $\gamma$ was taken as 0.8. For all early-type hot stars, $T_{e}$ was taken to be $0.5 \times T_{eff}$, where $T_{eff}$ is the effective temperature of the star. For cool stars, including the YHGs and RSGs, $T_{e}$ was taken to be 10,000 K. The distance, $D$, was assumed to be 5$\,$kpc, as justified in the Introduction.

$\dot{M}f_{cl}^{\frac{1}{2}}$ values for 9 of the 11 WRs detected, where spectral indices or index limits indicated optically thick or thin emission, in line with thermal stellar winds, are presented in Table \ref{table:mass_loss}. The absolute mass-loss rates are calculated for the \textsc{FullConcat}, \textsc{Full9}, and \textsc{Full5} datasets. The majority of these sources show consistency with the expected value for WRs, of $\sim$ 10$^{-5}$ M$_{\odot}$$\,$yr$^{-1}$. 

A slightly lower value is determined for WR S, though this could be due to the possibility that WR S is believed to be a post-interaction product of a previously close binary, and so has already experienced extra stripping of its outer layers during that phase. Our value for WR S, $\dot{M}f_{cl}^{\frac{1}{2}}$ 6.3$\,\pm\,$1.2$\,\times\,$10$^{-6}$ $M_{\odot}$$\,$yr$^{-1}$, is larger than determined by previous modelling of this object, where $\dot{M}f_{cl}^{\frac{1}{2}}$ was found to be 2.16$\,\times\,$10$^{-6}$ $M_{\odot}$$\,$yr$^{-1}$$\,$\citep[][]{Clark2014}, but is consistent with the measurement from millimetre observations of 5.18$\,\times\,$10$^{-6}$  $M_{\odot}$$\,$yr$^{-1}$$\,$\citepalias[][]{Fenech2018}. 

As WR B was found to have a negative spectral index, this indicates significant non-thermal emission associated with the source, and so the approximation for a thermally emitting stellar wind is not valid for this star. An estimate of $\dot{M}f_{cl}^{\frac{1}{2}}$ for the WR B flux in the \textsc{FullConcat} dataset was still made, found to be $\sim$ 1.90$\,\times\,$10$^{-4}$ M$_{\odot}$$\,$yr$^{-1}$. This is much higher than the rest of the cohort, supporting the conclusion that this source cannot be realistically approximated with a purely thermally emitting stellar wind. 

The radio flux of WR A gives an $\dot{M}f_{cl}^{\frac{1}{2}}$ value, 7.02$\,\times\,$10$^{-5}$ M$_{\odot}$$\,$yr$^{-1}$,  consistent within uncertainties to the value found in the millimetre \citepalias{Fenech2018} of 7.19$\,\times\,$10$^{-5}$  M$_{\odot}$\,$yr^{-1}$. As WR A was found to have extended emission, only the flux from its core component was used to calculate the mass-loss rate. As a core component could not be resolved for this source at 5.5$\,$GHz, $\dot{M}f_{cl}^{\frac{1}{2}}$ values were only determined for the \textsc{FullConcat} and \textsc{Full9} datasets. 

WR D was found to have a $\dot{M}f_{cl}^{\frac{1}{2}}$  value of 5.92$\,\pm\,$0.31$\,\times\,$10$^{-5}$ M$_{\odot}$$\,$yr$^{-1}$, larger than the mm value of 2.07$\,\times\,$10$^{-5}$ M$_{\odot}$$\,$yr$^{-1}$$\,$\citepalias{Fenech2018}. WR G was also found to have a larger mass-loss rate calculated from the radio fluxes, with a value of 3.41$\,\pm\,$0.30$\,\times\,$10$^{-5}$  M$_{\odot}$$\,$yr$^{-1}$ from the radio observations, in contrast to the mm value of 2.22$\,\times\,$10$^{-5}$ M$_{\odot}$$\,$yr$^{-1}$$\,$\citepalias{Fenech2018}. This was also the case for WR M, where the radio mass-loss rate was found to be  5.74$\,\pm\,$0.71$\,\times\,$10$^{-5}$ M$_{\odot}$$\,$yr$^{-1}$, with the mm mass-loss rate previously measured as 3.57$\,\times\,$10$^{-5}$ M$_{\odot}$$\,$yr$^{-1}$\citepalias{Fenech2018}.

WR I is found to give a lower value in the radio than in the millimetre. The mass-loss rate determined in the radio was 3.10$\,\times\,$10$^{-5}$ M$_{\odot}$\,$yr^{-1}$, in comparison to the $\dot{M}f_{cl}^{\frac{1}{2}}$ value calculated from the mm of 3.55$\,\times\,$10$^{-5}$ M$_{\odot}$$\,$yr$^{-1}$ $\,$\citepalias{Fenech2018}, however due to the uncertainties on the radio $\dot{M}f_{cl}$, the two values could still potentially be consistent. Higher values are also found in the millimetre for $\dot{M}f_{cl}^{\frac{1}{2}}$ for WR E. The $\dot{M}f_{cl}^{\frac{1}{2}}$ value from the radio observations was calculated to be 3.62$\,\pm\,$0.42$\times$10$^{-5}$ M$_{\odot}$$\,$yr$^{-1}$  for WR E, which can be compared to the mm mass-loss rate from the millimetre, of 5.24$\,\times\,$10$^{-5}$\citepalias{Fenech2018}.

We can also compare our mass-loss rates to the previous radio results in \citetalias{Dougher2010}.$\,$\citetalias{Dougher2010} find mass-loss values for two of the WR stars they detect - WR L and WR F. For WR L, they find a mass-loss rate of  2$\,\times\,$10$^{-5}$(v$_{\infty}$/1000km$\,$s$^{-1}$)M$_{\odot}$$\,$yr$^{-1}$. By taking the terminal velocity of this source to be v$_{\infty}$ $\sim$ 700$\,$km$\,$s$^{-1}$, this leads to a value of 1.4$\times$10$^{-5}$M$_{\odot}$$\,$yr$^{-1}$. In our observations, we determine a higher mass-loss rate of 2.72$\,\times\,$10$^{-5}$$\,$ M$_{\odot}$$\,$yr$^{-1}$. 

\citetalias{Dougher2010} found a mass-loss rate of 3.3$\,\times\,$10$^{-5}$(v$_{\infty}$/1000km$\,$s$^{-1}$) M$_{\odot}$$\,$yr$^{-1}$ for WR F. With an expected v$_{\infty}$ value of 1200$\,$km$\,$s$^{-1}$, this would lead to a mass-loss rate of 3.96$\,\times\,$10$^{-5}$ M$_{\odot}$$\,$yr$^{-1}$. Our mass-loss calculation of WR F results in a higher value, 6.62$\pm\,$0.34$\,\times\,$10$^{-5}$ M$_{\odot}$$\,$yr$^{-1}$.  We can also consider the predicted mass-loss rate derived for WR F of $\sim$ 3.16$\,\times\,$ 10$^{-5}$ M$_{\odot}$$\,$yr$^{-1}$ from \citep[][]{Clark2011}. These values can be reconciled if there are higher levels of clumping in the inner parts of the winds, as the \citep[][]{Clark2011} value is derived from optical and near-IR data. 

Mean mass-loss rates could be determined for the different classes of WR stars, with a mean mass-loss rate over the total ensemble calculated to be 4.61 $\times$ 10$^{-5}$ M$_{\odot}$$\,$yr$^{-1}$. The average mass-loss rate for the WN classes was determined as 4.26 $\times$ 10$^{-5}$ M$_{\odot}$$\,$yr$^{-1}$, and for WC subclasses, it was found to be 5.33 $\times$ 10$^{-5}$ M$_{\odot}$$\,$yr$^{-1}$. In general, the WC subtype can be seen to have higher values of mass-loss rates \textit{on average}, but there were only measurements from three WC stars, and one of these is a likely binary (WR F).

These values can be considered to be fairly consistent with previous mass-loss determinations of WR stars, with  a value from \citep[][]{Leitherer1997}, of $\dot{M} \sim$ 4 $\times$10$^{-5}$ M$_{\odot}$$\,$yr$^{-1}$ over all spectral sub-types. Measurements from \citep[][]{Cappa2004} gave $\dot{M} \sim$ 4$\,\pm\,$3$\times$10$^{-5}$ M$_{\odot}$$\,$yr$^{-1}$ for WN, so our WN mean value can be seen to be consistent, but their measurement for the WC subtype was $\sim$2$\,\pm\,$1$\times$10$^{-5}$ M$_{\odot}$$\,$yr$^{-1}$, lower than our average value found for this spectral subtype. The overall spread of the mass-loss rates is found to be similar to that seen in \citetalias{Fenech2018}, from 1.52 - 9.75 $\times$10$^{-5}$ M$_{\odot}$$\,$yr$^{-1}$ (ignoring the WN10-11h star/BHG, WR S). 

\subsubsection{Clumping ratios} 

\begin{table*}
	\caption{Clumping ratios of WRs.}
	\label{table:clumping_ratio}
	\begin{tabular}{p{1.5cm}p{2.3cm}*2{p{2.3cm}}p{3.5cm}*2{p{2cm}}}
		\hline \hline
		Source & Spectral Type  &\multicolumn{2}{c}{ATCA$\,$$\dot{M}f_{cl}^{\frac{1}{2}}$} (M$_{\odot}$$\,$yr$^{-1}$)  & ALMA $\dot{M}f_{cl}^{\frac{1}{2}}$ (M$_{\odot}$$\,$yr$^{-1}$)  & $\frac{f_{cl, 5.5}}{f_{cl, ALMA}}$ & $\frac{f_{cl,9}}{f_{cl, ALMA}}$ \\
		& & 5.5GHz & 9 GHz &&&   \\
		\hline
		WR A & WN7b + OB? & 6.06$\pm$0.59e-05 & 6.63$\pm$ 0.26e-05 & 6.63$\pm$0.14e-05 & 0.84 $\pm$0.21 & 1.00 $\pm$ 0.09 \\
		WR D & WN7o  & - & 1.84$\pm$0.41e-05 & 1.59$\pm$0.12e-05 & - &  1.34 $\pm$ 0.63  \\
		WR G & WN7o  & -  & 1.50$\pm$0.35e-05 &1.87$\pm$0.14e-05 & -  &  0.64 $\pm$ 0.31 \\
		WR I & WN8o    & 1.70$\pm$0.32e-05 & 3.06$\pm$0.20e-05 & 3.15$\pm$0.05e-05  & 0.29$\pm$0.11 & 1.47 $\pm$ 0.19 \\
		WR L & WN9h+OB? & 1.66$\pm$0.31e-05 & 2.67$\pm$0.16e-05 & 3.36$\pm$0.05e-05 & 0.24$\pm$0.09 & 0.63 $\pm$ 0.08 \\
		WR F & WC9d+OB?  & - & 3.07$\pm$0.25e-05 & 6.24$\pm$0.22e-05 & - & 0.24 $\pm$ 0.04 \\
		WR M & WC9d   & - & 1.23$\pm$0.25e-05 & 2.59$\pm$0.25e-05 & - & 0.23 $\pm$ 0.10 \\
		\hline
	\end{tabular}
\tablefoot{Clumping ratios were derived from calculated values of $\dot{M}f_{cl}^{\frac{1}{2}}$ from fluxes measured with the \textsc{Taper5} and \textsc{Taper9} datasets, to values found for $\dot{M}f_{cl}^{\frac{1}{2}}$ from fluxes given by \textsc{TaperALMA}. In order to calculate clumping ratios from different frequencies of emission, a constant mass-loss rate  is assumed throughout the wind (normalised to the $\dot{M}f_{cl}^{\frac{1}{2}}$ values given by\textsc{TaperALMA}).}
\end{table*}

Due to the different wavelength regimes utilised for the calculating $\dot{M}f_{cl}^{\frac{1}{2}}$, then different clumping factors may play a role in the value of $\dot{M}f_{cl}^{\frac{1}{2}}$ calculated, assuming a constant mass-loss rate throughout the wind. Unfortunately, the intrinsic mass-loss rate for these winds is unknown, so we can only consider the relative degrees of clumping for different wavelength regimes. This still gives an indication of the different levels of clumping present at varying radial geometries within the stellar winds.

Clumping ratios were determined by comparing stellar mass-loss rates at different frequencies. The datasets which contained tapered u-v visibilities were used, to allow for the comparison between the millimetre and radio observations as previously discussed. Clumping ratios were derived for 7 WR stars from the tapered datasets, shown in Table \ref{table:clumping_ratio}. These ratios were calculated assuming a normalisation of the mass-loss rate (using $\dot{M}f_{cl}^{\frac{1}{2}}$ values from \textsc{TaperALMA}) and clumping ratios were derived separately for the \textsc{Taper5} and the \textsc{Taper9} mass-loss rates against this normalised value.

A large spread of clumping ratios were found over the different Wolf-Rayet stars. The majority of the WR stars (WR A, WR G, WR L, WR F and WR M) demonstrate the expected behaviour of the clumping factor, decreasing as the wavelength increases. This is not seen for WR D, which has a clumping ratio of 1.34, significantly above 1. This could indicate that the wind is highly clumped even out at radio wavelengths, in the most extended part of the wind. The adverse clumping ratio (above the expected fraction for an outer part of the wind) could be related to the fact that WR D is a binary candidate. As the calculated mass-loss rate is reliant on assuming fluxes are dominated by thermal emission from stellar winds of a single star, non-thermal radio components due to binarity could have led to these unexpected results.

These sources don't give a clear indication of a uniform trend for clumping ratios between the mm and the radio-photosphere, but in general support the belief that in stellar winds, clumping should decrease within the wind as a function of radial distance from the star.
	   
	   \subsection{Cool supergiants and hypergiants}
	   
	   \begin{figure}
	   	\centering
	   	\resizebox{0.4\textwidth}{!}{\includegraphics{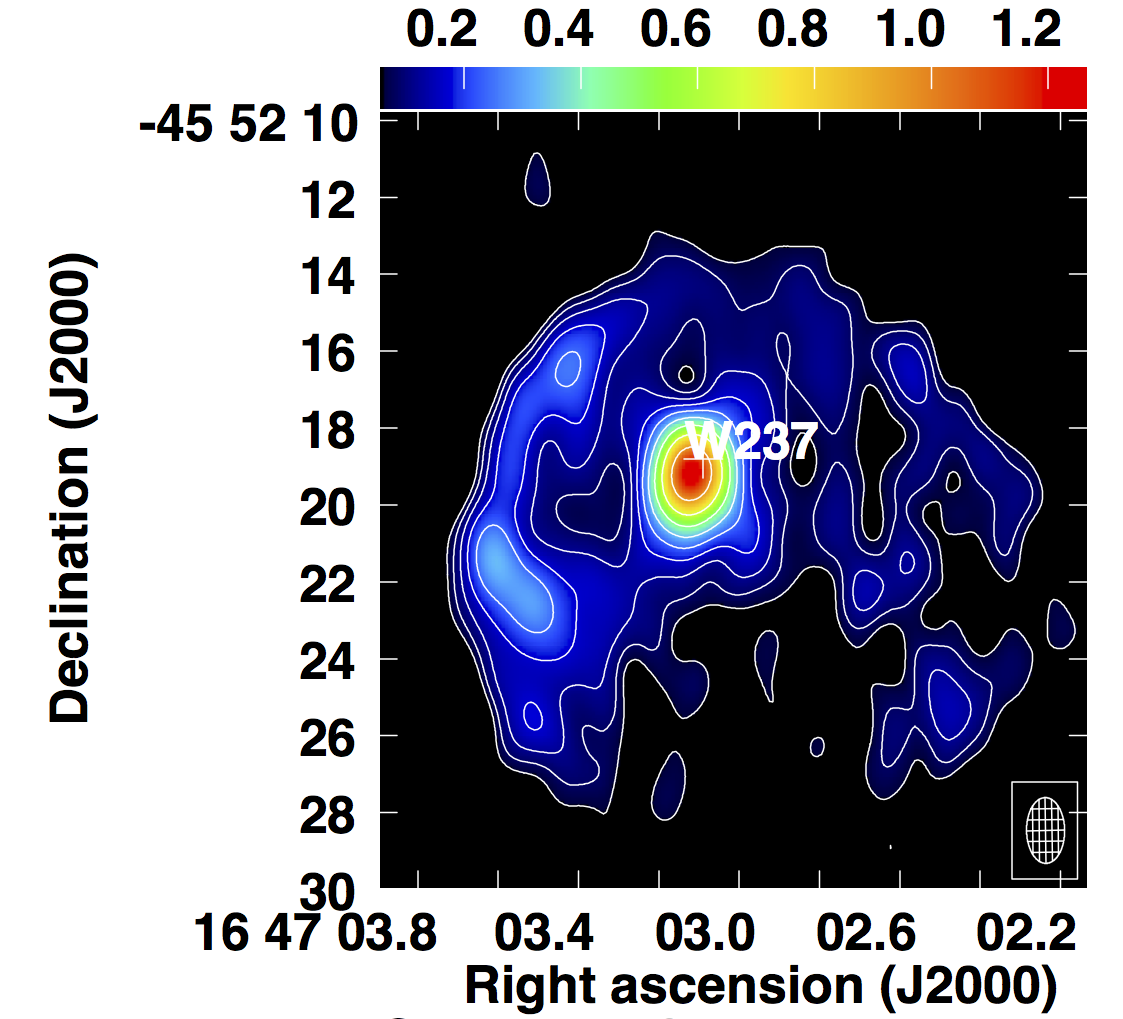}}
	   	\caption{Non-primary beam corrected image of RSG W237 from the \textsc{FullConcat} dataset. Contours are plotted at 1,1.4,2,2.8,4,5.7,8,11.3,16,22.6,32,45.25,64,90.50 $\times$ $\sigma$, the local rms value.}
	   	\label{fig:W237_postage}
	   \end{figure}
	   
	   \begin{figure}
	   	\centering
	   	\resizebox{0.4\textwidth}{!}{\includegraphics{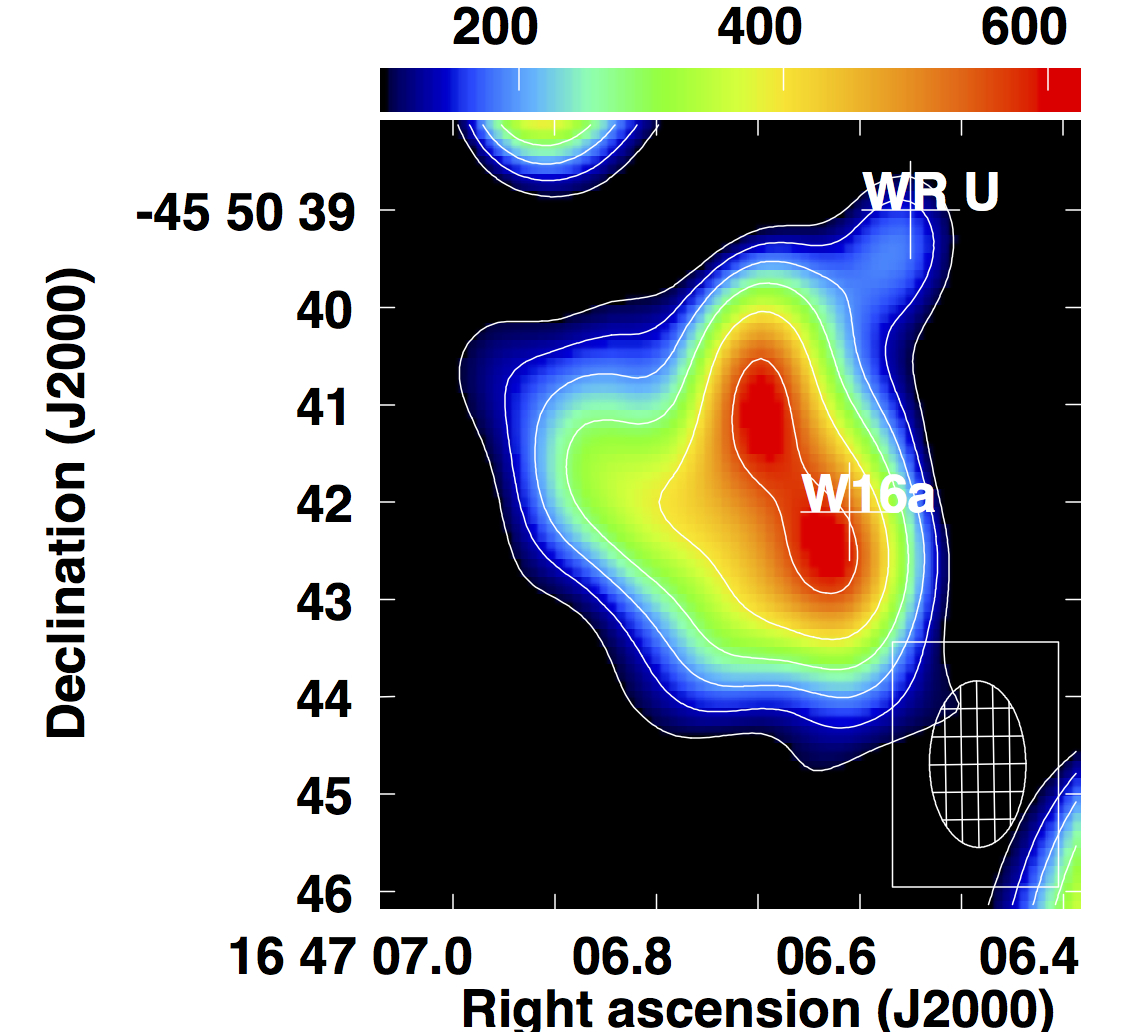}}
	   	\caption{Non-primary beam corrected image of YHG W16a from the \textsc{FullConcat} dataset. Contours are plotted at 1,1.4,2,2.8,4,5.7,8,11.3,16,22.6,32,45.25,64,90.50 $\times$ $\sigma$, the local rms value.}
	   	\label{fig:W16a_postage}
	   \end{figure}
	   
	   \begin{figure}
	   	\centering
	   	\resizebox{0.35\textwidth}{!}{\includegraphics{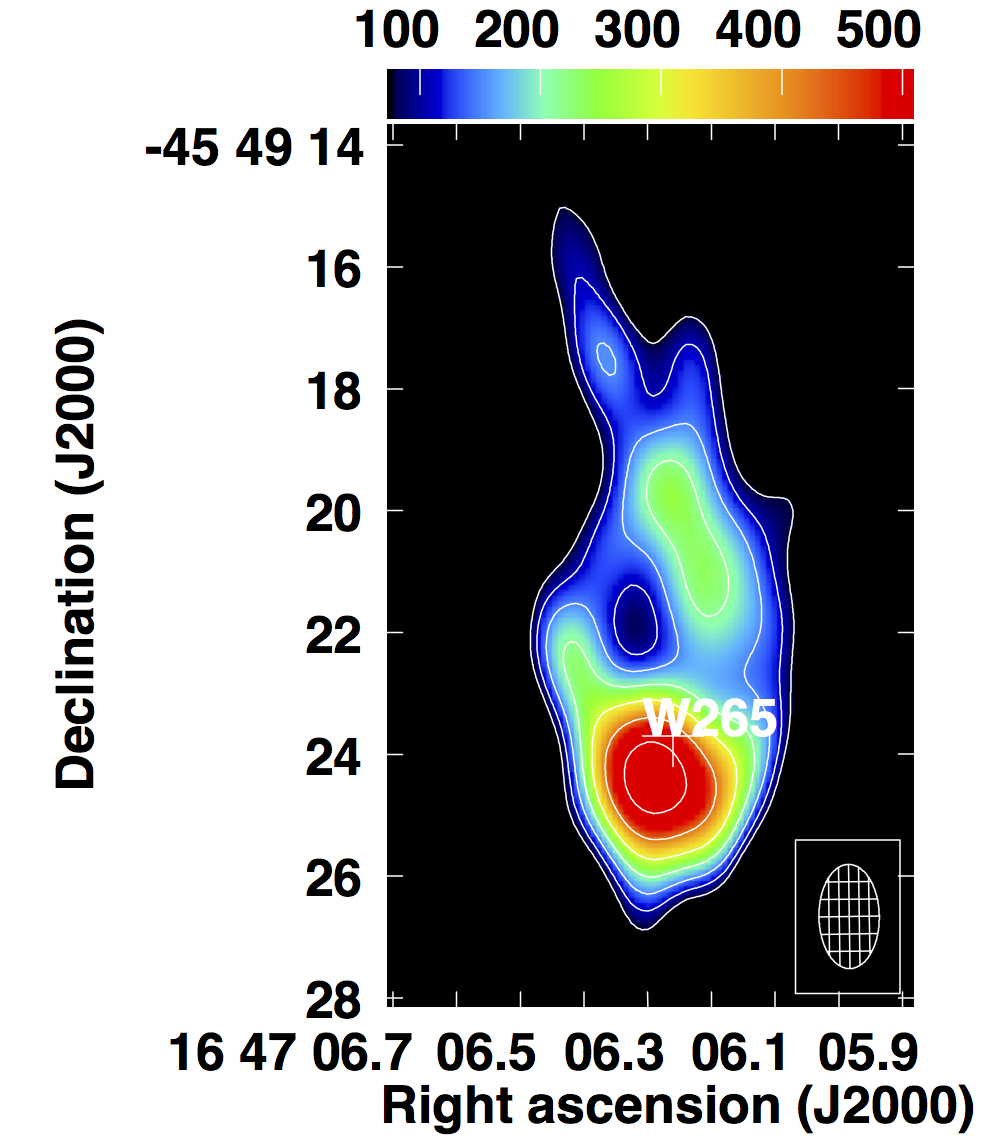}}
	   	\caption{Non-primary beam corrected image of YHG W265 from the \textsc{FullConcat} dataset. Contours are plotted at 1,1.4,2,2.8,4,5.7,8,11.3,16,22.6,32,45.25,64,90.50 $\times$ $\sigma$, the local rms value.}
	   	\label{fig:W265_postage}
	   \end{figure}
	   
	   \begin{table}
	   	\caption{Mass-loss estimates for the YHGs and RSGs.}
	   	\label{table:mass_loss_sgs}
	   	\centering
	   	\begin{tabular}{p{1.5cm}p{2.3cm}p{3.2cm}}
	   		\hline \hline
	   		Source & Spectral Type & ATCA $\dot{M}f_{cl}^{\frac{1}{2}}$ (M$_{\odot}$$\,$yr$^{-1}$)  \\
	   		\hline
	   		W12a & F1Ia$^{+}$ & 1.66$\pm$0.02$\times$10$^{-5}$ \\
	   		W4a & F3Ia$^{+}$ & 2.59$\pm$0.02$\,$$\times$10$^{-5}$  \\
	   		W32 & F5Ia$^{+}$ & 3.01$\pm$0.50$\times$ 10$^{-6}$  \\
	   		W265 & F5Ia$^{+}$ & 1.10$\pm$0.03$\times$ 10$^{-5}$ \\
	   		W237 & M3Ia  & 2.00$\pm$0.02 $\times$$\,$10$^{-6}$ \\
	   		W75 & M4Ia & 7.27$\pm$0.48$\times$$\,$$10^{-7}$ \\
	   		W20 & M5Ia & 3.28$\pm$0.03 $\times$$\,$10$^{-6}$  \\
	   		\hline
	   	\end{tabular}
   	\tablefoot{Fluxes were used from the core components of the emission attributed to the YHGs and RSGs, given by the \textsc{FullConcat} dataset. These mass-loss rates are calculated for objects where an approximation of a thermally emitting stellar wind is considered valid, using the Wright and Barlow equation \citep[][]{WrightBarlow1975}.}
	   \end{table}
	   
	   Nine of the ten RSGs and YHGs present in Westerlund 1 were detected, with the singular exception of the YHG W8a. 8 of these sources are resolved, with extended nebulae detected around all of these sources. As these are all cool supergiants and hypergiants, the typical photospheric temperatures ($\sim$ 4000K for RSGs, 4000 - 8000K for YHGs) would not be enough to cause the ionisation of the nebulae material and lead to radio emission. Instead, a diffuse radiation field caused by the hundreds of O and B stars present in the cluster was postulated as the source of the ionisation, allowing for the stellar nebulae to be observed. 
	   
	   The structures of the extended emission around these sources were seen to be asymmetric and complex in nature. A mixture of morphologies were seen around several of the sources, including the bow shock structure as seen around W237 in Figure \ref{fig:W237_postage} and the cometary morphology around W265, as shown in Figure \ref{fig:W265_postage}.  
	    
	   Spectral indices were determined for all detected RSGs and YHGs except the source W32. A marginal detection could be found from the \textsc{Taper9} dataset at $\sim$ 0.13$\,$mJy, (with a SNR of only $\sim$ 2.69), alongside a marginal detection of $\sim$0.08$\,$mJy in the \textsc{Taper5} dataset (at a SNR of $\sim$ 1.26). The ALMA flux of W32 from \citetalias{Fenech2018}, 0.38$\,\pm\,$0.07$\,$mJy, in combination with marginal detections suggest that the emission is of a thermal nature. 
	   
	   Core components of the source emission were used to calculate spectral indices. For most of the cool supergiants and hypergiants, spectral indices were found to be extremely flat, such as 0.00$\,\pm\,$0.04 for W237,  -0.01$\,\pm\,$0.12 for W75, and -0.01$\,\pm\,$0.04 for W4a, consistent with composite emission of both optically thin and thick components, as seen previously in the millimetre \citepalias{Fenech2018}.  Amongst the YHGs, some sources were found to have more negative spectral indices, including W265 and W12a, which had spectral indices of -0.13$\,\pm\,$0.23 and -0.30$\,\pm\,$0.07 respectively. Due to the relatively high uncertainty on these values, this is consistent with the presence of non-thermal emission, but also consistent with optically thin emission surrounding these sources.  
	   
	   W16a, shown in Figure \ref{fig:W16a_postage}, was found to have a spectral index -0.10$\,\pm\,$0.05, which would suggest the presence of non-thermal or optically thin emission components. This object may actually have a much flatter spectral index, as there is possible source confusion due to extremely close presence of WR U, which is not readily distinguished from the emission associated with W16a. 
	   
	   Mass-loss rates were calculated for sources for which the emission present could be attributed to optically thin and thick components in the stellar wind, excluding sources with potential non-thermal emission. The core component of flux detected for these stellar sources were used in the mass-loss calculation, with the exception of W75, which was found to be far fainter than the rest of the sources. W26 and W16a were not considered as no core component could be isolated within from the emission associated with these sources. The mass-loss estimates are presented in Table \ref{table:mass_loss_sgs}. 
	   
	   We can compare the range of mass-loss rates found for RSGs, $\sim$ 10$^{-7}$ - 10$^{-6}$  M$_{\odot}$$\,$yr$^{-1}$, to that found for field RSGs, which vary from 10$^{-6}$ - 10$^{-4}$ M$_{\odot}$$\,$yr$^{-1}$ \citep{Jura1990,Sylvester1998}. An RSG with similar morphology, VY CMa, has been found to have mass-loss values of $\dot{M}f_{cl}^{\frac{1}{2}} \sim$ 10$^{-3}$ M$_{\odot}$$\,$yr$^{-1}$ or greater$\,$\citep[][]{Shenoy2016}. For W237, a time-averaged mass-loss rate was previously calculated in \citetalias{Fenech2018}, $\sim$ 10$^{-5}$ M$_{\odot}$$\,$yr$^{-1}$, which is significantly higher than the value determined here, but uses the measure of the ionised mass of the extended nebula surrounding the source. 
	   
	   There is a clear caveat in comparing the mass-loss estimates determined here to those calculated in the literature, as the methods used to determine the $\dot{M}f_{cl}^{\frac{1}{2}}$ values here are derived from the free-free emission in the wind, whereas $\dot{M}f_{cl}^{\frac{1}{2}}$ values in the literature are typically calculated from fitting models to the metal lines of stars, as seen for YHGs such as $\rho$ Cas, \citep{Lobel1998, Lobel2003}, or from the modelling of low excitation CO lines and using models to extrapolate the mass-loss rates, as has been carried out for the YHG IRC+10420 \citep{Oudmaijer1996,CastroCarrizo2007}, and the RSG VY CMa \citep[][]{Decin2006}. These models often require many assumptions to be made about the velocity and temperature structure of the stellar wind, as well as the heating and cooling processes occurring within the wind, and the relative gas and dust densities. These methods also use information from the emission of different wavelength regimes, and so the effect of clumping and structure within the wind could lead to inconsistent results between our estimates and the literature. 
	   
	   The range of $\dot{M}f_{cl}^{\frac{1}{2}}$  $\sim$ 10$^{-6}$ - 10$^{-5}$  M$_{\odot}$$\,$yr$^{-1}$ is found for the YHGs present in Westerlund 1. Some direct comparisons can be made to previous observations, including W4a, where a mass-loss rate estimate was inferred from millimetre observations, $\sim$ 10$^{-5}$M$_{\odot}$$\,$yr$^{-1}$$\,$\citepalias{Fenech2018}. This is found to be consistent with our estimate. \citetalias{Dougher2010} suggest an estimate of $\sim$ 10$^{-5}$ M$_{\odot}$$\,$yr$^{-1}$ for W265. Our mass-loss rate for this source, 1.10$\,\pm\,$0.03$\times$ 10$^{-5}$ M$_{\odot}$$\,$yr$^{-1}$, is therefore also consistent with this earlier estimate, even though the \citetalias{Dougher2010} mass-loss rate was derived from more the extended region of emission.  
	   
		\subsection{OB stars, supergiants, hypergiants, LBVs, and sgB[e] stars}

		Of the radio detections made from the survey of Westerlund 1, 11 of these are OBA stars. This includes the LBV A2 Ia star W243, the sgB[e] star Wd1-9, 7 confirmed OB supergiants and 2 radio sources previously discovered in Do10, D09-R1 and D09-R2, which have the assigned spectral type of blue supergiants (BSGs). The detections cover a range of spectral types, from late type O supergiants (O9Ib) to early type B supergiants (B1Ia). No stars from a later spectral type were detected. This indicates that the majority of stars within the cluster exhibit typical thermal spectral behaviour, and so their expected radio fluxes fall well below the noise level for these observations \citepalias[as was indicated by the lack of detections for most OB stars in the cluster from previous millimetre observations;][]{Fenech2018}. 

		\subsubsection{OB super and hypergiants}

		In total, 9 OB stars were detected in the cluster with 5 new radio detections. W10 is a known double-lined spectroscopic binary (SB2) binary in the cluster, with stellar components B0.5I + OB, and was found to have an associated flux density of 0.16$\,\pm\,$0.05$\,$mJy. W18 and W19, both early type B supergiants, with spectral types B0.5Ia and B1Ia, were found to have fluxes 0.36$\,\pm\,$0.06$\,$mJy and 0.08$\,\pm\,$0.03$\,$mJy respectively. Additionally, two late type O supergiants were newly detected, W1031 of spectral type 09III with a flux density of 0.90$\,\pm\,$0.05$\,$mJy, and W1056, a 09.5II star, with a flux density measured of 0.09$\,\pm\,$0.02$\,$mJy. 

		Cluster members D09-R1 and D09-R2, which have the current spectral type designation of BSG, were previously discovered in the radio \citepalias{Dougher2010}. These sources currently lack an accurate spectral classification. Fluxes were found for D09-R1 for both a core component and total value of 1.29$\,\pm\,$0.03$\,$mJy and 5.44$\,\pm\,$0.10$\,$mJy. D09-R1 is embedded in an area of diffuse radio emission, as was previously indicated in \citetalias{Dougher2010}. A flux was also determined for the source D09-R2, and found to be 0.96$\,\pm\,$0.06$\,$mJy.

		As these sources were also detected in the millimetre, spectral indices could be determined. They were found to have fairly flat, slightly negative spectral indices of $\sim$ -0.01 and -0.03 for D09-R1 and D09-R2 respectively. This suggests that these objects experience a composite behaviour. This could include a combination of partially optically thick, optically thin and non-thermal emission components. The spectral indices are roughly consistent with previous radio-mm spectral indices, with the spectral index determined  for D09-R1 slightly flatter in comparison to a previous radio-mm measurement of $\sim$ 0.17$\,$\citepalias{Fenech2018}. 
	
		W17 and W15 are both OB stellar sources with previous radio detections \citepalias{Dougher2010}. W15 is currently assigned the spectral classification O9Ib, and was found to have a flux density of 1.65$\,\pm\,$0.09$\,$mJy. W17, a late O supergiant, was found to have a radio flux of 0.98$\,\pm\,$0.04$\,$mJy for the core component of the source, surrounded by an extended component with a total flux density of 2.04$\,\pm\,$0.08$\,$mJy. W17 was also detected in the millimetre, and so a spectral index was derived for this source of $\sim$ -0.03, suggesting a composite spectra, which may be made up of optically thin and thick, as well as non-thermal components. This revised spectral index was found to be significantly flattened in comparison to the previous radio-mm spectral index found of $\sim$ 0.33$\,$\citepalias{Fenech2018}. 
	
		As the emission from W17 could be attributed to optically thick and thin components, an approximation could be made to a thermal stellar wind in order to determine the mass-loss rate. A mass-loss rate was determined for this source of 1.38$\,\pm\,$0.03$\,\times\,$10$^{-4}$ M$_{\odot}$$\,$yr$^{-1}$. This is higher than would be expected for a late O supergiant. This suggests that the approximation of a thermally emitting stellar wind is not suitable, and may indicate the presence of binarity, supported by the fact that W17 has already been found to exhibit variability \citep{Bonanos2007}.  
		
		6 of the OB supergiants did not have counterpart millimetre detections. This included a range of spectral types, over B1Ia - 09Ib. W15 was the only one of these sources previously detected in the radio in \citetalias{Dougher2010}. 
	
		By using the flux limits of the ALMA dataset presented in \citetalias{Fenech2018}, spectral index limits could be determined. These limits involved the use of the \textsc{FullConcat} dataset, rather than the tapered datasets, due to the intrinsic faintness of the sources. All of the stars were found to have spectral index limits consistent with non-thermal emission, with especially strongly negative spectral indices for W15 and W1031. We considered the approximation of the radio fluxes to thermal stellar wind emission, to determine mass-loss rate estimates. The range of mass-loss rates found for the OB supergiants varied from 4.8$\,\times\,$10$^{-6}$ - 1.31$\,\times\,$10$^{-4}$ M$_{\odot}$$\,$yr$^{-1}$. 
	
		Although the lower end of this range is in line with expectations, in general these values are much higher than would be expected for late O/early B type supergiants, $\sim$ 10$^{-6}$ - 10$^{-7}$ M$_{\odot}$$\,$yr$^{-1}$. The high mass-loss rates, along with the negative spectral index limits derived, indicate that attributing the radio emission of these stars to stellar winds from \textit{single stars} is not valid. It is then worth considering that these stars may be binaries. 
	
		W15 has no prior evidence from hard X-ray detections or RV variability, but is most clearly constrained by its spectral index limit of $<$ -1.15, strongly suggesting non-thermal emission. If W15 is a binary system, then this system must be either in an eccentric or a very wide orbit in order to be consistent with other observational results for this star. A wide system would allow for the WCZ to occur outside of the radio photosphere, so non-thermal emission would not be obscured by surrounding thermal emission from a stellar wind, and the proximity between the WCZ and the star would be insufficient for shocks to produce hard X-rays, as would typically be expected for binaries \citep[][]{Pittard2018}. This is also supported by recent observational evidence that binary stars have been seen to present X-ray signatures that are consistent with single stars of a similar spectral type \citep[][and references within]{Clark2019}.
	
		This system can be compared to the WR star, WR 140. WR 140 is considered to be the archetypal wind colliding binary (WCB), consisting of a WC7 WR and an O5 companion. W140 has been found to have variable emission due to the high eccentricity of the orbit \citep[$\sim$ 0.90,][]{Fahed2011}, with different variabilities measured at various radio wavelengths$\,$\citep{Dougherty2011}. WR 140 has been detected to have X-ray variability, including in the hardness of its X-ray spectra over the orbital phase$\,$\citep{Corcoran2011}. If W15 is a similar system, then this scenario could explain  the lack of a hard X-ray detection. 
	
		Another piece of evidence supporting this hypothesis is the variability seen in the radio flux between the \citetalias{Dougher2010} 8.6$\,$GHz and ATCA 9$\,$GHz observations - where the increase in flux is almost of a factor 2. This variance could be due to a change in orbital phase between the two measurements. W15 is an object that would benefit from a further time-domain study, to map out possible changes of orientation over orbital phase, in order to provide constraints on the properties of the system. 
	
		W10 is the only previously confirmed binary of the detected OB supergiants, first presented as an SB2 in \cite{Clark2008}. W10 has also been previously found to be an RV variable. W10 was recently found to experience softer and dimmer X-ray emission than other SB2s in Westerlund 1, suggesting an eccentric binary where the emission from a WCZ was less present in observations taken in 2005 \citep{Clark2019}. This may be supported by the weak constraint on the nature of emission for W10, with a spectral index limit of $<$-0.04, consistent with a composite spectra. 
	
		W18 was found to have a negative spectral index limit, $<$ -0.43, strongly indicative of non-thermal emission. This stellar source has no current associated evidence for binarity from X-ray observations and no RV detections. This doesn't rule out binarity for this source, as bias in the observational limits of these methods could mean that a binary system is present that was not easily detected by RV variations. The surrounding dense cluster wind and material from other stars in the cluster could also obscure X-ray emission, as there is diffuse X-ray emission known to be present throughout the cluster \citep[][]{Muno2006}. W19 is also a cluster member with detected radio emission but no prior evidence of binarity from RV or X-ray measurements, although it has previously been found to be H$\alpha$ variable. It has the weakest constraint on its spectral index limit, allowing for a flat composite spectrum of emission. It is also of note that W18 and W19, alongside W10 and W15, have previously been found to be associated with long period/aperiodic variability, supportive of the conclusion of binarity \citep{Bonanos2007}.
	
		The other two stellar cluster members detected, W1031 and W1056, are both confirmed RV variables. The period for W1031 is not currently determined. W1056 with a candidate period of less than 10 days. It's of note that W1056 has a much looser constraint on the spectral index, $<$ 0.02, in comparison to the other stars with late O spectral classification, which may be consistent with a flat composite spectra, which would be expected if stellar winds are obscuring some of the radio emission from the colliding wind region. 
	
		We suggest that we would not expect to detect any emission from purely thermal stellar wind emitters in the radio. This is supported by the lack of millimetre detections in \citetalias{Fenech2018} for the majority of O/B supergiants in Westerlund 1. With this in mind, along with the previously observed binarity characteristics for many of these objects in the UV and X-ray, we can infer that \textit{all} of the OB supergiants we detect are either binaries or binary candidates. We believe most of these objects are highly likely to be long period binaries, where the separation between the stellar components is sufficiently high for non-thermal emission from colliding wind regions to occur outside the region of the stellar wind radio photospheres. Therefore, we have managed to detect 3 new binary candidates, that are likely long period binaries where RV variations could not be detected. 
		
		\subsubsection{W243}
		
		W243, a LBV with assigned spectral type A2 Ia, is detected in the ATCA observations. The spectral index was found to be $\sim$ 0.82, consistent with previous measurements. This is indicative of a thermal spectrum, with a slight steepening in comparison to the canonical stellar wind value. Our mass-loss determination for this object was found to be 3.12$\,\times\,$10$^{-5}$ M$_{\odot}$$\,$yr$^{-1}$.  
		
		\citetalias{Dougher2010} determined a mass-loss estimate for W243 on the order 1.1$\times$10$^{-5}$(v$_{\infty}$/200km$\,$s$^{-1}$)M$_{\odot}$$\,$yr$^{-1}$. From an assumed v$_{\infty}$ $\sim$ 550$\,$km$\,$s$^{-1}$ for W243, the mass-loss rates therefore can be considered to be consistent. We can also compare this mass-loss rate to the previous millimetre observations, which gave a $\dot{M}$ of $\sim$ 2.5 $\times$10$^{-5}$ M$_{\odot}$$\,$yr$^{-1}$ $\,$\citepalias{Fenech2018}. The radio and mm values are both higher than the expected mass-loss rate, $\dot{M} \sim$ 8 $\times$ 10$^{-7}$ M$_{\odot}$$\,$yr$^{-1}$, given from modelling in$\,$\citet{Ritchie2009}. Clear limitations for this modelling included the issue of fitting to all H and He emission features prominent in the spectra, suggesting that the spectral type approximation used was not suitable in order to generate an accurate fit. 
		
		This LBV $\dot{M}f_{cl}^{\frac{1}{2}}$ value is also larger than a mass-loss rate of $\sim$ 1.35$\,\times\,$10$^{-5}$ M$_{\odot}$$\,$yr$^{-1}$, found from radio observations of a post-RSG LBV HD160529 $\,$\citep{LeithererRobert1995}. It can also be compared to other LBV measurements from sources including AFGL2338, AG Car, FMM362 and the Pistol Star$\,$\citep{ClarkLBV,Najarro2009,Groh2009}, which have mass-loss rates over the range of 3 -  6$\,\times\,$10$^{-5}$M$_{\odot}$$\,$yr$^{-1}$, all larger than the mass-loss rate calculated for W243.  
		
		An estimate of the clumping ratios are also found for this object, between the millimetre and radio data. The clumping ratios are found to be $\sim$ 0.027 and $\sim$ 0.07 when comparing 5.5$\,$GHz and 9$\,$GHz respectively to the millimetre observations, suggesting a much higher level of structure and clumping within the inner regions of the wind sampled by the mm observations, that then decreases with increasing distance from the star.
		
		\subsubsection{W9}
	
		\begin{figure}
			\centering
			\resizebox{0.5\textwidth}{!}{\includegraphics{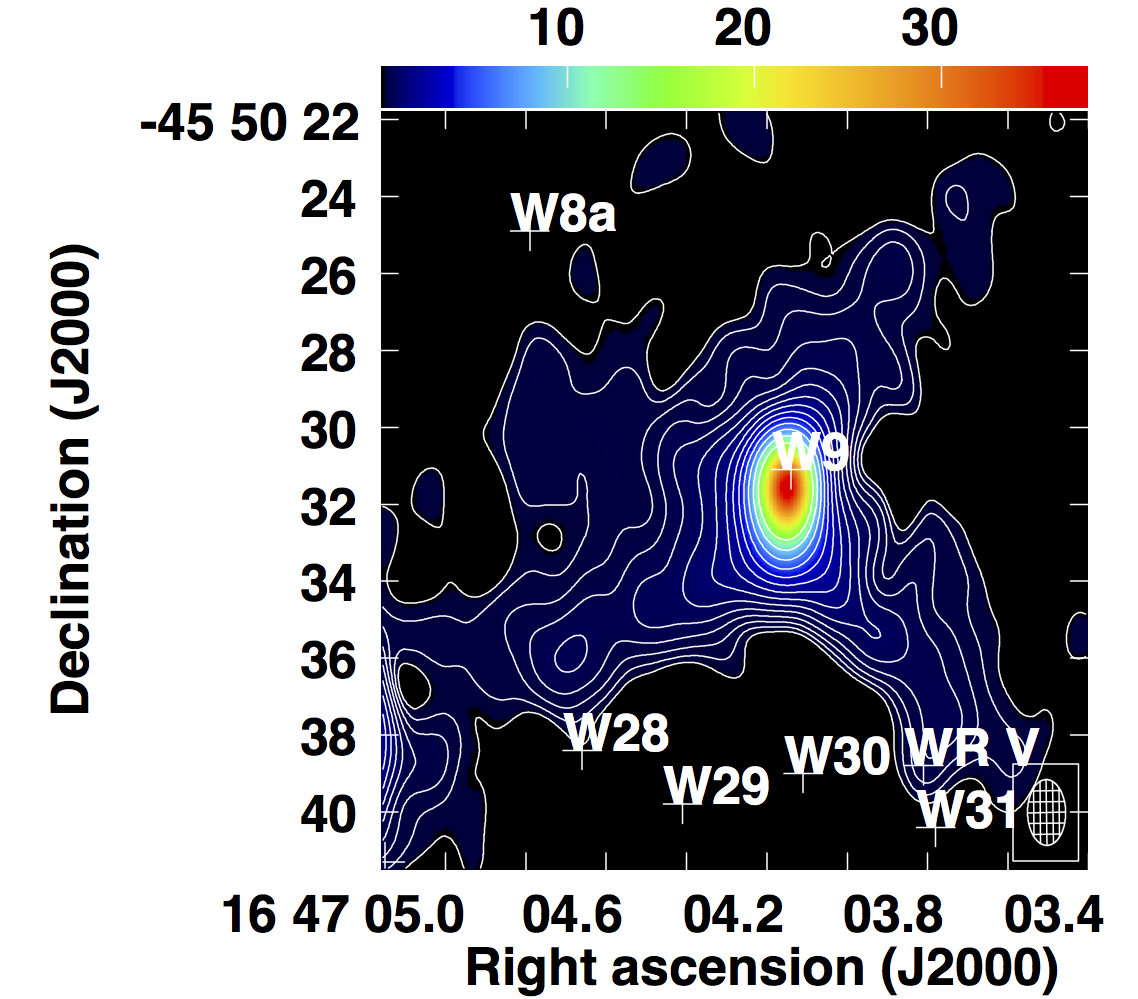}}
			\caption{Non-primary beam corrected image of stellar source sgB[e] Wd1-9 from the \textsc{FullConcat} dataset. Contours are plotted at 1,1,1.4,2,2.8,4,5.7,8,11.3,16,22.6,32,45.25,64,90.50 $\times$ 150$\,$$\mu$Jy.}
			\label{fig:Wd19_postage}
		\end{figure} 
		
		W9 can be seen in Fig \ref{fig:Wd19_postage}, where the two filaments of emission to the south-east and south-west of the source as observed previously in millimetre observations$\,$\citepalias[\citealt{Fenech2017};][]{Fenech2018} can be seen at much lower resolution in the radio, as well as an additional third filament to the north-west.  The flux of the core compact source region of W9 was found to be 27.38$\pm$0.06$\,$mJy, and the extended region around it gives a total flux value of 83.14$\,\pm\,$0.41$\,$mJy. The spectral index of W9 indicates thermal emission, in line with the canonical value for a stellar wind, with a value of $\alpha$ = 0.68$\,\pm\,$0.02. This value is consistent with the previous measurements.  
		
		As W9 was found to have a thermal nature derived from the spectral index, we could then make an estimate of the approximate mass-loss rate for this source, despite the  evidence that this source is known to be a binary from its X-ray emission \citep[][]{Clark2013}. The mass-loss rate was calculated to be $\dot{M} f_{cl}^{\frac{1}{2}} \sim$ 13.03$\,\pm\,$0.01$\,\times\,$10$^{-5}$ M$_{\odot}$$\,$yr$^{-1}$, similar to the value found from Model 2 of Wd1-9 in \citetalias{Dougher2010}. 
		
		\section{Origin of the extended emission within the cluster}
		
		As was previously introduced in Section 4, there is a pervasive level of diffuse radio emission that can be seen throughout the cluster. This can be seen most clearly by looking at Figure \ref{fig:fig1}, which shows the ATCA observations separated by configuration. The spatial scales measured by the shortest configuration with a maximum baseline of 750m demonstrate a significant amount of diffuse radio emission present throughout the cluster. Much of this extended radio emission appears to be centred on the radio bright sources, such as W9, W20, or even the WR star WR B, but some of the extended emission cannot be attributed to any of the known stellar sources. These areas still contain compact sources with no stellar counterpart, as discussed in section 5.2. These areas of concentrated emission that appear throughout the cluster indicate the high level of varying structure present, and that the overall structure of the diffuse emission is clumped and non-spherical. 
		
		\subsection{Stellar and cluster wind interaction} 
		
		From the morphology of this diffuse background, as seen clearly in all the figures displaying the whole cluster region (Figures \ref{fig:fig1} - \ref{fig:fig3}, \ref{fig:figB1}), we can see this diffuse emission is more concentrated towards the centre of the cluster. At extended distances from stellar sources, a large level of radio emission is detected, indicating interactions between the stellar winds of the sources and a cluster wind. The interaction of stellar winds with the surrounding environment has previously been discussed in \citet{Mackey2015} as an origin of the sculpted nebulae surrounding Wd1-26. Typically, this diffuse emission would be expected to arise as a result of interaction between young massive stars and remnants of their parent giant molecular cloud (GMC), but at an expected age of 5$\,$Myr, this is not believed to be the case for Wd1. 
		
		This is also supported by the general lack of evidence to support any ongoing star formation within the cluster. This is currently supported by results from many mid- and near-infrared surveys (2MASS, GLIMPSE, MIPSGAL) that have included the cluster within their field of view \citep[][]{Clark2015}, as well as a lack of any detections of massive young stellar objects (YSOs) in the vicinity of the cluster, or the presence of any methanol masers, a typical signature of star forming regions (SFRs). We can also compare the general geometry of the diffuse emission to stellar clusters that are known to be located in star forming regions, such as the clusters Danks 1 and 2, located in G305. The structure of the diffuse emission around these clusters are seen to follow a much simpler classical wind blown shell geometry \citep{ClarkPorter2004,Davies2012}.
		
		Some future observations that could help us to constrain the origin of the gas within this cluster would be to search more explicitly for spectral signatures that constrain the origin to natal emission (with linked evidence of ongoing SFRs present), or from the outflows of the stellar population. A signature of recent star formation (and hence, diffuse emission from the natal remnants) already mentioned is that of methanol masers. Future studies of transition frequencies, such as 12.2$\,$GHz for methanol, or 1.7$\,$GHz for OH masers, would provide evidence of star formation and hence constrain the origin of the gas; or a lack of detection would help suggest that the current conclusion of stellar emission is more valid.
		
		A more explicit method to constrain the nature of the gas due to stellar emission would be to search for He and N emission lines, using optical/near-IR spectroscopy. N [II] emission lines have previously been detected around some cluster members, including the RSG population \citep[][]{Clark2010}. N [II] emission lines are known to occur in the ejecta of other cool evolved stars with nearby stellar companions, where the external ionisation of the ejecta is due to material from the stellar companion's wind, as in the case of the YHG HR 8752 \citep[][]{Stickland1978}. Observational evidence of these emission lines within the diffuse regions of the gas would therefore provide further evidence that the extended emission can be related to stellar winds of massive stars in the cluster.

		Previous evidence of a cluster wind in Wd1 has been found in the X-ray$\,$\citep{Muno2006}, radio$\,$\citepalias{Dougher2010} and millimetre$\,$\citepalias{Fenech2018}. This cluster wind is also indicated by the common orientation of several of the resolved extended nebulae around the cool super and hypergiants in the cluster, where bow shocks and cometary tails are indicative of a general outward direction, originating from the centre of the cluster$\,$\citep{Andrews2018}. The origins of cluster winds are believed to be related to the feedback of stellar winds that become thermalized and then drive the cluster winds. The properties of the mechanisms of cluster wind propagation are not fully determined, due to uncertainties in the conversion of stellar kinetic energy to thermal energy$\,$\citep{Chevalier1985,Canto2000,StevensHartwell2003}. The origin of the possible cluster wind in Westerlund 1 is not confirmed, but is believed to be due to a diffuse radiation field arising in the cluster from the radiation pressure of the hundreds of OB stars present. 
		
		The overall flux of the cluster wind can be considered. A value of the total flux in the cluster was determined for the central 1.5' of the cluster (as carried out in \citetalias{Dougher2010}). This led to a value of the total flux present of 635$\,$mJy at 9$\,$GHz and 640$\,$mJy at 5.5$\,$GHz. This was found to be significantly larger than the flux determined by \citetalias{Dougher2010} at 8.6$\,$GHz of 422$\,$mJy and 461$\,$mJy at 4.8$\,$GHz, as well as the single dish measurements of the flux of the cluster from \citet{Kothes2007}, of 450$\,$mJy and 499$\,$mJy at 4.8$\,$ and 8.6$\,$GHz respectively. From a total flux in the \textsc{FullConcat} dataset of 647$\,$mJy, this leads a to flux for the extended emission (with the stellar components subtracted by considering flux from known detections), of 408$\,$mJy over the \textsc{FullConcat} dataset. A general estimate of ionised mass present in the cluster can be considered from $\sim$ 408$\,$mJy flux found for the extended emission at 9$\,$GHz over a assumed spherical region of diameter 1.5$\,$arcminutes), gave a value of 29$\,$M$_{\odot}$, (assuming a general plasma temperature of 10kK), almost twice the ionised mass previously determined for the extended emission in the cluster$\,$\citepalias[\citealt{Kothes2007};][]{Dougher2010}. 
		
		By characterising the cluster wind, we can hope to learn more about the stellar ejecta, and how it is impacted not only by the physics of stellar outflows, but also by the differing surrounding environments of these stars, and how it may sculpt and shape the ejecta as is clearly seen here. The presence of this cluster wind indicates the important role that an ionisation field may play on stellar populations within large clusters and associations. 
		
		Understanding the diffuse emission and its origin is important not only for determining accurately the level of material that is associated with each stellar object, but also in starting to consider how the stellar ejecta are affected by a cluster wind. Characterising the cluster wind and the interaction between an intra-cluster medium and stellar winds is important for many areas of astrophysics. The processes involved have an impact on the complete life-cycles of stars, from start to end. Cluster winds may play a role in both inhibiting and triggering the process of star formation, and may also lead to the disruption of natal giant molecular clouds. It also helps in formalising the feedback from stars into their surrounding environment, understanding resultant galactic superwinds as well as the production of cosmic rays, for which the production mechanisms require a significant level of interstellar material, or a strong radiation field, evidence for both of which can be seen in Westerlund 1. 
		
		A dense background may also affect the resulting SNe that will occur from many stars when they reach their stellar endpoint. Interactions between stars and their surrounding circumstellar nebulae can impact the resultant SNe light curve, and so affect the relative ratios observed of each SNe type. This may in turn impact the expected ratios of observed SNe in the early universe. This shows a clear wider implication of this environment for a wide range of fields of study within astrophysics. 
		
		\section{Conclusion}
		
		This paper has presented the full results of a radio census of the massive stellar cluster Westerlund 1.  Observations were made using ATCA, over 4 different pointings of varying maximum baselines from 750$\,$m to 6$\,$km, and over two spectral windows centred at 5.5$\,$GHz and 9$\,$GHz. The census aimed to present a direct follow up to previous observations of the rich co-eval population of Westerlund 1, especially the previous radio ATCA observations$\,$\citepalias{Dougher2010}. 30 stellar sources were detected in the radio, over a range of spectral types. A strong and diffuse radio background was also clearly seen to be present in the cluster. 5 new detections were made in the radio for WR stars in the cluster, and 5 new radio detections were made of OB supergiants in the cluster. For sources that had previously been detected in the radio, new details of surrounding extended structures of emission have come to light, due to the better resolution and sensitivity now possible with ATCA. From the 5 new OB radio detections, we were able to determine the presence of 3 new binary candidates. 
		
		Many of the sources have been found to exhibit emission that has a spectral index fully consistent with the canonical value for thermal emission from a single stellar wind. The use of the Wright and Barlow formalism, \citep[][]{WrightBarlow1975}, allowed for mass-loss rates to be determined for a number of the sources. This was carried out for sources which were found to have spectral indices indicative of thermal emission. Clumping ratios could then be determined for many of the WR stars, with comparison between the radio observations and previously published millimetre observations$\,$\citepalias{Fenech2018}. 
		
		The majority of sources detected were found to be resolved. The origin of extended emission around most of these sources is not clear. One possible explanation is due to binarity, but many of sources detected with extended emission are not binary candidates. Many of the sources are also not expected to have experienced large levels of variability or pulsations, which could have been another solution to explain the presence of the extended nebulae. The majority of the sources are post-Main Sequence objects, RSGs, YHGs or WRs. There is a possibility that all these stars may have experienced previous epochs of extreme mass-loss, but the degree of structure surrounding sources is not seen to vary uniformly by spectral type, which indicates it may not be due to the evolutionary status of the star.
		
		Understanding the origin of the diffuse background and the extent of its interaction with the stellar wind material is of importance. The boundary between the stellar wind material and the surrounding diffuse emission is not clear, and shows a clear need for more observations of similar environments as well as more developed models, taking into account the interaction between stellar winds of massive stars and their parent clusters. Future observations, especially of the neutral material present in the cluster, would allow for additional consideration of the intra-cluster material present as well as the extended nebulae of each source, allowing for updated considerations to be made of the nebulae masses and geometries around many of the stars, especially the population of the YHGs and RSGs present in Wd1. 
		
		Higher resolution, longer baseline, radio observations of these objects would also be of great use. The use of higher resolution observations would  allow for constraints to be applied on how to distinguish between the diffuse cluster material and the outflows from the stellar sources. Such observations may allow for any possible CWB regions to finally be resolved for many of the binary candidates present in the cluster. This includes the non-thermal emitters Wd1-15 and the other OB supergiants, W10, W18, and W1031.
		
	 More tailored modelling of mass-loss rates and consideration of clumping at different radial geometries of the wind that quantifies the radial stratification taking place would also help to provide better constraints on the observations of the stellar winds across a large variety of different stellar spectral types, as seen to populate Westerlund 1.

	\begin{acknowledgements}
	This paper makes use of the following ALMA data: ADS/JAO.ALMA/2013.1.00897.S. ALMA is a partnership of ESO (representing its member states), NSF (USA) and NINS (Japan), together with NRC (Canada), NSC and ASIAA (Taiwan), and KASI (Republic of Korea), in cooperation with the Republic of Chile. The Joint ALMA Observatory is operated by ESO, AUI/NRAO and NAOJ. H. Andrews wishes to acknowledge STFC for the funding of a PhD Studentship. D. Fenech wishes to acknowledge funding from a STFC consolidated grant (ST/M001334/1).
	\end{acknowledgements}
	%-------------------------------------------------------------------
		\bibliographystyle{aa}
		\bibliography{ref.bib} 

	\onecolumn
	\appendix
		\section{Tables}
		\begin{table*}[h!]
			\caption{Flux densities for known stellar sources from the \textsc{Full9} dataset.}
			\label{table:atca_do10comp}
			\centering
			\begin{tabular}{p{1.2cm}p{2.2cm}p{2.2cm}p{2cm}}
				\hline \hline
				Source  & Flux$_{Full9}$  & Flux$_{Do10}$ & Notes \\
				\hline
				WR A  & 0.48 $\pm$ 0.04 & 0.5 $\pm$ 0.06 & \\
				& 1.34 $\pm$ 0.09 & - & 	\\
				WR D  & 0.34 $\pm$  0.06 & - &  \\
				WR B  &3.24 $\pm$ 0.06$^{c}$& - &  Crowded\\
				& 9.50$\pm$ 0.18$^{t}$ &  4.3 $\pm$ 0.4$^{r}$ & \\
				WR I  & 0.45 $\pm$ 0.03 & - & Isolated \\
				WR V  & - & 0.4 $\pm$ 0.06 & Crowded \\
				WR L  & 0.40 $\pm$ 0.04 & 0.4 $\pm$ 0.06  & Isolated \\
				WR S  & 0.08 $\pm$ 0.03 & 0.3 $\pm$ 0.06 &  \\
				WR E  & 0.11 $\pm$ 0.02 & -  & Isolated \\
				WR F & 0.34 $\pm$ 0.03 & 0.3 $\pm$ 0.06 & Isolated \\
				WR M & 0.16 $\pm$ 0.04 & - & \\
				\hline 
				W16a  & 1.79 $\pm$ 0.09 & 1.6 $\pm$ 0.3  & Crowded \\
				W12a   & 1.66 $\pm$ 0.04$^{c}$ & - &  \\
				& 3.28 $\pm$ 0.08 & 2.9 $\pm$ 0.3$^{r}$& \\
				W4a & 1.79 $\pm$ 0.04$^{c}$  & 0.8 $\pm$ 0.08  & \\
				& 4.19 $\pm$ 0.11$^{t}$ & 2.2 $\pm$ 0.2$^{r}$ & \\
				W32 & 0.16 $\pm$ 0.04 & 0.4 $\pm$ 0.06 & Crowded \\
				W265  & 0.94 $\pm$ 0.06$^{c}$ &  - & Isolated  \\
				& 2.72 $\pm$ 0.16$^{t}$ & 2.3 $\pm$ 0.3$^{r}$ & \\
				W237 & 1.26 $\pm$ 0.04$^{c}$ & 1.8 $\pm$ 0.2 & Isolated  \\	
				& 7.01 $\pm$ 0.22$^{t}$ & 5.6$\pm$ 2.2$^{r}$ & \\
				W75  & 0.26 $\pm$ 0.04 & 0.3 $\pm$ 0.06 & Isolated \\
				W20  & 2.39 $\pm$ 0.05$^{c}$ &  - &  \\
				& 16.62 $\pm$ 0.27$^{t}$ & 3.8 $\pm$ 0.4$^{r}$ & \\
				W26  & 152.15 $\pm$ 0.33 &  20.1 $\pm$ 2.0$^{r}$ & \\ 
				\hline
				W17   & 0.98$\pm$ 0.05  & - & Crowded\\
				&1.27 $\pm$ 0.07 & 1.7$\pm$0.2 & \\
				W243  &  1.65 $\pm$ 0.05 & 1.5 $\pm$ 0.2 & Isolated \\
				\hline
				W9 &  30.47$\pm$0.09$^{c}$ & 24.9 $\pm$ 2.5 & Crowded \\
				& 80.8 $\pm$ 0.5$^{t}$ &  30.5 $\pm$ 3.0$^{r}$ & \\
				D09-R1 & 1.20 $\pm$ 0.04$^{c}$ & 0.7 $\pm$ 0.07 & Crowded \\
				& 4.66 $\pm$ 0.14$^{t}$ & 6.5 $\pm$ 1.2$^{r}$ &\\ 
				D09-R2 &  0.88 $\pm$ 0.08 & 0.7 $\pm$ 0.06 & Crowded \\
				\hline	
				W15 & 0.84 $\pm$ 0.10 & 0.6 $\pm$ 0.06  & \\
				\hline
			\end{tabular}
		\tablefoot{This takes in data from the ATCA image of all configurations at 9 GHz, and applies the SEAC tool with consideration of segmentation where necessary. Flux densities are given in $mJy$. This is all assuming $\sigma_{f}$ = 3, $\sigma_{s}$ = 5. Core and total flux densities are given for relevant sources, noted by $^{c}$, $^{t}$ respectively. These values can be compared to the fluxes found in \citetalias{Dougher2010}, at 8.6$\,$GHz, where the superscript $^{r}$ refers to objects that were found to be spatially resolved.}
		\end{table*}
	
		\begin{table*}[h!]
		\caption{Flux values of source detections from \textsc{FullConcat}, for ALMA detected sources discovered in \citetalias{Fenech2018}.}
		\label{table:alma_uncat}
		\centering
		\begin{tabular}{p{3cm}p{2.5cm}*2{p{2.8cm}}p{3.5cm}}
			\hline \hline
			FCP18 Source & RA & DEC  &  Flux$_{\textsc{FullConcat}}$ (mJy)  \\
			\hline
			2  & 16 46 58.60265  &   -45 50 31.4920  & 1.09 $\pm$ 0.02 \\ 
			3  &  16 46 58.88981 &     -45 50 28.7929 & 1.88$\pm$0.02$^{c}$ \\
			&&& 3.92$\pm$0.03$^{t}$ \\
			6 &  16 46 59.40638   &  -45 50 37.1942 & 2.11$\pm$0.02$^{c}$  \\
			&&& 4.88$\pm$0.04$^{t}$ \\
			8 &  16 46 59.57887  &  -45 50 26.6947 & 0.31 $\pm$ 0.04 \\
			12 &  16 47 1.01416   &  -45 50 36.2975 & 0.75 $\pm$ 0.04 \\
			16 &  16 47 1.81753 &    -45 51 23.0987 & 2.11	$\pm$	0.04  \\
			22 &  16 47 2.33444  &   -45 51 21.5992 &  0.27 $\pm$ 0.02$^{c}$ \\
			&& & 0.66 $\pm$ 0.04$^{t}$  \\
			27  & 16 47 2.85135  &   -45 51 19.1996  & 1.12 $\pm$ 0.05  \\	
			35  &  16 47 3.68413  &   -45 51 16.5000 & 0.31 $\pm$ 0.04  \\
			37 & 16 47 3.74162   &  -45 50 24.0000  & 0.38 $\pm$ 0.06 \\
			39   & 16 47 3.88514  &   -45 51 6.3000 & 1.80$\pm$0.03$^{c}$\\
			&&& 4.63 $\pm$0.09$^{t}$ \\
			49   & 16 47 4.22972 &     -45 51 16.2000 & 	0.35 $\pm$ 0.05 \\
			60 & 16 47 5.06241  &   -45 51 2.3997 &  1.59	$\pm$ 0.02  \\ 
			64  &  16 47 5.75158  &   -45 51 7.1991 & 2.81 $\pm$ 0.03 \\ 
			72  & 16 47 6.29690  &   -45 50 44.9985 &  0.60 $\pm$ 0.03 \\ 		
			73  & 16 47 6.38337 &    -45 51 13.7984 & 1.56 $\pm$ 0.02 	\\  
			77 & 16 47 6.46965  &  -45 51 23.9983 & 0.33	$\pm$ 0.03 \\	
			80 & 16 47 6.81406 &    -45 51 10.4978  & 1.24 $\pm$ 0.05 \\ 	
			82  & 16 47 7.30222  &  -45 51 11.0970 & 0.45 $\pm$ 0.04 \\
			85  & 16 47 7.87660   &  -45 51 15.2959 & 0.09 $\pm$ 0.02 \\
			86 & 16 47 8.30665  &   -45 50 43.4949 &   0.32 $\pm$ 0.03 \\	
			87   & 16 47 8.33615  &    -45 51 19.7949  &  0.40 $\pm$ 0.03 \\ 
			90   & 16 47 8.50825  &   -45 51 11.0944 & 2.13 $\pm$ 0.05 \\ 
			91   & 16 47 8.53623 &   -45 50 38.6944 & 0.30 $\pm$ 0.05  \\	
			93  & 16 47 8.70934  &    -45 51 14.6939  & 2.28 $\pm$ 0.04  \\ 
			95 & 16 47 8.87997  &    -45 50 6.2935  & 0.48 $\pm$ 0.02  \\ 
			97  & 16 47 8.93797 &    -45 50 30.2933 &  0.28 $\pm$ 0.03  \\ 
			99   & 16 47 9.16849  &    -45 51 3.2927 & 0.28 $\pm$ 0.03 \\ 
			100  & 16 47 9.36973  &   -45 51 12.2921  & 0.76 $\pm$ 0.04 \\	
			101  &  16 47 10.77535   &  -45 50 30.2875 & 0.35 $\pm$ 0.04 \\ 
			\hline
		\end{tabular}
	\tablefoot{The source determinations are gathered from SEAC. The use of the additional segmentation tools were applied, with thresholds set at $\sigma_{f}$ = 3, $\sigma_{s}$ = 5. For any significantly extended sources, core and total regions are specified with the use of superscripts $^{c}$ and $^{t}$ respectively. Segmentation parameters were set to 0.8333 as the smoothing parameter in the \textsc{Gauss} segmentation, and noise thresholds in the \textsc{Noise} segmentation were set with a top threshold of 50\% and a bottom threshold of 30\%.}
	\end{table*}	

	\begin{table*}[h!]
	\caption{Spectral index values found for ALMA detected sources discovered in \citetalias{Fenech2018}.}
	\label{table:spec_index_almaonly}
	\centering
	\hspace*{-1cm}
	\begin{tabular}{p{3cm}*3{p{3.2cm}}p{3cm}} 
		\hline \hline
		FCP18 Source & Flux$_{\textsc{Taper5}}$ (mJy) & Flux$_{\textsc{Taper9}}$ (mJy) & Flux$_{\textsc{TaperALMA}}$ (mJy) &  Spectral Index ($\alpha$) \\
		\hline
		3 & 2.69 $\pm$ 0.04 & 1.91 $\pm$ 0.06 & 1.18  $\pm$ 0.14 & -0.26 $\pm$ 0.05  \\
		6  & 3.98$\pm$0.05 & 3.75 $\pm$ 0.08 &  1.04 $\pm$ 0.12 & -0.49 $\pm$ 0.05  \\
		16 &  1.66 $\pm$ 0.07 &  1.91 $\pm$ 0.07 &  0.96 $\pm$ 0.06 & -0.23 $\pm$ 0.03	\\
		22 & -  & 0.30 $\pm$ 0.05 & 0.10 $\pm$ 0.03 & -0.81 $\pm$ 0.12 \\
		27 & 0.54 $\pm$ 0.09 &  0.85 $\pm$ 0.06 & 0.34 $\pm$ 0.05 &  -0.27 $\pm$ 0.07 	\\	
		39  & 2.04 $\pm$ 0.08 & 2.28 $\pm$ 0.07 & 1.57 $\pm$ 0.15 & -0.10 $\pm$ 0.04  \\ 
		60 &  1.42 $\pm$ 0.07 & 1.05 $\pm$ 0.06 & 0.68 $\pm$ 0.12 & -0.23 $\pm$ 0.07  \\
		64 &  2.33 $\pm$ 0.08  & 1.89 $\pm$ 0.07 & 1.67 $\pm$ 0.17 & -0.09 $\pm$ 0.04  \\	
		72  &  0.40 $\pm$ 0.08 & 0.49 $\pm$ 0.09 & 0.26 $\pm$ 0.07 & -0.19 $\pm$ 0.13  \\
		73  & 1.20 $\pm$ 0.05 & 1.26 $\pm$ 0.06 & 0.72 $\pm$ 0.12 & -0.19 $\pm$ 0.07  \\	
		82  & - & 0.34 $\pm$ 0.05 &  0.16 $\pm$ 0.05 & -0.82 $\pm$ 0.13  \\	
		90  & 2.00 $\pm$ 0.15 &   1.81 $\pm$ 0.09 &  0.74 $\pm$ 0.12 & -0.35 $\pm$ 0.07 \\
		93 & 2.00 $\pm$ 0.14 & 2.28 $\pm$ 0.10 & 0.82 $\pm$ 0.13 &  -0.34 $\pm$ 0.07\\
		95 & 0.57 $\pm$ 0.06 & 0.62 $\pm$ 0.06 &  0.33 $\pm$ 0.06 &  -0.21 $\pm$ 0.08  \\
		99 &-  &  0.25 $\pm$ 0.07 & 0.11 $\pm$ 0.04 & -0.50 $\pm$ 0.16 \\
		\hline
	\end{tabular}
\tablefoot{Flux values were measured from \textsc{Taper5}, \textsc{Taper9} and \textsc{TaperALMA}. Source fluxes were detected from running the datasets through SEAC, set with thresholds of $\sigma_{f}$ = 3, $\sigma_{s}$ = 5. Spectral indices are derived by comparing the ATCA data, spilt into the two spectral windows, 5.5$\,$GHz and 9$\,$GHz, to data from the ALMA 100$\,$GHz observations. All data are tapered to contain the same range of u-v visibilities. When possible, core components of the sources are used in order for better source comparison between the differently resolved radio and millimetre observations.}
\end{table*}  

	\begin{table*}[h!]
	\caption{Table showing values of radio sources from the \textsc{FullConcat} dataset with no literature or ALMA counterparts.} 
	\label{table:atca_uncat}
	\centering
	\begin{tabular}{*4{p{3cm}}}
		\hline \hline
		HA19 Source ID & RA & DEC  & Flux$_{ATCA}$  (mJy)  \\ 	\hline
		3 &  16 46 55.96080 & -45 50 46.1824  & 0.12 $\pm$ 0.03 \\
		4 & 16 46 56.10430 & -45 50 47.6830 & 0.05 $\pm$ 0.01 \\
		5 & 16 46 56.22091 & -45 50 2.3835 & 0.27 $\pm$ 0.04 \\
		6 & 16 46 56.47741 & -45 50 51.5846 & 0.31 $\pm$ 0.03 \\
		8 & 16 46 56.88056 & -45 50 18.2862 & 0.08 $\pm$ 0.02 \\
		10 & 16 46 57.16625 & -45 50 58.7872 & 0.06 $\pm$ 0.02 \\
		11 & 16 46 57.59801 &  -45 50 25.7888 & 0.25 $\pm$ 0.04 \\
		12 & 16 46 59.20651 & -45 49 50.9937 & 0.21 $\pm$ 0.03 \\
		14 & 16 46 59.52162 & -45 50 19.1945 & 0.09 $\pm$ 0.03 \\
		16 & 16 47 0.38207 & -45 51 2.3964 & 0.25 $\pm$ 0.04 \\
		17 & 16 47 0.66916 & -45 51 5.3970 & 0.06 $\pm$ 0.02 \\
		23 & 16 47 1.61678 & -45 51 0.5984 & 0.10 $\pm$ 0.03 \\
		31 & 16 47 3.10981 & -45 51 17.3998 & 0.12 $\pm$ 0.03 \\
		32 & 16 47 3.25367 & -45 49 59.9998 & 1.21 $\pm$ 0.07 \\
		33 & 16 47 3.42563 & -45 51 36.5999  &0.13 $\pm$ 0.02 \\
		43 & 16 47 5.52210 & -45 51 37.7993 & 0.38 $\pm$ 0.04 \\
		44 & 16 47 5.69449  & -45 51 47.0992 & 0.07 $\pm$ 0.02 \\
		48 & 16 47 6.64016 & -45 49 8.6981 & 0.11 $\pm$ 0.03 \\
		50 & 16 47 7.04369 & -45 51 4.7975 & 0.32 $\pm$ 0.04 \\
		51 & 16 47 7.27263 & -45 50 17.6971 & 0.12 $\pm$ 0.02 \\
		54 & 16 47 7.67579  & -45 51 25.7963 & 0.32 $\pm$ 0.03 \\
		55 & 16 47 7.76142 & -45 50 58.4961 & 0.10 $\pm$ 0.03  \\
		56 & 16 47 7.87526 & -45 50 5.9959  & 0.49 $\pm$ 0.04 \\
		57 & 16 47 7.90541 & -45 51 19.7958 & 0.67 $\pm$ 0.05 \\
		58 & 16 47 7.93390 & -45 51 8.3958 & 0.29 $\pm$ 0.04 \\
		59 & 16 47 7.99108 & -45 50 56.0956 & 0.35 $\pm$ 0.04 \\
		60 & 16 47 8.07573 & -45 49 42.8954 & 0.10 $\pm$ 0.02 \\
		61 & 16 47 8.07736 & -45 51 2.9954 & 0.55 $\pm$ 0.04 \\
		62 & 16 47 8.33501 & -45 50 27.2949 & 0.61 $\pm$ 0.08 \\
		65 & 16 47 8.70722 & -45 49 44.3939 & 0.09 $\pm$ 0.02 \\
		67 & 16 47 8.90891 & -45 50 16.1934 & 0.07 $\pm$ 0.02  \\
		69 & 16 47 9.05354 & -45 50 59.3930 & 0.07 $\pm$ 0.02 \\
		70 & 16 47 9.08041 & -45 49 46.7929 & 0.19 $\pm$ 0.03 \\
		72 & 16 47 9.34005 & -45 50 35.9922 & 0.94 $\pm$ 0.05 \\
		73 & 16 47 9.56953 & -45 50 28.7915 & 0.63 $\pm$ 0.04 \\
		74 & 16 47 9.68496 & -45 50 49.7912 & 0.39 $\pm$ 0.03 \\
		76 & 16 47 10.17335 & -45 50 59.0896 & 0.21 $\pm$ 0.03  \\
		77 & 16 47 10.28717 & -45 50 26.0892  & 0.47 $\pm$ 0.04 \\
		78 & 16 47 10.43165 & -45 50 55.4887 & 1.63 $\pm$ 0.06 \\
		79 & 16 47 10.46117 & -45 51 20.3886 & 0.19 $\pm$ 0.03 \\
		81 & 16 47 10.83417 & -45 51 11.3872 & 0.08 $\pm$ 0.02 \\
		82 & 16 47 10.94981 & -45 51 33.8868 & 1.20 $\pm$ 0.06 \\
		83 & 16 47 11.17729 & -45 50 30.5859 & 0.55 $\pm$ 0.05 \\
		85 & 16 47 11.34994 & -45 50 41.3853 & 0.05 $\pm$ 0.01 \\
		86 & 16 47 11.52155 & -45 50 23.9846 & 0.32 $\pm$ 0.04 \\
		87 & 16 47 11.55173 &  -45 51 2.9844 & 0.14 $\pm$ 0.03 \\
		88 & 16 47 11.75329 & -45 51 17.3836 &0.14 $\pm$ 0.02 \\
		89 & 16 47 11.78273  & -45 51 35.9835 & 0.14 $\pm$ 0.02 \\
		90 & 16 47 11.86639 & -45 50 32.6831 & 0.11 $\pm$ 0.02 \\
		91 & 16 47 12.03810 & -45 50 18.8824  & 0.36 $\pm$ 0.04 \\
		92 & 16 47 12.09525 & -45 50 12.2821 & 0.84 $\pm$ 0.07 \\
		93 & 16 47 12.38490 & -45 51 14.0808 & 0.14 $\pm$ 0.03 \\
		94 & 16 47 14.33810 & -45 51 24.8708 & 0.24 $\pm$ 0.04 \\	
		\hline	
	\end{tabular}
	\tablefoot{These fluxes have been found from a run-through of the SEAC algorithm, without the use of the segmentation tool, with thresholds of $\sigma_{f}$ = 3 and $\sigma_{s}$ = 5. Source numbers are assigned to the definitive radio catalogue output of 94 radio sources detected in \textsc{FullConcat}, ordered by increasing RA.}
\end{table*} 

\clearpage
	\section{Figures}
	
	 \begin{figure*}[h!]
	\includegraphics[width=\textwidth]{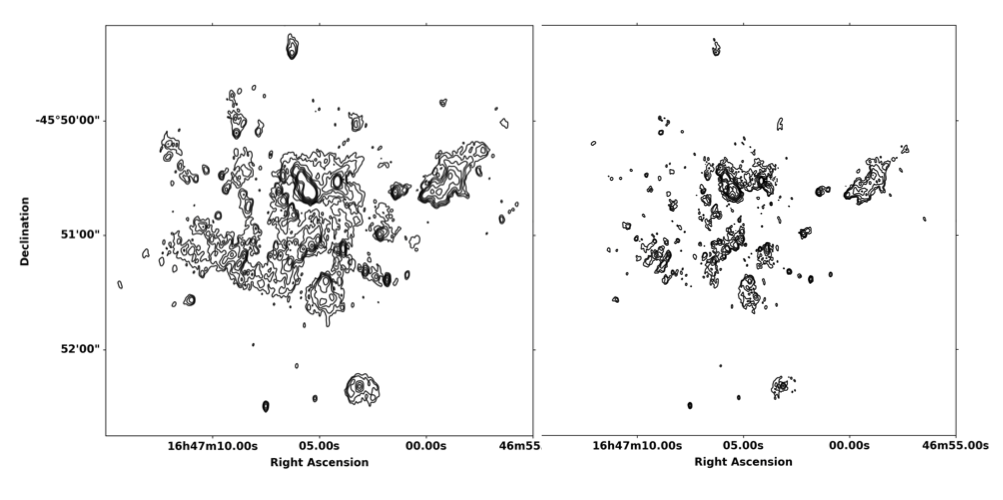}
	\caption{This figure shows an image of ATCA contours from the \textsc{Full5} dataset (\textit{left}), and the \textsc{Full9} dataset (\textit{right}), using non-primary beam corrected images. The contour levels are set at -3, 3, 6, 9, 12, 24, 28 and 192 $\times$ $\sigma$,where $\sigma$ is the average rms over the whole image. This value, $\sigma$, is set at 0.03$\,$mJy for the 5.5$\,$GHz data and 0.04$\,$mJy for the 9$\,$GHz data.}
	\label{fig:figB1}	
	\end{figure*}

	\begin{figure*}[h!]
	\includegraphics[width=\textwidth]{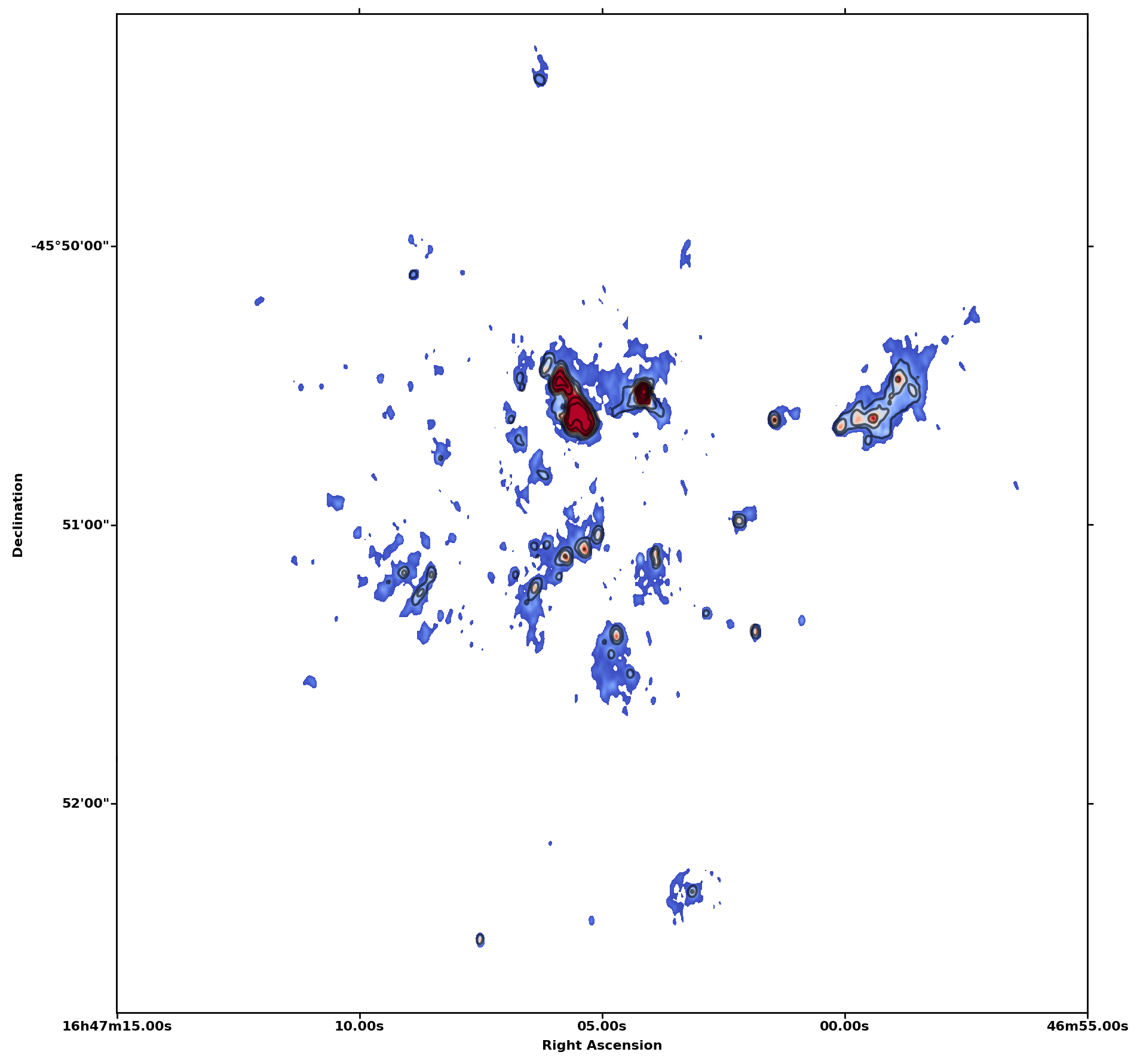}
	\caption{ATCA colour-scale of the \textsc{Full9} dataset (from the non-primary beam corrected image). Contours are overlaid from the 8.6$\,$GHz \citetalias{Dougher2010} observations. The colour scale range is from 0.12 - 2.0$\,$mJy beam$^{-1}$, and the contour levels are set at -3, 3, 6, 9, 12, 24, 48, 96, 192 $\times$ $\sigma$, where $\sigma$ is set at 0.6 mJy beam$^{-1}$, as used in \citetalias{Dougher2010}.}
	\label{fig:figB2}
	\end{figure*}
 		
    \begin{figure*}[h!]
     	\includegraphics[width=\textwidth]{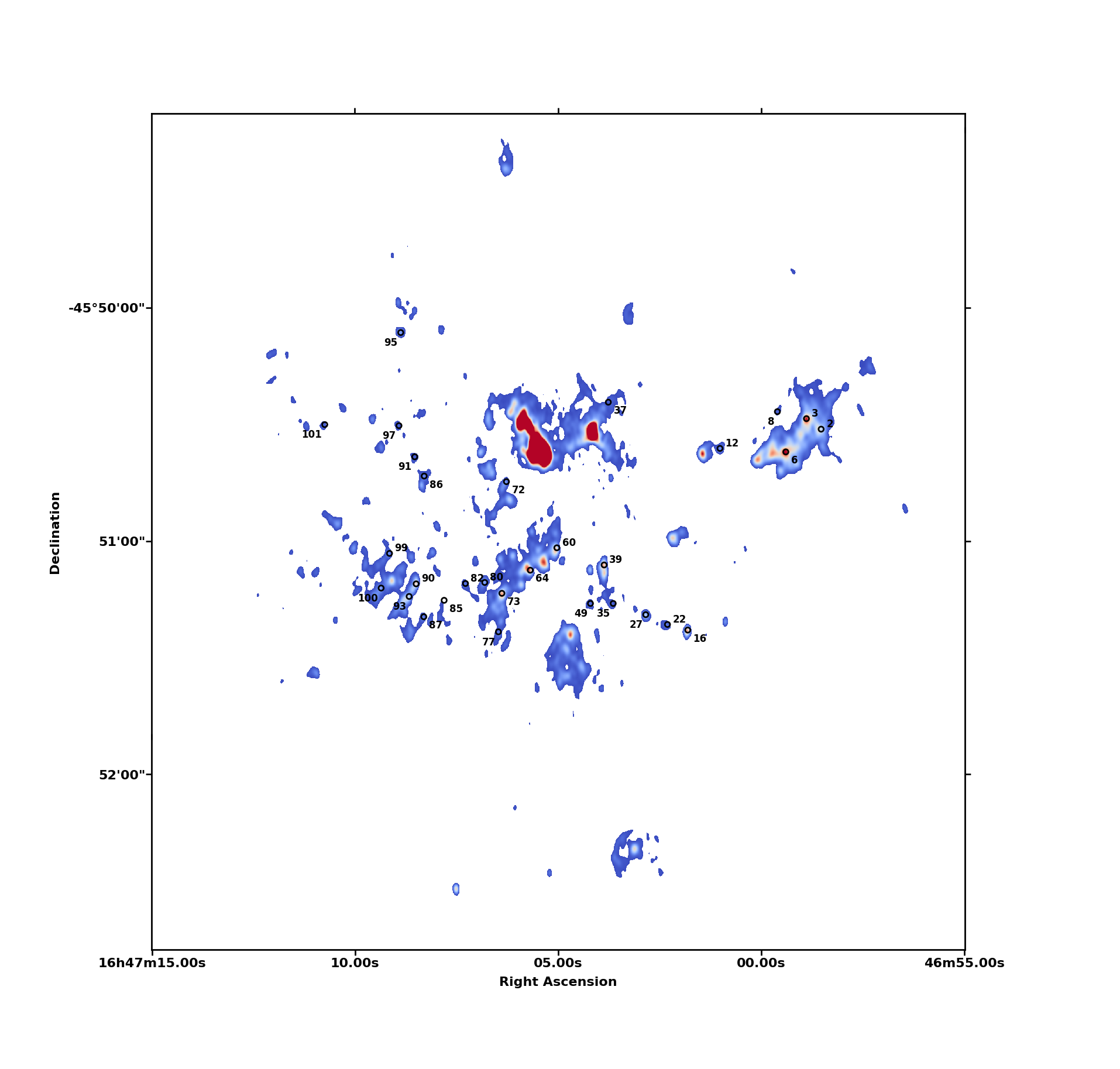}
     	\caption{ATCA colour-scale  image (from the non-primary beam corrected image) of the \textsc{FullConcat} dataset. The colour-scale is set from 0.1 - 2.0 mJy beam$^{-1}$. Unknown sources identified in the ALMA observations \citepalias{Fenech2018} were compared to SEAC detections from the ATCA dataset and sources that were detected in both are indicated with circles and labelled with the source number as given in \citetalias{Fenech2018}. Positions and fluxes are listed in \ref{table:alma_uncat}.}
     	\label{fig:figB3}
     \end{figure*}
    
   \begin{figure*}[h!]
     	\includegraphics[width=\textwidth]{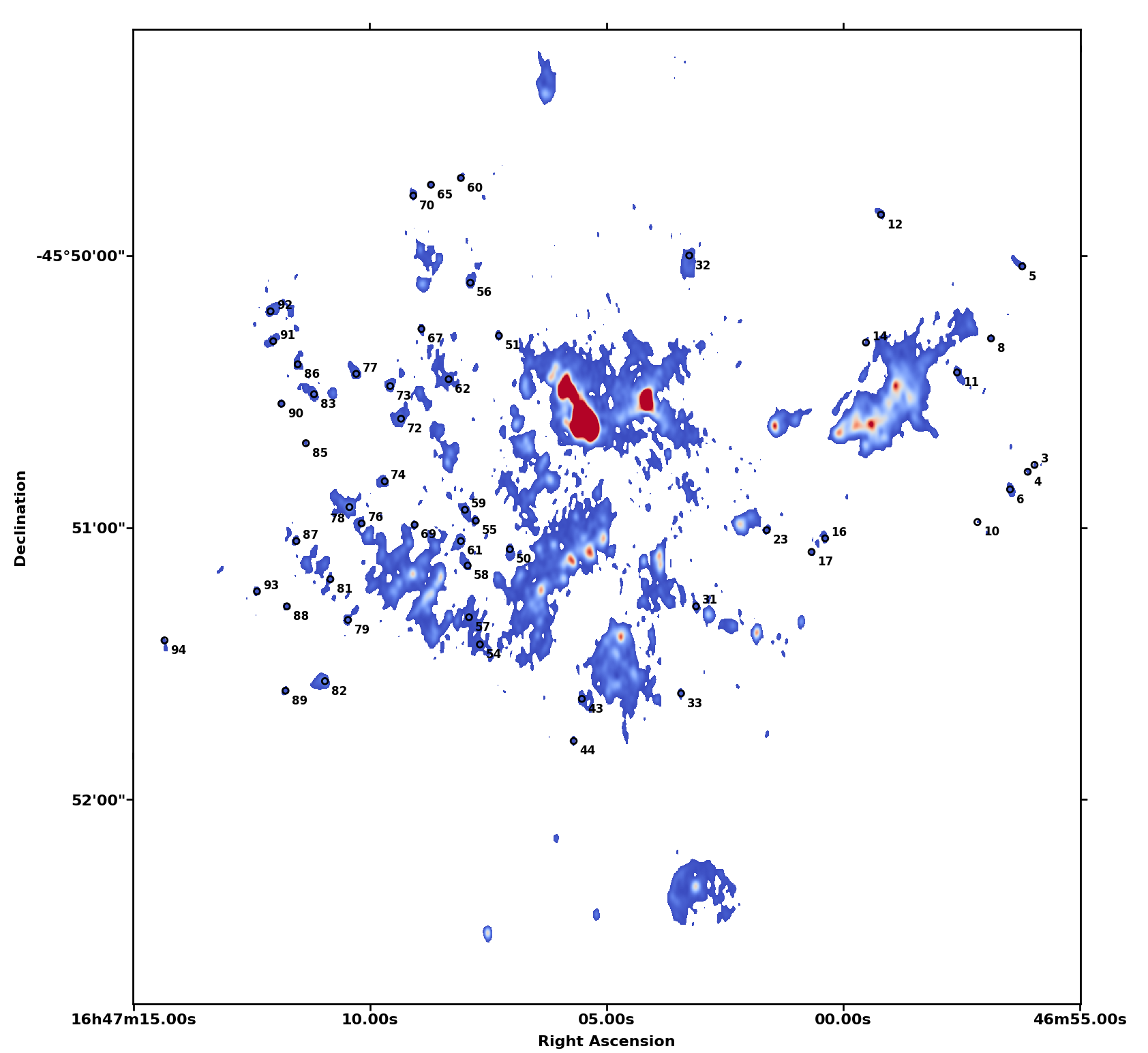}
     	 \caption{ATCA colour-scale image (from the non-primary beam corrected image) of the \textsc{FullConcat} dataset. The colour-scale is set from 0.1 - 2.0 mJy beam$^{-1}$. Unknown sources with no counterparts in the optical or the millimetre are are indicated with circles, and labelled. Positions and fluxes are listed in \ref{table:atca_uncat}.}
     	\label{fig:figB4}
    \end{figure*}

\onecolumn
	\begin{figure}
	\begin{subfigure}{0.32\textwidth}
		\resizebox{\hsize}{!}{\includegraphics{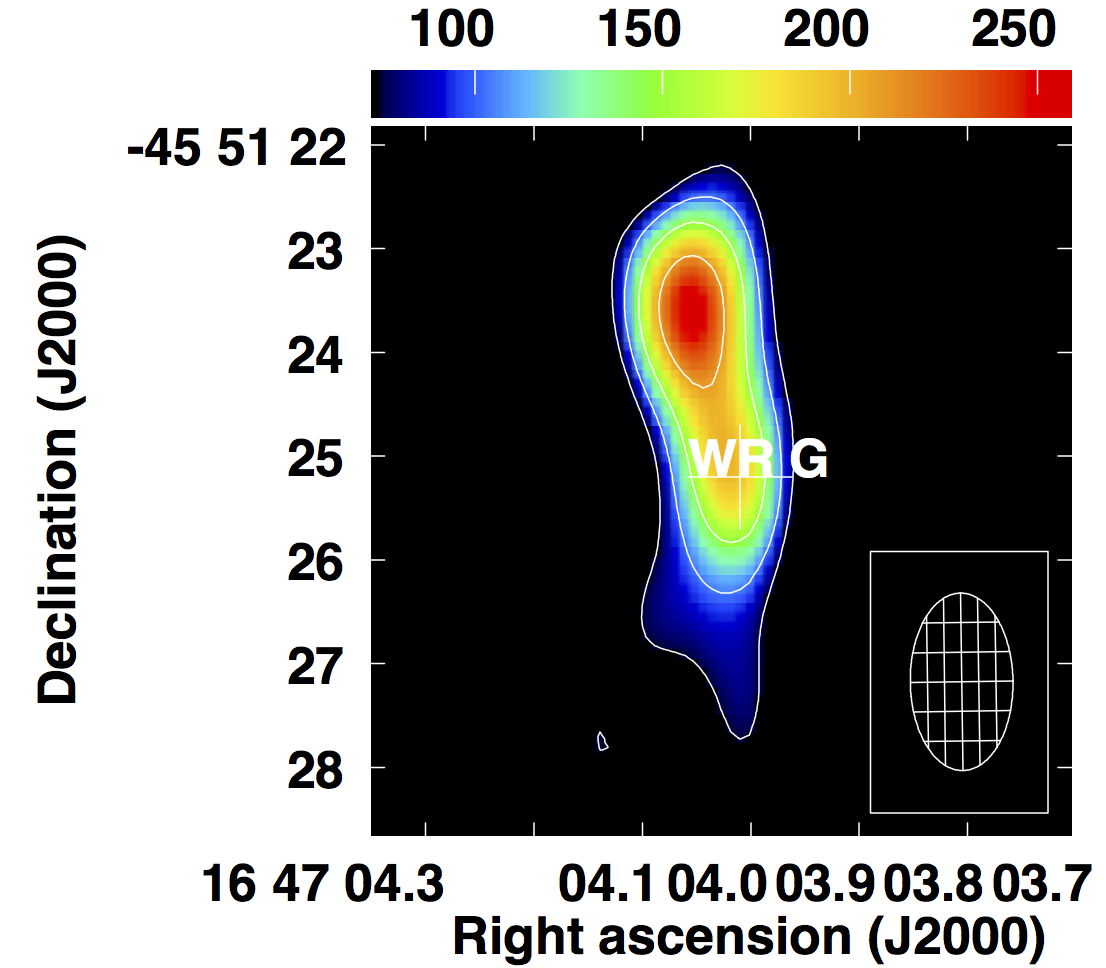}}
		\caption{WR G.}
		\label{fig:WRG_postage}
	\end{subfigure}
	\begin{subfigure}{0.34\textwidth}
		\resizebox{\hsize}{!}{\includegraphics{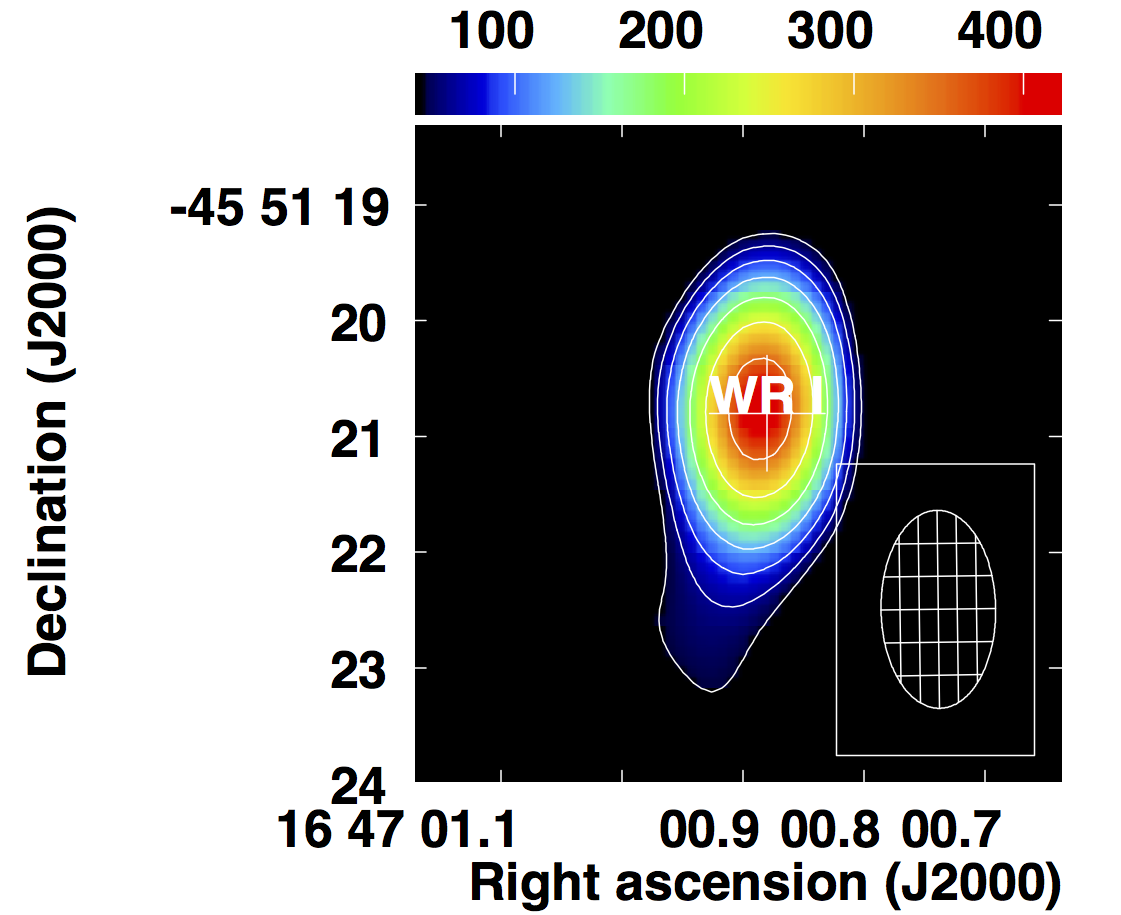}}
			\caption{WR I.}
	\label{fig:WRI_postage}
\end{subfigure}	
	\begin{subfigure}{0.34\textwidth}
	\resizebox{\hsize}{!}{\includegraphics{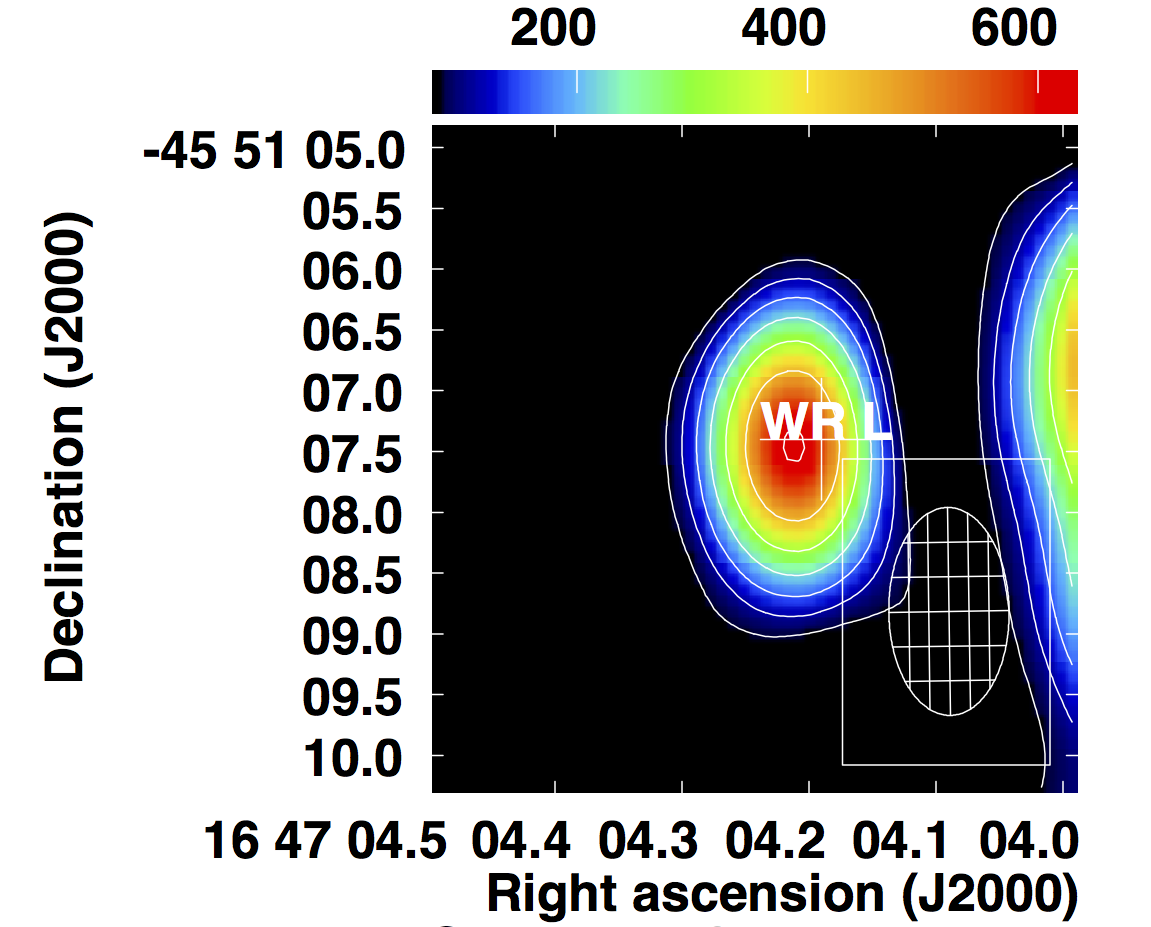}}
	\caption{WR L.}
	\label{fig:WRL_postage}
\end{subfigure}	
	\begin{subfigure}{0.34\textwidth}
	\resizebox{\hsize}{!}{\includegraphics{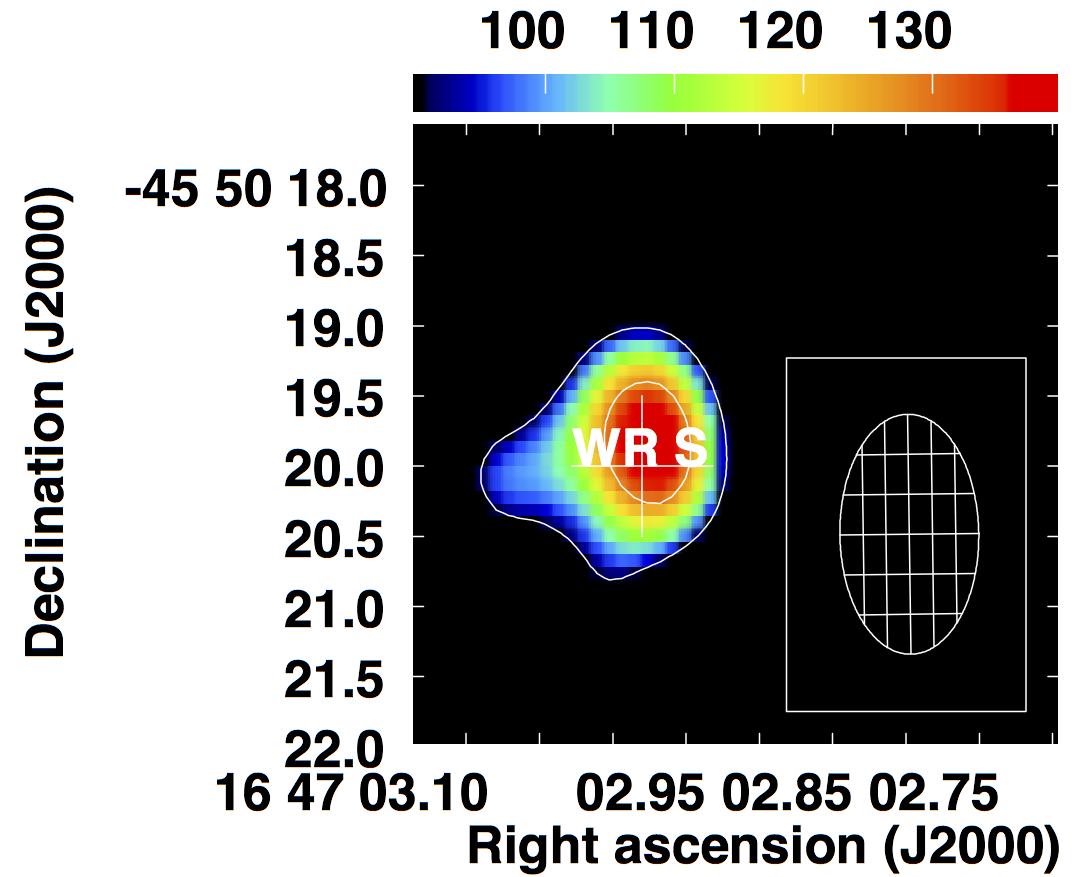}}
	\caption{WR S.}
	\label{fig:WRS_postage}
\end{subfigure}	
	\begin{subfigure}{0.32\textwidth}
	\resizebox{\hsize}{!}{\includegraphics{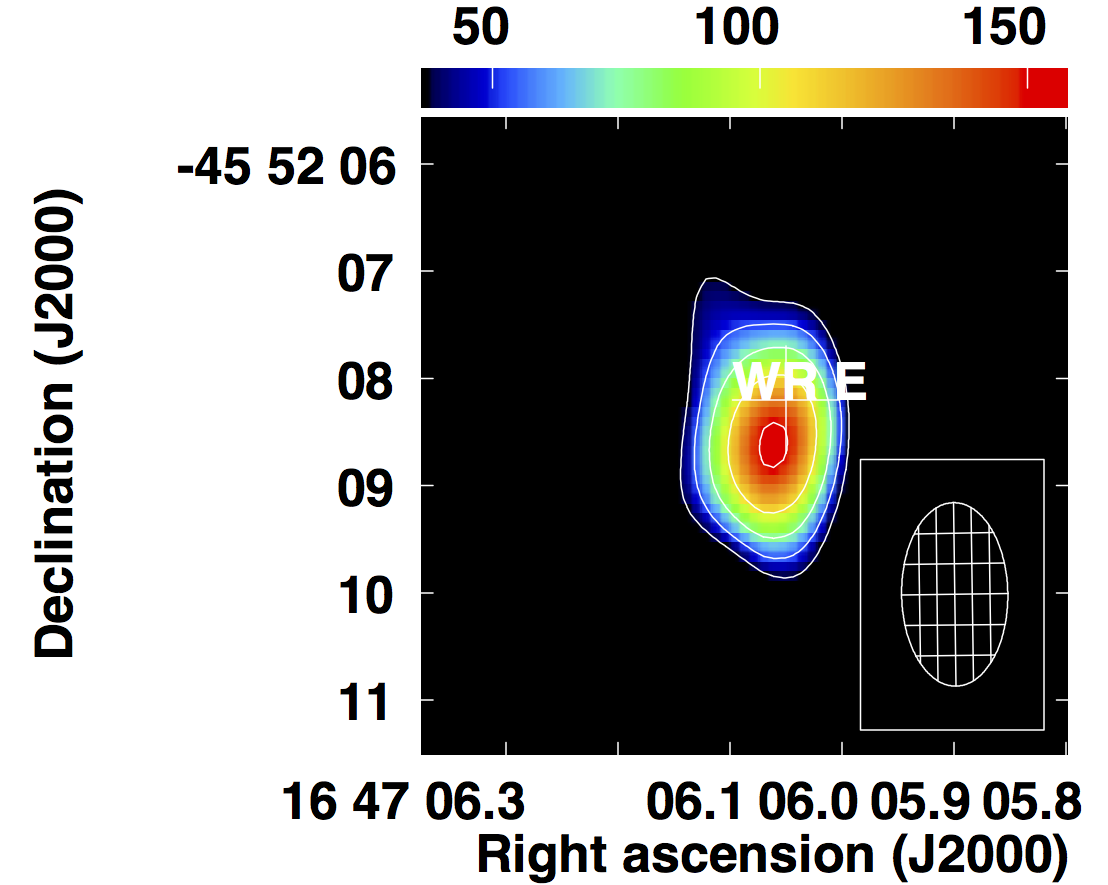}}
	\caption{WR E.}
	\label{fig:WRE_postage}
\end{subfigure}	
	\begin{subfigure}{0.32\textwidth}
	\resizebox{\hsize}{!}{\includegraphics{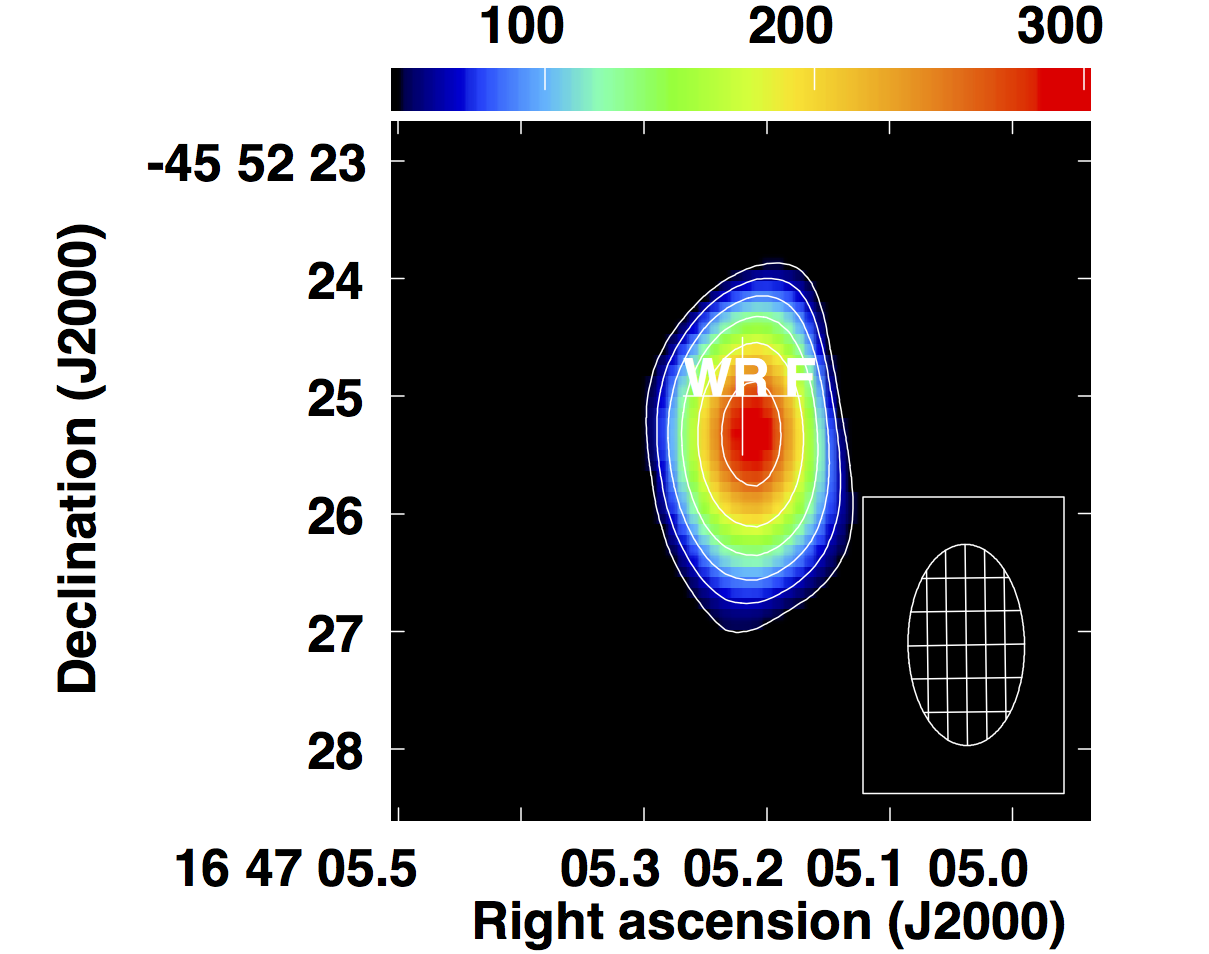}}
	\caption{WR F.}
	\label{fig:WRF_postage}
\end{subfigure}	
	\begin{subfigure}{0.33\textwidth}
	\resizebox{\hsize}{!}{\includegraphics{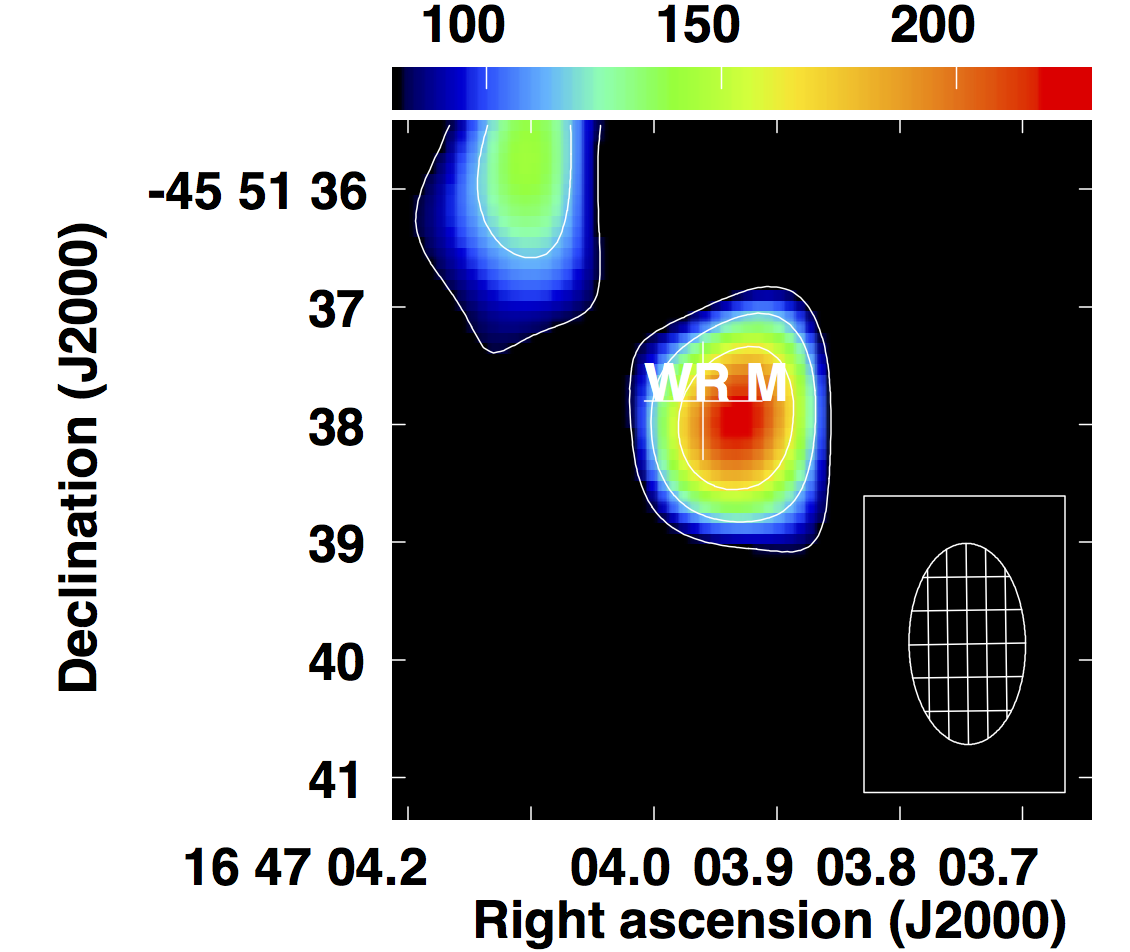}}
	\caption{WR M.}
	\label{fig:WRM_postage}
\end{subfigure}
	\begin{subfigure}{0.33\textwidth}
	\resizebox{\hsize}{!}{\includegraphics{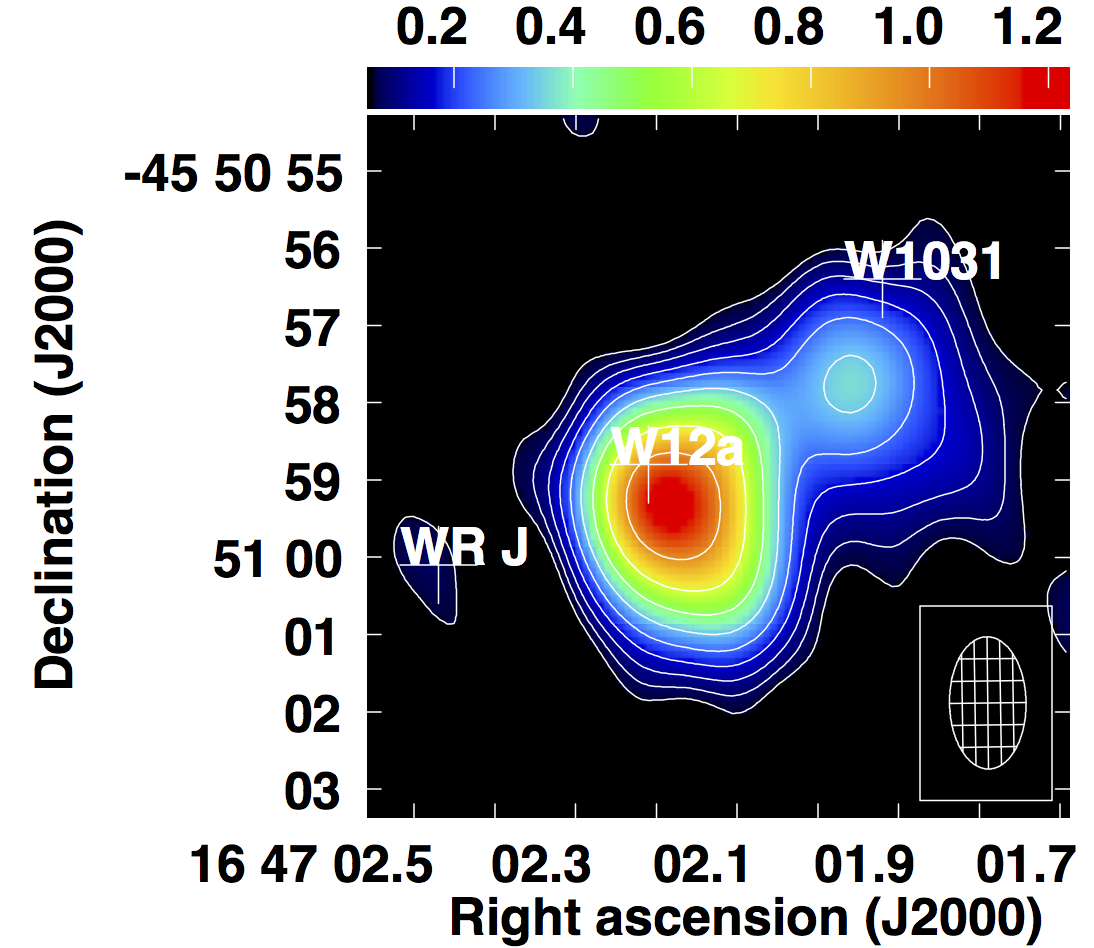}}
	\caption{W12a.}
	\label{fig:W12a_postage}
\end{subfigure}
	\begin{subfigure}{0.33\textwidth}
	\resizebox{\hsize}{!}{\includegraphics{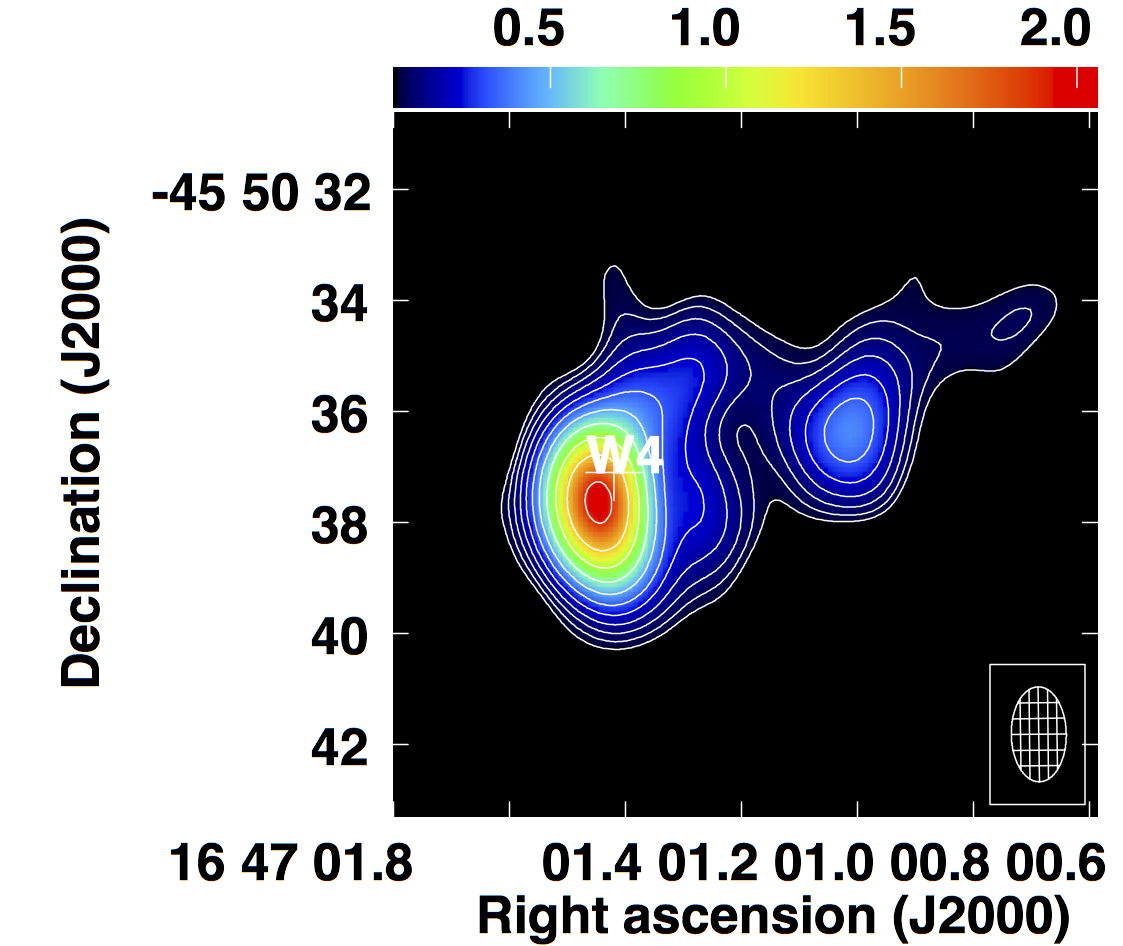}}
	\caption{W4a.}
	\label{fig:W4a_postage}
\end{subfigure}
	\begin{subfigure}{0.33\textwidth}
	\resizebox{\hsize}{!}{\includegraphics{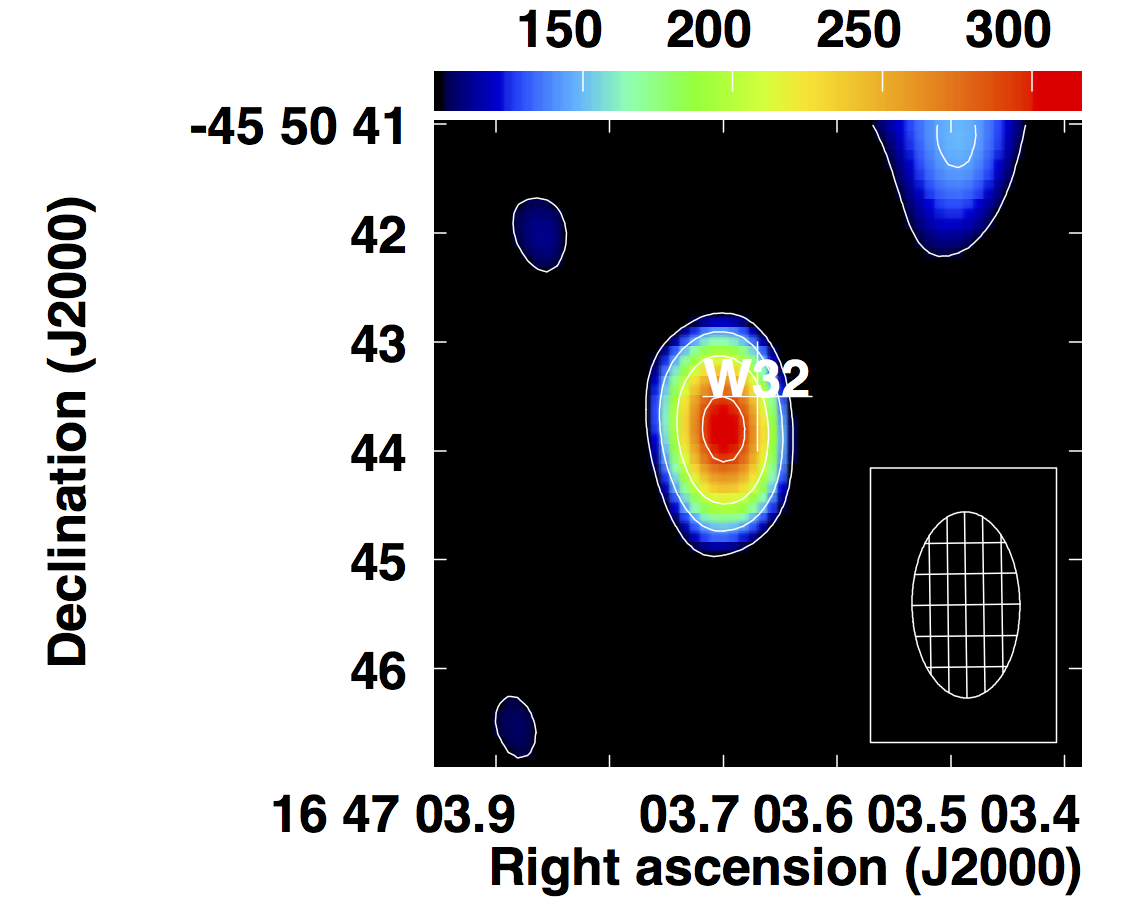}}
	\caption{W32.}
	\label{fig:W32_postage}
\end{subfigure}
	\begin{subfigure}{0.33\textwidth}
	\resizebox{\hsize}{!}{\includegraphics{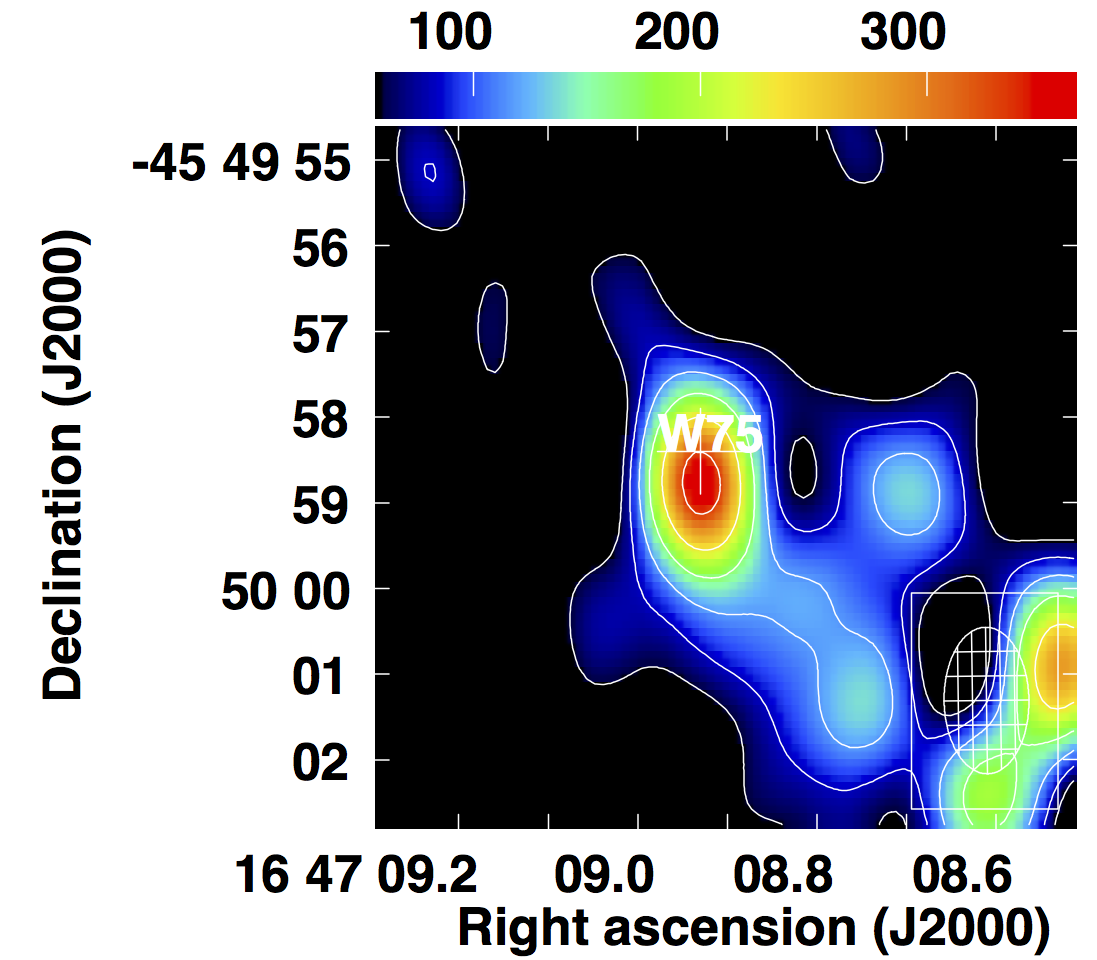}}
	\caption{W75.}
	\label{fig:W75_postage}
\end{subfigure}
\begin{subfigure}{0.33\textwidth}
	\resizebox{\hsize}{!}{\includegraphics{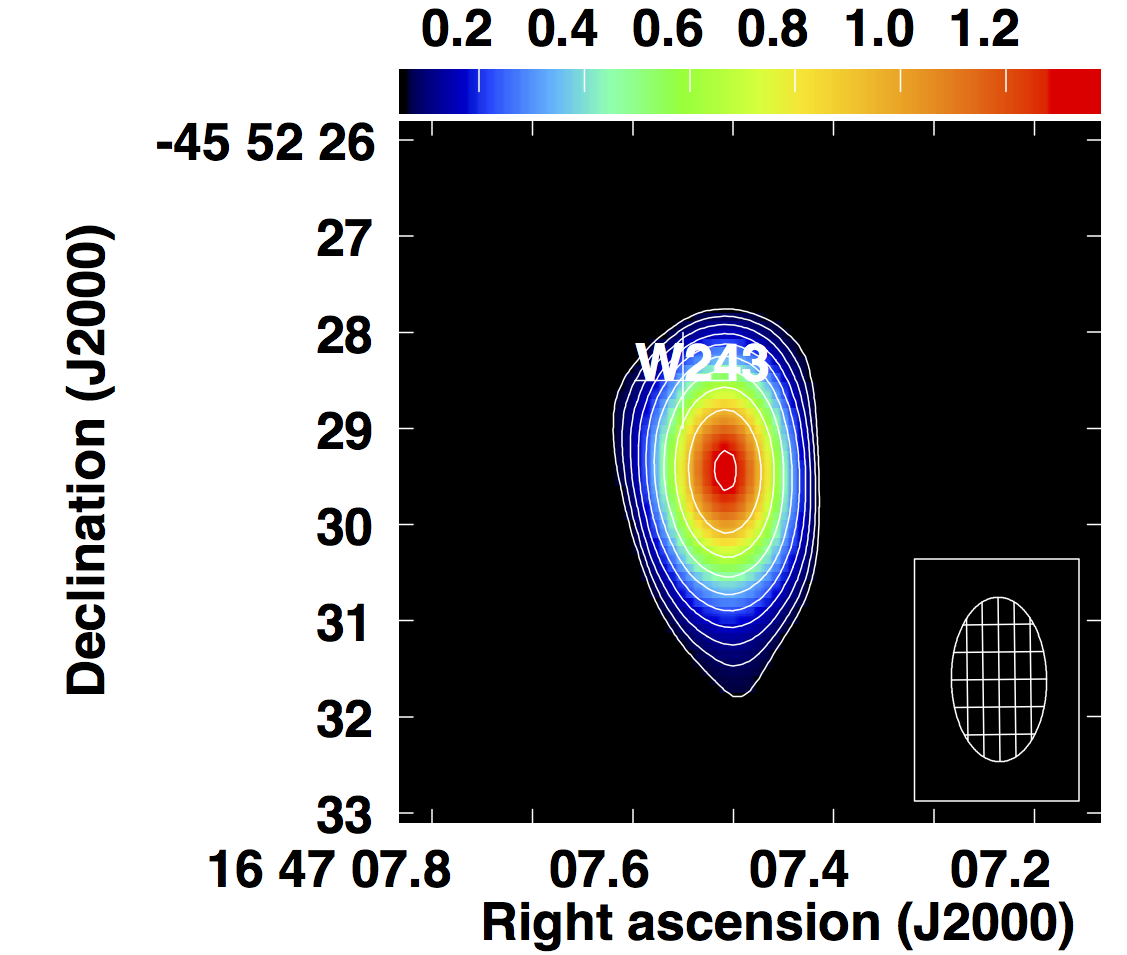}}
	\caption{W243.}
	\label{fig:W243_postage}
\end{subfigure}
	\label{fig:post_images1}
\end{figure}
\begin{figure}\ContinuedFloat
		\vspace*{-1cm}
	\begin{subfigure}{0.33\textwidth}
	\resizebox{\hsize}{!}{\includegraphics{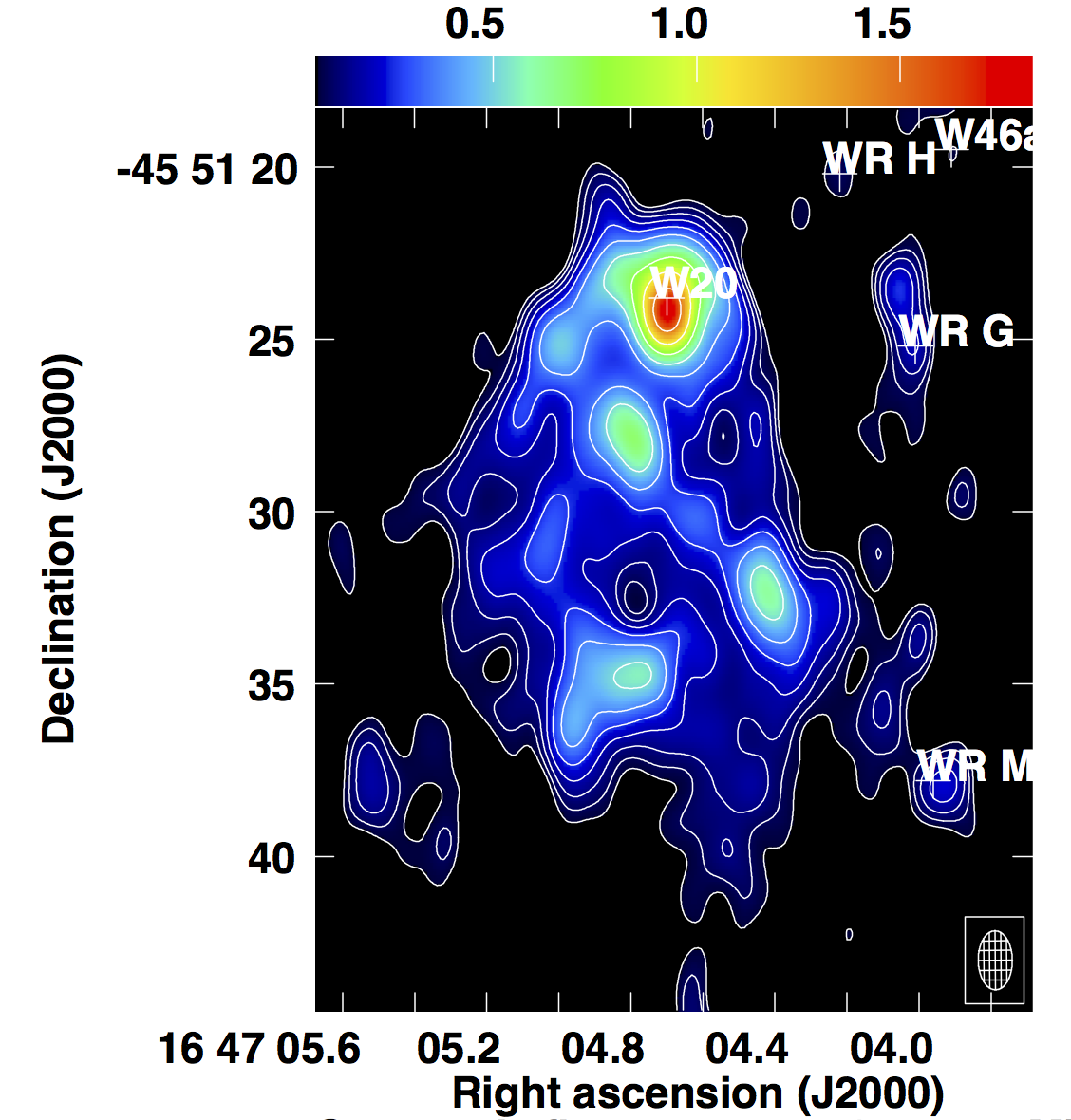}}
		\caption{W20.}
		\label{fig:W20_postage}
	\end{subfigure}
	\begin{subfigure}{0.32\textwidth}
		\resizebox{\hsize}{!}{\includegraphics{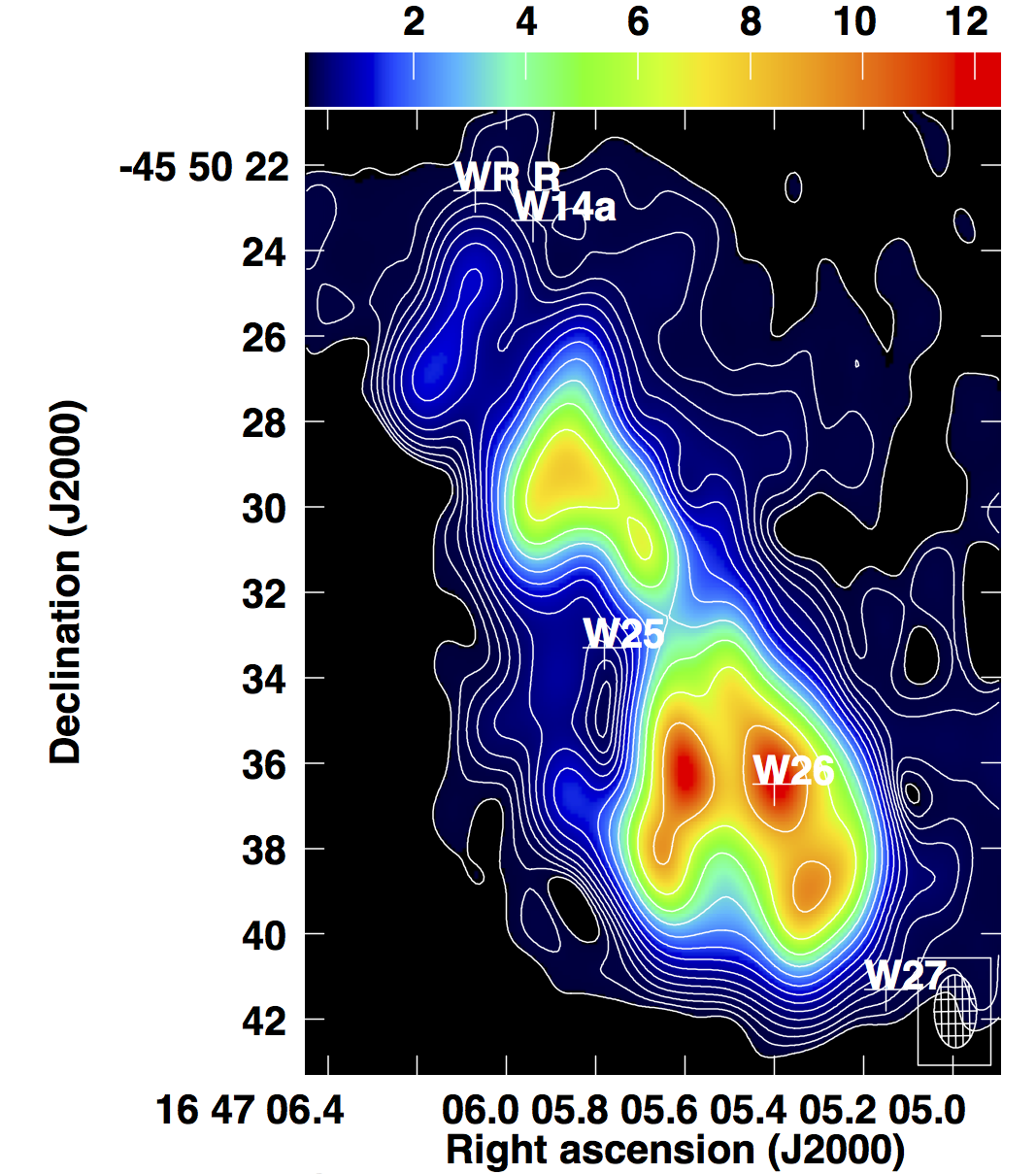}}
		\caption{W26}
		\label{fig:W26_postage}
	\end{subfigure}
	\begin{subfigure}{0.33\textwidth}
	\resizebox{\hsize}{!}{\includegraphics{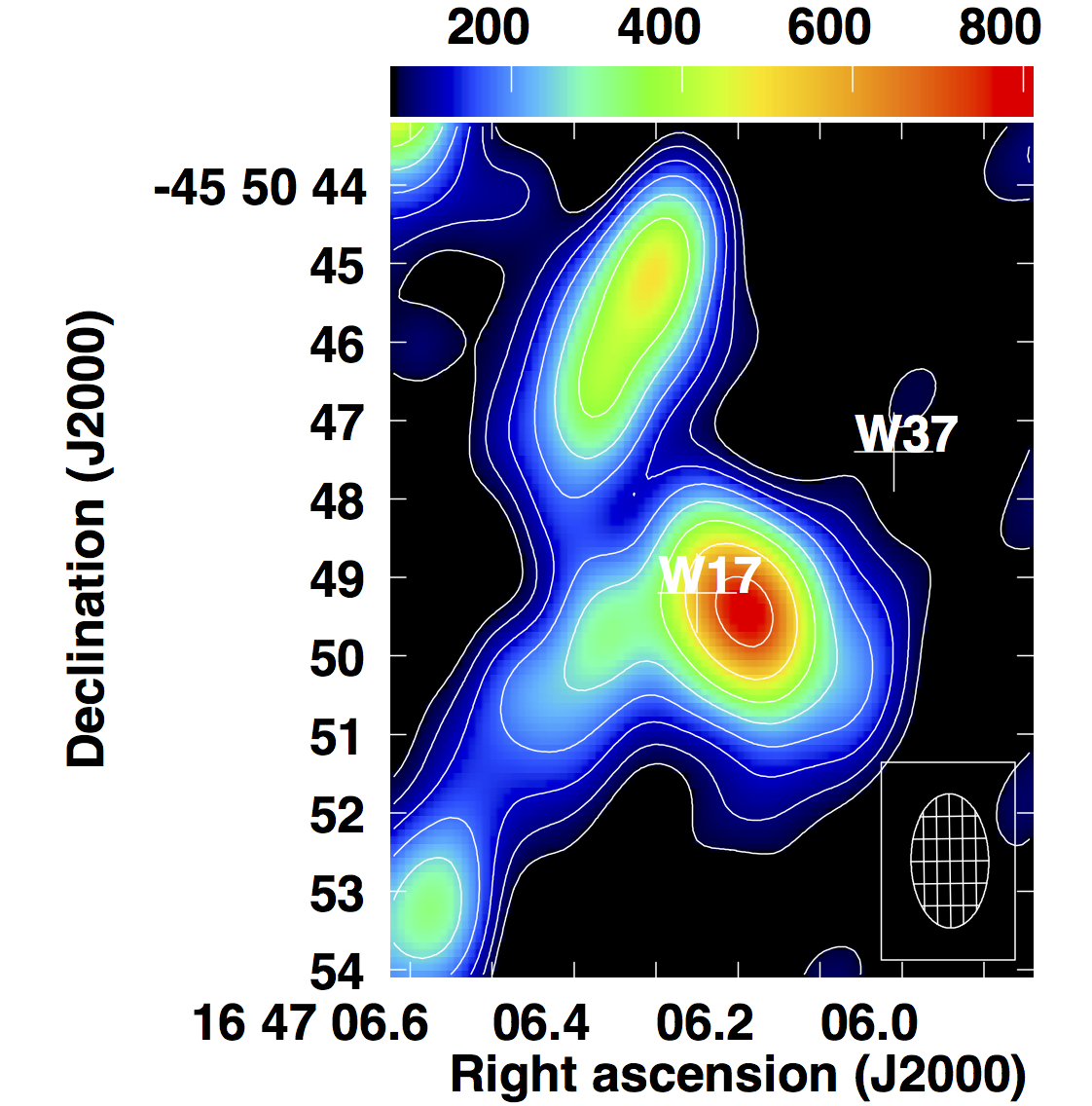}}
	\caption{W17}
	\label{fig:W17_postage}
\end{subfigure}
	\begin{subfigure}{0.33\textwidth}
	\resizebox{\hsize}{!}{\includegraphics{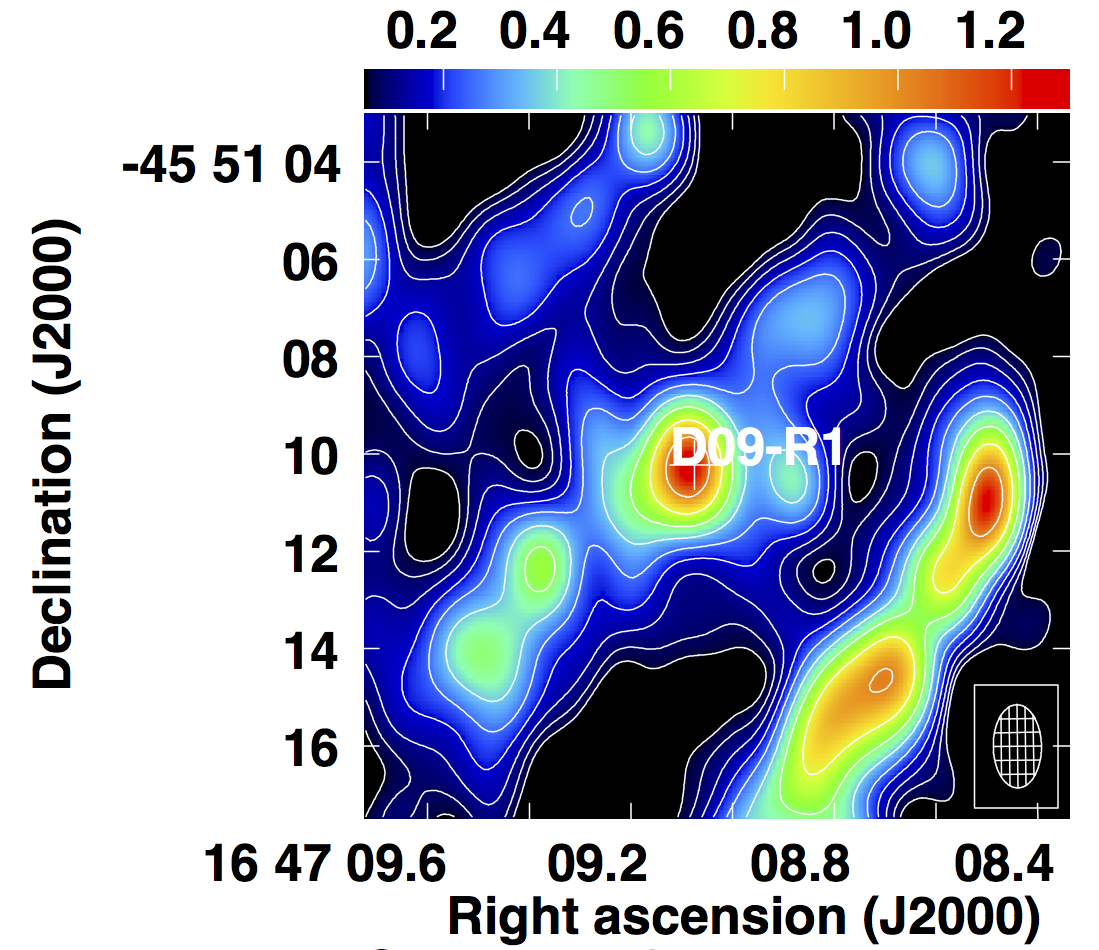}}
	\caption{D09-R1.}
	\label{fig:D09-R1_postage}
\end{subfigure}
		\begin{subfigure}{0.33\textwidth}
		\resizebox{\hsize}{!}{\includegraphics{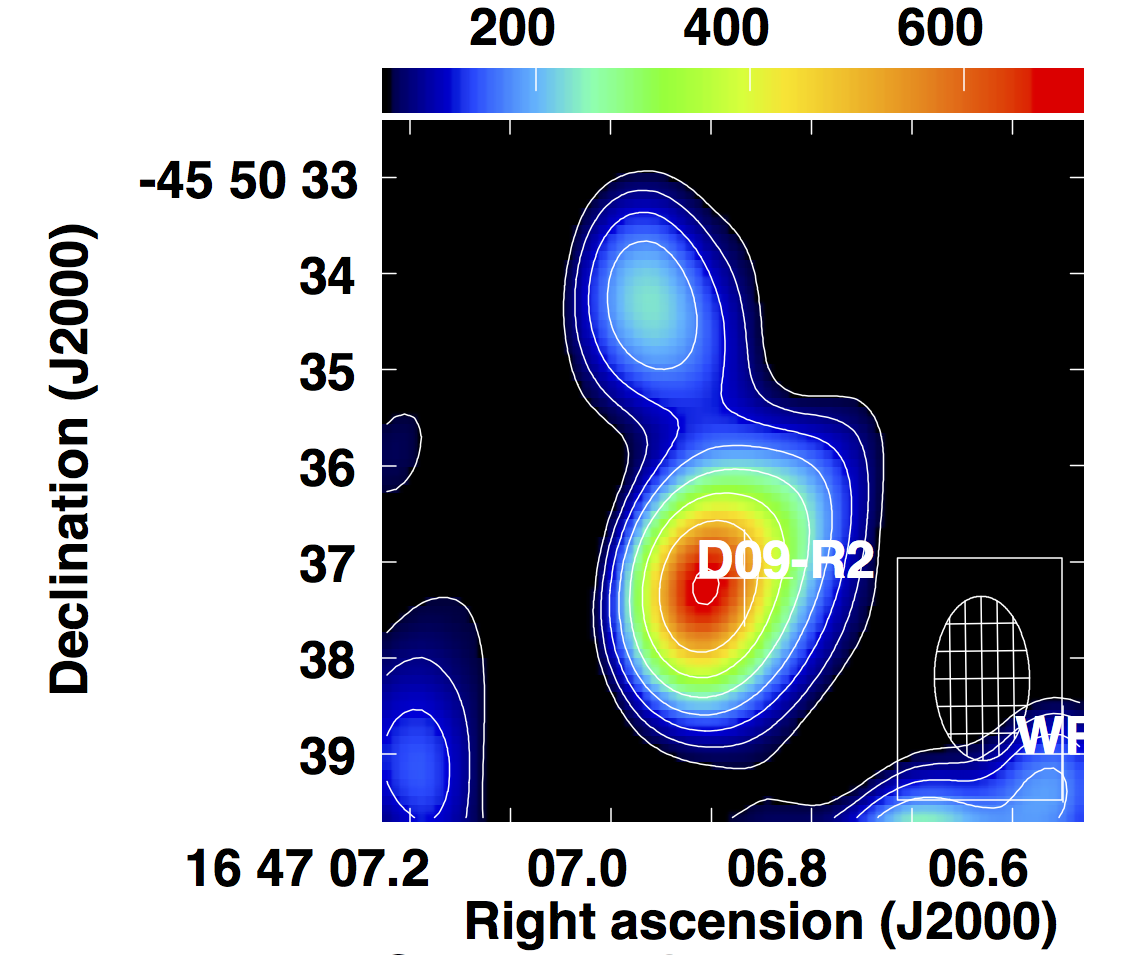}}
		\caption{D09-R2.}
		\label{fig:D09-R2_postage}
\end{subfigure}
		\begin{subfigure}{0.33\textwidth}
	\resizebox{\hsize}{!}{\includegraphics{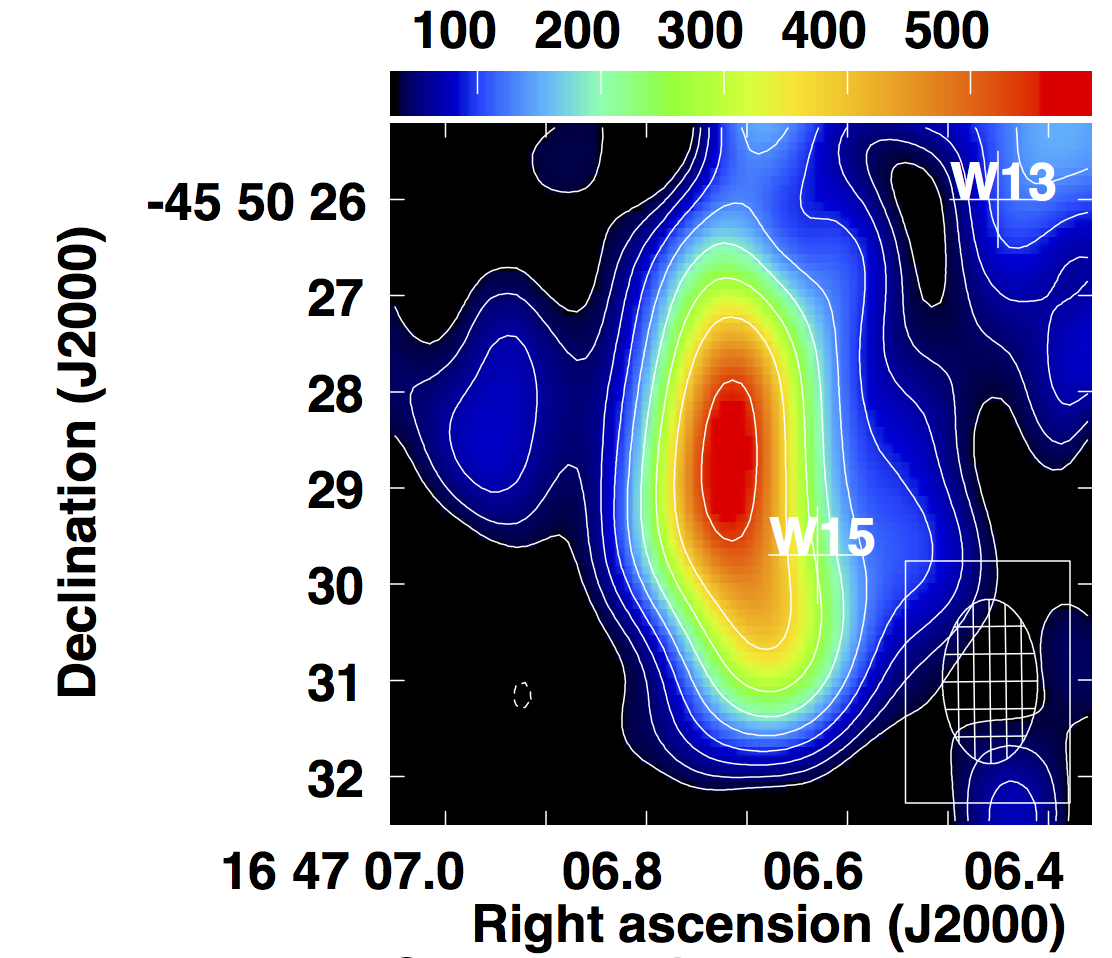}}
	\caption{W15.}
	\label{fig:W15_postage}
\end{subfigure}
		\begin{subfigure}{0.33\textwidth}
	\resizebox{\hsize}{!}{\includegraphics{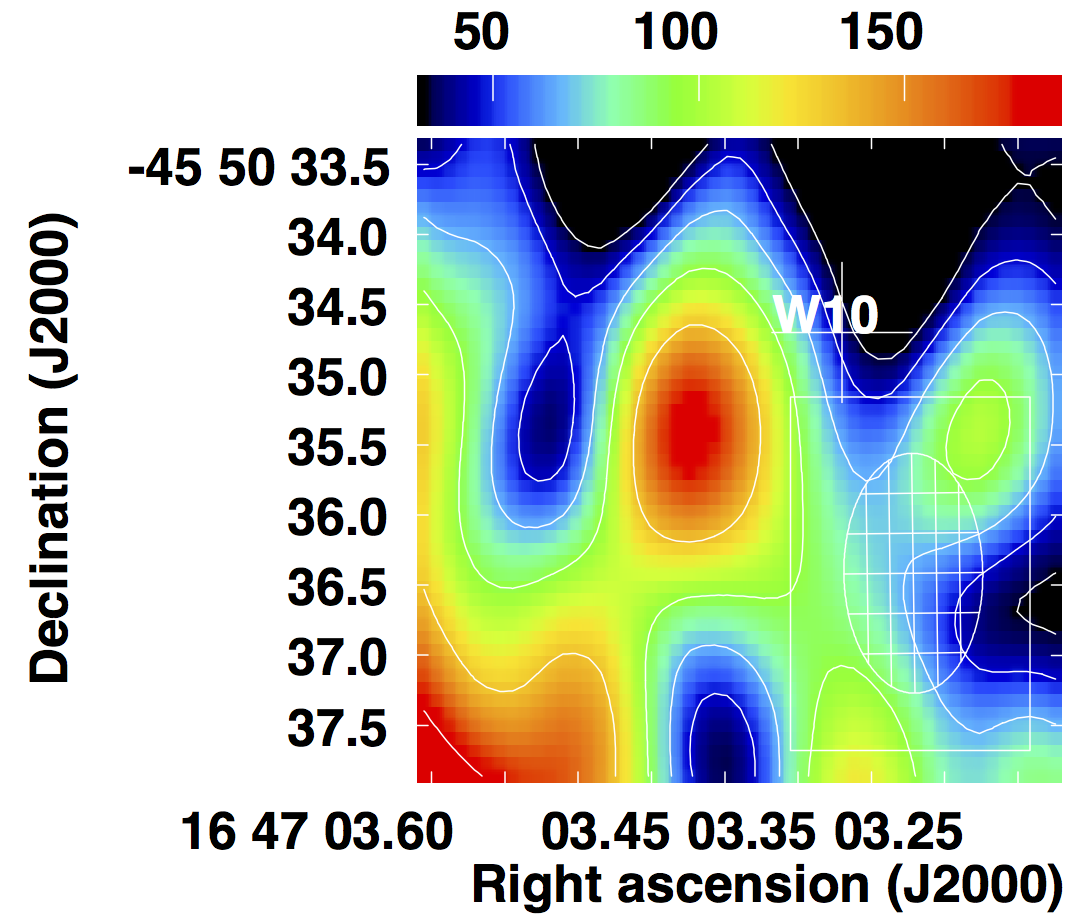}}
	\caption{W10.}
	\label{fig:W10_postage}
\end{subfigure}
		\begin{subfigure}{0.33\textwidth}
	\resizebox{\hsize}{!}{\includegraphics{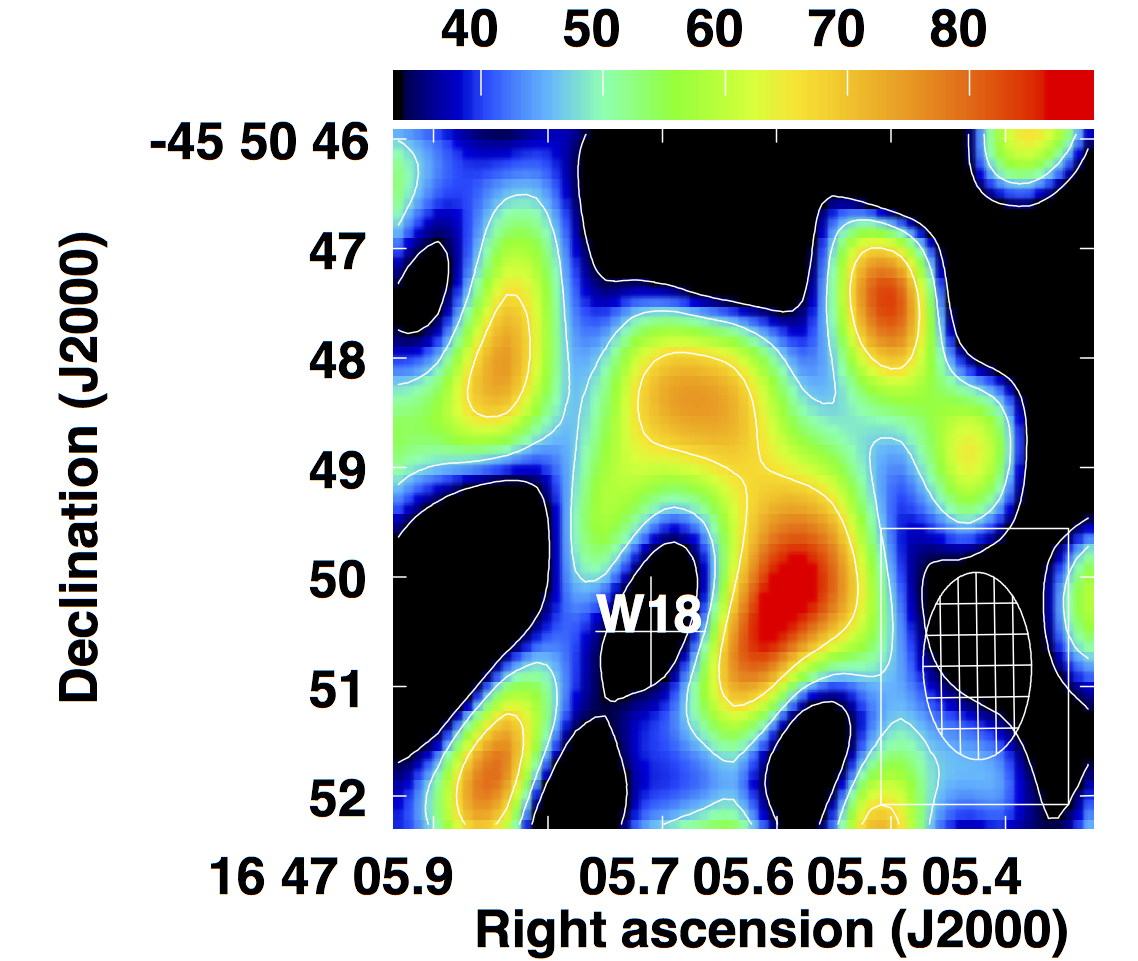}}
	\caption{W18.}
	\label{fig:W18_postage}
\end{subfigure}
		\begin{subfigure}{0.33\textwidth}
	\resizebox{\hsize}{!}{\includegraphics{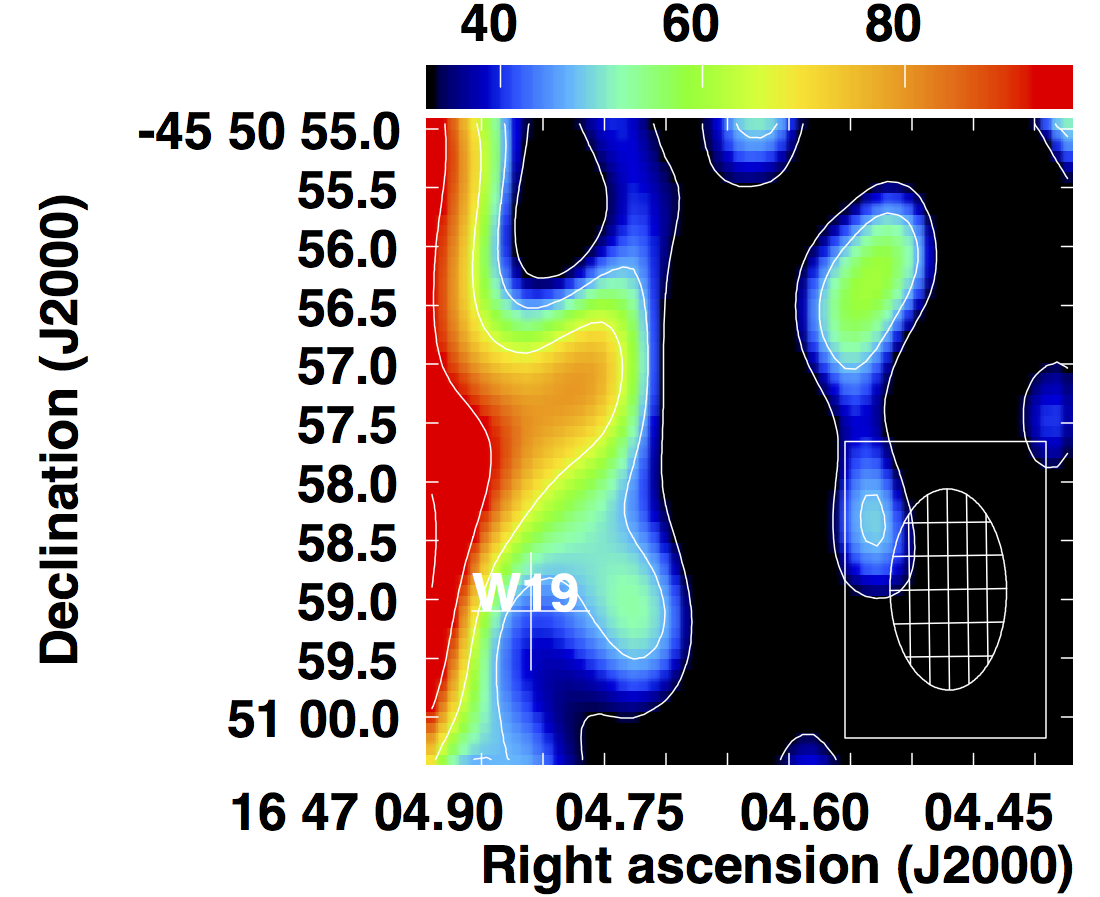}}
	\caption{W19.}
	\label{fig:W19_postage}
\end{subfigure}
		\begin{subfigure}{0.33\textwidth}
	\resizebox{\hsize}{!}{\includegraphics{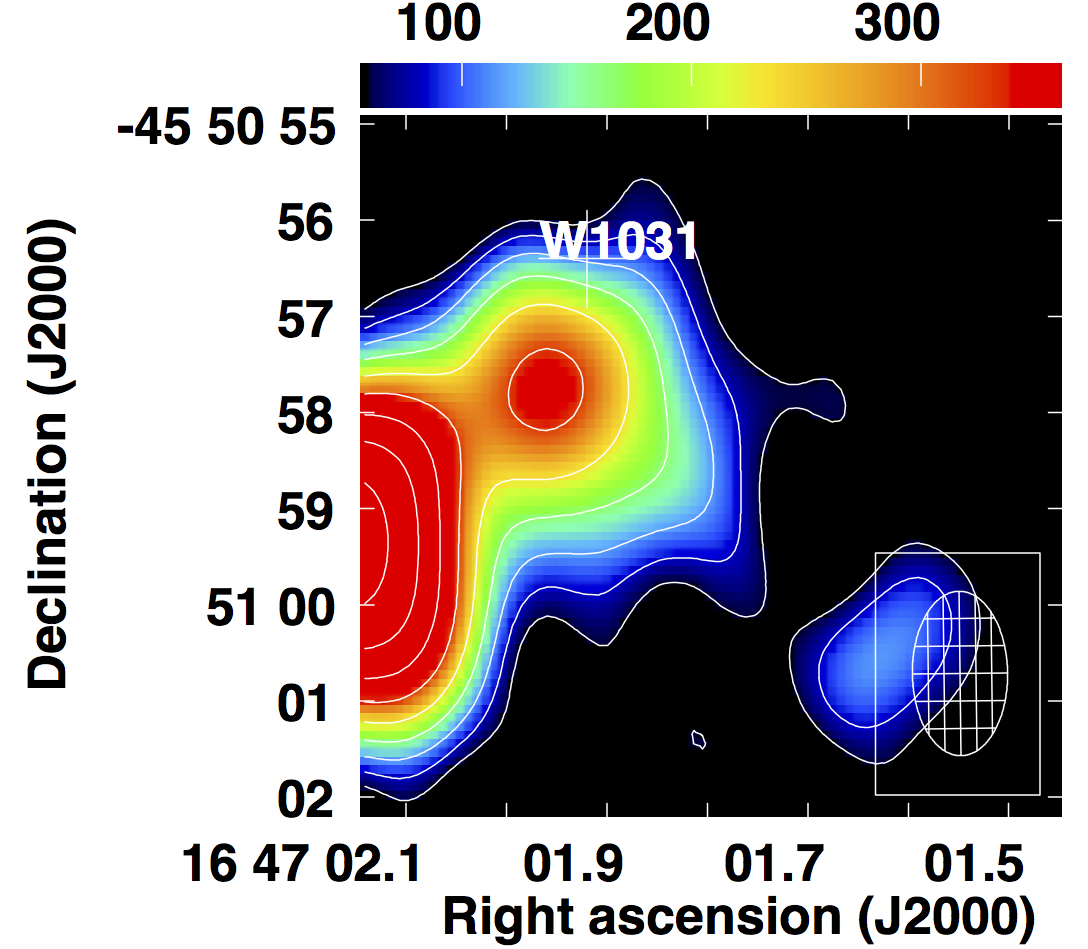}}
	\caption{W1031.}
	\label{fig:W1031_postage}
\end{subfigure}
		\begin{subfigure}{0.36\textwidth}
	\resizebox{\hsize}{!}{\includegraphics{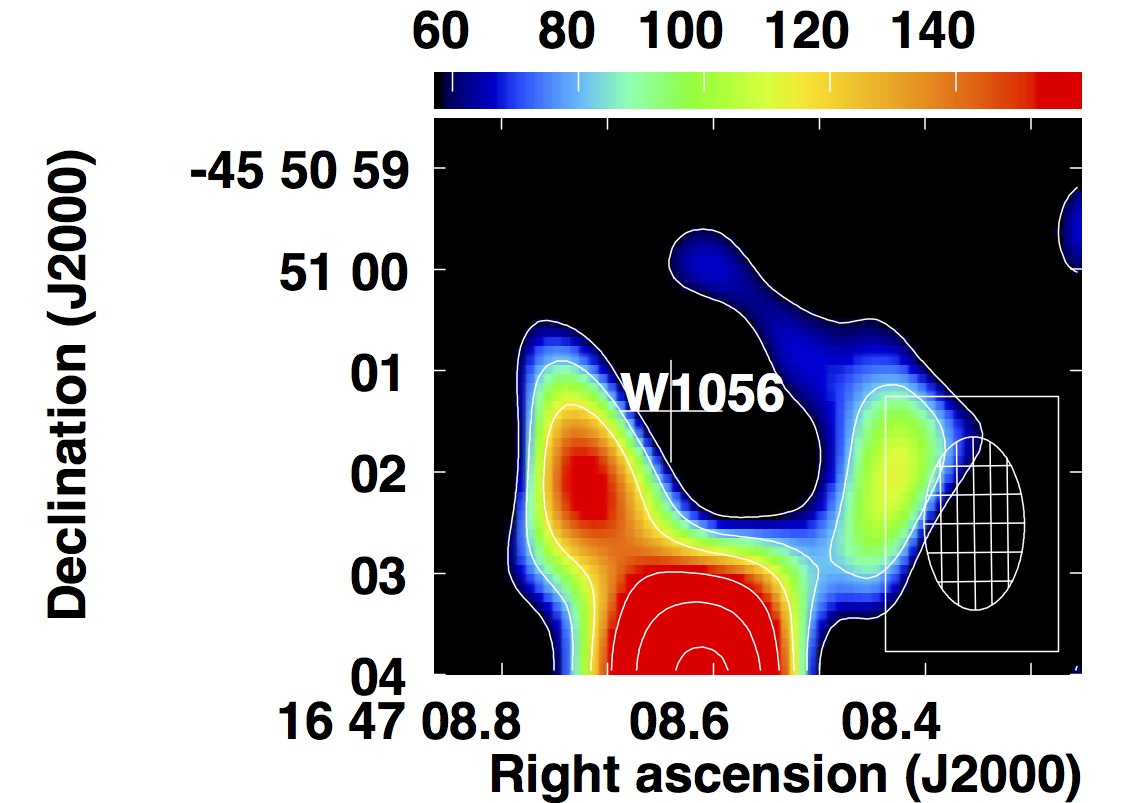}}
	\caption{W1056.}
	\label{fig:W1056_postage}
\end{subfigure}
	\label{fig:post_images2}
	\caption{ATCA colour-scale images of the stellar sources detected, from the non-PB corrected image of the \textsc{FullConcat} dataset. Contours are plotted at 1,1,1.4,2,2.8,4,5.7,8,11.3,16,22.6,32,45.25,64,90.50 $\times$ $\sigma$. }
\end{figure}
 
\end{document}